\documentclass{article}
\usepackage{blindtext}
\usepackage{multicol}
\usepackage{amsmath, amsthm,graphicx,amsfonts,amssymb,mathrsfs,showidx, multicol}

\usepackage{scalerel}

\usepackage{rotating}
\usepackage{pdflscape}
\usepackage{tabulary}

\usepackage{array}
\usepackage{booktabs}  

\newcolumntype{C}[1]{>{\centering\let\newline\\\arraybackslash\hspace{0pt}}m{#1}}

\usepackage{booktabs}
\usepackage{blkarray}
\usepackage{hyperref}
\hypersetup{
    colorlinks=true,
    linkcolor=blue,
    filecolor=blue,      
    urlcolor=cyan,
    citecolor=blue,
}

\usepackage{graphicx}
\graphicspath{ {images/} }

\usepackage[]{todonotes}

\usepackage{tcolorbox}
\definecolor{mycolor}{rgb}{0.122, 0.435, 0.698}

\newtcbox{\mybox}{on line,
  colframe=mycolor,colback=mycolor!10!white,
  boxrule=0.5pt,arc=4pt,boxsep=0pt,left=6pt,right=6pt,top=6pt,bottom=6pt}

\usepackage{relsize}
\usepackage{color}
\definecolor{rojo}{rgb}{1,0,0}
\usepackage[all]{xy}
\makeindex
\usepackage{hyperref}
\usepackage{color}
\definecolor{abelian}{cmyk}{0.50,0,1,.4}
\definecolor{noabelian}{cmyk}{0.94,0.54,0,0}
\definecolor{rojo}{cmyk}{0,1,1,0}
\definecolor{verde}{cmyk}{0.91,0,0.88,0.12}
\xyoption{arc}

\usepackage{tikz}
\usetikzlibrary{tqft}

\newtheorem{thm}{Theorem}
\newtheorem{theorem}{Theorem}

\newtheorem{proposition}[thm]{Proposition}
\newtheorem{lemma}[thm]{Lemma}
\newtheorem{cor}[thm]{Corollary}

\theoremstyle{definition}
\newtheorem{definition}[thm]{Definition}
\newtheorem{example}[thm]{Example}

\newtheorem{assum}[thm]{Assumption}

\theoremstyle{remark}

\newtheorem{remark}[thm]{Remark}

\newcommand{\op}{\operatorname}
\newcommand{\ma}{\mathcal}
\newcommand{\ben}{\begin{equation}}
\newcommand{\een}{\end{equation}}
\newcommand{\bena}{\begin{equation*}}
\newcommand{\eena}{\end{equation*}}
\newcommand{\To}{\longrightarrow}

\title{Petri nets in epidemiology}
\author{Carlos Segovia\thanks{Instituto de Matem\'aticas UNAM-Oaxaca: csegovia@matem.unam.mx}}

\begin{document}
\maketitle

\begin{abstract}
This work provides a geometric version of the next-generation matrix method for obtaining the basic reproduction number of an epidemiological model. 
We exhibit a certain correspondence between any system of ODEs and Petri nets. 
We observe that any epidemiological model has the basic structures found in the SIR model of Kermack-McKendrick.
This means that the basic reproduction number depends only on three substructures inside the Petri net, which are also given by three Petri nets inside, representing the susceptible population, the infection process, and the infected population. 
The five assumptions of the next-generation matrix method given by van den Driessche-Watmough can be described geometrically using Petri nets. Thus, the next-generation matrix results in a matrix of flows between the infection compartments with a dominant eigenvalue given by the basic reproduction number. 
\end{abstract}

\section{Introduction}\label{sec1}
The present work is a geometric point of view of the next-generation matrix method \cite{DHM90,DW02,DW08}. Instead of analyzing a specific epidemiological model described by a system of ordinary differential equations (ODEs), we consider an underlying geometric structure that encodes the system of ODEs by the diagrammatic structure of a Petri net \cite{Reisig}. 
The theory of stochastic Petri nets in the book of 
Baez--Biamonte \cite{Ba}, associates to any Petri net, together with a rate function, a unique system of ODEs by means of what is called the rate equation. 
In this work, we develop an inverse map for this assignment. 
More precisely, we can construct a Petri net and a rate function for any system of ODEs such that the rate equations recover the original system of ODEs.

A  Petri net consists of compartments or places  (represented by circles), transitions  (represented by squares), and arrows between compartments and transitions, together with a rate function that assigns a parameter to each transition.
We are interested in Petri net modeling the course of a pandemic, which can be highly sophisticated. The more important contribution of our work is interpreting the basic reproduction number $R_0$ in terms of Petri nets.
This comprises identifying three structures inside any Petri net in epidemiology, called the Kermack--McKendrick modules: susceptible, infection-process, and infection (shortly the KM modules). The five assumptions developed by van den Driessche--Watmough \cite{DW02,DW08} to show that $R_0$ can be defined as the dominant eigenvalue of the next generation matrix, are interpreted using the KM modules. We obtain a clearer form of the functions that define the matrices $F$ and $V$ of the next-generation matrix $FV^{-1}$. The initial motivation was a geometric interpretation of the coordinates of $FV^{-1}$ mentioned in \cite[pag. 33]{DW02}: the $(i,k)$ entry of the product $FV^{-1}$ is the expected number of new infections in compartment $i$ produced by the infected individual originally introduced into compartment $k$.

We can use Petri nets to interpret the coordinates matrix $FV^{-1}$ by summing along all the paths of flows in the Petri net satisfying certain conditions between the KM modules.
The basic reproduction number is the dominant eigenvalue of $FV^{-1}$, which can be found analytically, or there is a geometric procedure \cite{Camino-Lewis,Camin-Lewis-Driessche} where a geometric Gaussian elimination is performed. If $FV^{-1}$ has rank one, the basic reproduction number is the trace, which sums along all the expected secondary infections for loops inside the infection module that traverse once the infection-process module. 
This corresponds to the fertility loops of the life cycle graph in \cite{Rueffler-Metz} in the case $FV^{-1}$ has a single non-zero eigenvalue.

The proposal for working with Petri nets in epidemiology is to give a proposed set of disease dynamics with the following components: first, the distinction of the different populations in compartments. Second, we identify the various interactions between the populations where two (or more) populations get into a transition to transform into another population. In epidemiology, these interactions are the coexistence of a vulnerable population with sick populations transforming into another population (usually, they get into the infection compartment). 
Third, we identify the bifurcation of a population in a transition to different types of populations (sometimes they have assigned weights denoting the proportion or the probability).
From there, we identify which compartments are endowed with a demographic model (with birth and death transitions) and which have an extinction transition.   

The present paper is written for two types of persons: 1) people who want to calculate the basic reproduction number quickly and must verify the five assumptions of the next-generation matrix method. The five geometric assumptions proposed in our work can be an easier alternative to check.  
Two directions make our approach very attractive. We can start by developing a Petri net model of our epidemiological phenomenon, and after the application of the rate equation, we end with the associated system of ODEs. However, we can start with a specific system of ODEs, then we apply the algorithm developed in this work and obtain a representable Petri net. 2) Experts in mathematical epidemiology search in different directions for developing essential concepts, such as the basic reproduction number, herd immunity, among others, in terms of other mathematical structures. Petri nets are a promising direction using possible tools of graph theory and stochastic methods, and our work could be the starting point of a new theory where several problems, not only in epidemiology, can be understood in terms of Petri nets. The present work differs from the numerical equivalence between Petri net-based models and their classical ODE counterparts studied in \cite{trevor1}. Also, the proposal of our work differs from the approach in \cite{Heinerr,HeinerrArx}, where the Petri nets modeling construction uses the spatio-geographic information among other stakeholders. Our work was built in the foundational spirit of \cite{DW02,DW08}.

This article is organized as follows: in Section \ref{secPN}, we give the prerequisites of the theory of Petri nets, where we introduce the rate equation of a Petri net to obtain the associated system of ODEs. We present several Petri nets for plenty of epidemiological models, such as SIR, SIS, SEIR, SEAIR, SCIR, SIWR, SIQR, and models for Malaria and vaccinations.
In Section \ref{inverseODE} we present an algorithm that finds a representable Petri net for any system of ODEs. 
In Section \ref{secNGM}, we explain the next-generation matrix method for a system of ODEs of an epidemiological model. The five assumptions of van den Driessche-Watmough are presented, and we give examples of calculating the basic reproduction number using this analytic method. 
Section \ref{R0PN} is the most relevant part of this work. 
In Section \ref{GeoNGM}, we define the KM modules and provide the geometric versions of the first four assumptions of the next generation matrix method. We delay the five assumptions for Section \ref{threshold}, where we show in Theorem \ref{teoremon} that the reproduction number computed provides a threshold for an outbreak in the Petri net model. 
We provide examples of applying the geometric procedure for the SEAIR, Malaria, and the two vaccination models. 
In Section \ref{Graph}, we compare our work with the graph-theoretic method in \cite{Camino-Lewis,Camin-Lewis-Driessche}. We found a very useful formula in \cite{Rueffler-Metz} for a next-generation matrix with a single non-zero eigenvalue that confirms our interpretation of $R_0$ as the sum of expected secondary infections along loops inside the infection module that traverses once the infection module.  
At the end of the paper, we include the $F$ and $V$ matrices, the next-generation matrix, and the basic reproduction number for all the models studied in Section \ref{secPN}.


\section*{Acknowledgment} 
We thank Professor Jorge X. Velasco for relevant conversations regarding this work and Professor Pauline van den Driessche for comments and suggestions. 
We especially thank the software Snoopy \cite{Snoopy} for its data structures and software dependability, which enabled the diagrammatic structures of Petri nets. We thank the anonymous reviewer for suggesting the work of de Camino-Beck-Lewis, which confirms our geometric interpretation of $R_0$ and the suggestion of including the disease dynamics previous the construction of the Petri net.
The author is supported by Investigadores por M\'exico SECIHTI.

\section{Petri nets}
\label{secPN}
Petri nets are promising methods for modeling and simulating biological systems \cite{Reisig,Ba}.
Carl Adams Petri designed them \cite{Smith}, a German mathematician dedicated to computer science who lived in the time of general relativity of Einstein. For instance, the locality of time is one of the most essential advantages of Petri nets.
The Ph.D. thesis of Petri \cite{Petri} was forward-thinking and innovative, initiating the theory of Petri nets with fundamental ideas such as sequence, conflict, and concurrency. Indeed the translation to English of this thesis \cite{Petri2} was classified information by the United States Air Force. 


Generalizations of different types of Petri nets exist in the literature depending on the needs of the problem; see \cite{DPetriNet}. To mention a few of them, we have colored Petri nets with tokens taken values in different data types; see \cite{Jensen}. Continuous Petri nets have the feature that the marking of a place is a real (positive) number and no longer an integer representing the number of tokens. Nested Petri nets are an extension of the classic Petri nets where each compartment, transition, and token can be a Petri net.  
The precise definition of a Petri net is as follows:

\begin{definition}
A {\it Petri net} consists of a set of compartments $S=\{x_1,\cdots,x_k\}$ and a set of transitions $T=\{z_1,\cdots,z_l\}$, together with a  finite set of arrows $\ma{A}$ with a matching function 
$f:\ma{A}\longrightarrow S\times T\sqcup T\times S\,.$
The arrows are divided into two disjoint subsets $A=A_1\sqcup A_2$ with $f(A_1)\subset S\times T$ (arrows from compartments to transitions) and $f(A_2)\subset T\times S$ (arrows from transitions to compartments). For $x_i\in S$ and $y_j\in T$, we set:
\begin{itemize}
    \item by $m_{i,j}=|f^{-1}(x_i,z_j)|$ the number of arrows with source $x_i$ and target $z_j$, and 
    \item by $n_{i,j}=|f^{-1}(z_j,x_i)|$ the number of arrows with source $z_j$ and target $x_i$.
\end{itemize}
\end{definition}

\subsection{The rate equation}

An important application of the theory of Petri nets is the connection with ordinary differential equations (ODEs). 
ODEs are used in many contexts in sciences describing real-life models like physics, medicine, etc., see \cite{arnold}.
The connection of Petri nest with ODE starts with the assumption of a {\it rate function} for the transitions
\begin{equation}
r:\{z_1,\cdots,z_l\}\To(0,\infty)\,.
\end{equation}
These types of Petri nets are called {\it stochastic}; see \cite{Ba}. Denote by $r_j:=r(z_j)$; every variable $x_i$ has associated a {\it rate equation} given by the formula 
\begin{equation}\label{rateeq}
    x'_i(t)=\sum_{j=1}^l r_j\left(n_{ij}-m_{ij}\right)x_1^{m_{1j}}\cdots x_k^{m_{kj}}\,.
\end{equation}

\begin{remark}\label{remarkweight}
    In certain circumstances, there are associated weights between compartments and transitions $w_{i,j}$, with $x_i\in S$ and $z_j\in T$, representing proportions or probabilities. Mostly all weights are one, and we locate them between the arrows that connect the compartments with the transitions. The variation of the rate equation is as follows:
\begin{equation}\label{rateeq1}
    x'_i(t)=\sum_{j=1}^l w_{ij}r_j\left(n_{ij}-m_{ij}\right)x_1^{m_{1j}}\cdots x_k^{m_{kj}}\,.
\end{equation}
In the diagrammatic representation of a Petri net, in the case where we do not locate any label of the weight between a compartment and a transition, we assume the weight is one.     
\end{remark}

We can use Petri nets to model infectious diseases. For this purpose, we need to describe the epidemic processes using a proposed set of disease dynamics to construct the associated Petri net. 
In epidemiology, this study might be a difficult task. However, we have identified that to define the Petri net, at least for simple models (e.g., SIR, SIS, SIRS, SEIR, SEAIR, SCIR, etc.), we can identify the following components:
\begin{enumerate}
    \item The identification of the different populations located in the compartments.
    \item The interaction between different types of populations in a transition into another population at a specific rate per unit of time. 
    \item The bifurcation from a transition to different types of populations (sometimes they have assigned weights denoting the proportion or the probability).
    \item Moving a population into another with a waiting time $1/\lambda$. This is equivalent to moving individuals from one population to another with a rate $\lambda$ per unit of time.
    \item Identification of demographic models associated with specific populations and extinction transitions for certain compartments. 
\end{enumerate}
 Usually, the susceptible population is endowed with the Malthusian models (simplified logistic model) for slow diseases.

Now, we illustrate the construction of Petri nets in plenty of examples, mostly all in epidemiology. The rate equations of these Petri nets will provide some well-known ODEs associated with models in epidemiology. In our figures of Petri nets, we represent the compartments with yellow circles and the transitions with blue squares. 


\begin{example}
The formation of water is the basis of life. A water molecule has an oxygen atom with two hydrogen atoms connected by covalent bonds. 
We can take a model manufacturing system for creating water molecules with two populations of hydrogen and oxygen atoms. They enter into a transition of ``reaction" where the number of hydrogen atoms doubles the number of oxygen atoms. Thus, we have the compartments $\{H,O,W\}$.
The water molecules are manufactured at a rate of $\alpha$ in a unit of time, and we can represent this process by the Petri net of Figure \ref{formwater}.  
\begin{figure} \centering
 \begin{tikzpicture}[scale=0.3,x=1pt,y=-1pt]

\definecolor{BLACK}{RGB}{0,0,0}
\definecolor{YELLOW}{RGB}{255,255,0}
\draw[BLACK, solid, line join=round, line cap=round, line width=1, fill=YELLOW]
	(120,100) ellipse[x radius=33, y radius=35];
\draw (120,100) node {$H$};
\draw[BLACK, solid, line join=round, line cap=round, line width=1, fill=YELLOW]
	(120,280) ellipse[x radius=30, y radius=30];
\draw (120,280) node {$O$};
\draw[BLACK, solid, line join=round, line cap=round, line width=1, fill=YELLOW]
	(440,180) ellipse[x radius=30, y radius=30];
\draw (440,180) node {$W$};
\definecolor{CYAN}{RGB}{0,255,255}
\draw[BLACK, solid, line join=round, line cap=round, line width=1, fill=CYAN]
	(229,146) rectangle +(62,68);
\draw (262,179) node {$\alpha$};
\draw[BLACK, solid, line join=round, line cap=round, line width=1]
	(149,117) -- (229,162);
	\draw (194,120) node {$2$};
\draw[BLACK, solid, line join=round, line cap=round, line width=1, fill=BLACK]
	(229,162) -- (219,160) -- (222,154) -- (229,162) -- cycle;
\draw[BLACK, solid, line join=round, line cap=round, line width=1]
	(144,263) -- (229,202);
\draw[BLACK, solid, line join=round, line cap=round, line width=1, fill=BLACK]
	(229,202) -- (223,211) -- (219,205) -- (229,202) -- cycle;
\draw[BLACK, solid, line join=round, line cap=round, line width=1]
	(291,180) -- (410,180);
\draw[BLACK, solid, line join=round, line cap=round, line width=1, fill=BLACK]
	(410,180) -- (400,183) -- (400,177) -- (410,180) -- cycle;
\end{tikzpicture}
    \caption{The formation of water.}
    \label{formwater}
\end{figure}
The following ODE gives the rates equations associated with the compartments of the Petri net: 
$$\begin{array}{l}
        H'(t)=-2\alpha H^2O,\\
        O'(t)=-\alpha H^2O,\\
        W'(t)=\alpha H^2O\,.
    \end{array}$$
\end{example}

\begin{example}\label{SIR1}
Kermack-McKendrick \cite{KM} developed a model that, even at present, is the basic principle used to simulate a pandemic. 
For a historical account of this model and its essential features, the reader can consult the references \cite{Br08,Ma}. 
This structure is known as the {\bf SIR model} by the compartments {\it Susceptibles} $S$, {\it Infected} $I$, and {\it Recovered} $R$. The disease dynamics consists of one interaction: a susceptible individual comes into contact with an infected individual and becomes infected. Assume that the number of infected individuals produced in a unit of time by the interaction between the susceptible and infected populations is given at a transmission rate $\beta$.
Also, consider moving infected individuals into the recovery compartment to become resistant. Assume the infected individuals pass to the recovery compartment at a recovery rate $\alpha$ in a unit of time, equivalent to the disease in an infected individual taking about $1/\alpha$ of time to end. This epidemic process is modeled by the Petri net in Figure \ref{figu2}.
\begin{figure} \centering
     \begin{tikzpicture}[scale=0.3,x=1pt,y=-1pt]

\definecolor{BLACK}{RGB}{0,0,0}
\definecolor{YELLOW}{RGB}{255,255,0}
\draw[BLACK, solid, line join=round, line cap=round, line width=1, fill=YELLOW]
	(120,300) ellipse[x radius=30, y radius=30];
	\draw (120,300) node{$S$};
\draw[BLACK, solid, line join=round, line cap=round, line width=1, fill=YELLOW]
	(400,300) ellipse[x radius=30, y radius=30];
	\draw (400,300) node{$I$};
\draw[BLACK, solid, line join=round, line cap=round, line width=1, fill=YELLOW]
	(720,300) ellipse[x radius=30, y radius=30];
	\draw (720,300) node{$R$};
\definecolor{CYAN}{RGB}{0,255,255}
\draw[BLACK, solid, line join=round, line cap=round, line width=1, fill=CYAN]
	(526,266) rectangle +(68,68);
	\draw (559,299) node{$\alpha$};
\draw[BLACK, solid, line join=round, line cap=round, line width=1, fill=CYAN]
	(226,86) rectangle +(68,68);
	\draw (259,119) node{$\beta$};
\draw[BLACK, solid, line join=round, line cap=round, line width=1]
	(143,270) -- (234,154);
\draw[BLACK, solid, line join=round, line cap=round, line width=1, fill=BLACK]
	(234,154) -- (230,164) -- (225,160) -- (234,154) -- cycle;
\draw[BLACK, solid, line join=round, line cap=round, line width=1]
	(377,270) -- (286,154);
\draw[BLACK, solid, line join=round, line cap=round, line width=1, fill=BLACK]
	(286,154) -- (295,160) -- (290,164) -- (286,154) -- cycle;
\draw[BLACK, solid, line join=round, line cap=round, line width=1]
	(430,300) -- (526,300);
\draw[BLACK, solid, line join=round, line cap=round, line width=1, fill=BLACK]
	(526,300) -- (516,303) -- (516,297) -- (526,300) -- cycle;
\draw[BLACK, solid, line join=round, line cap=round, line width=1]
	(594,300) -- (690,300);
\draw[BLACK, solid, line join=round, line cap=round, line width=1, fill=BLACK]
	(690,300) -- (680,303) -- (680,297) -- (690,300) -- cycle;
\draw[BLACK, solid, line join=round, line cap=round, line width=1]
	(294,109) -- (320,100) -- (460,260) -- (430,280);
\draw[BLACK, solid, line join=round, line cap=round, line width=1, fill=BLACK]
	(430,280) -- (436,272) -- (440,277) -- (430,280) -- cycle;
\draw (430,150) node{$2$};	
\end{tikzpicture}
    \caption{The SIR model.}
    \label{figu2}
\end{figure} 

We list the parameters and the variables in the Table \ref{SIR}.
\begin{table}[h!]
    \centering
\begin{tabular}{ l l }
\hline
Notation & Meaning\\
\hline
 $\alpha$ & recovery rate \\ 
 $\beta$ &  transmission rate\\  
 $S(t)$ & susceptible individuals\\
 $I(t)$ & infected individuals\\
 $R(t)$ & recovered individuals
\end{tabular}
    \caption{List of parameters, variables, and their meanings for the SIR model.}
    \label{SIR}
\end{table}

From the Petri net, we can organize the rate equations of the ODE:
\begin{equation}\label{SIR-model}\begin{array}{l}
S'(t)=-\beta SI,\\
I'(t)=\beta SI-\alpha I,\\
R'(t)=\alpha I\,.
\end{array}\end{equation}
\end{example}

\begin{example}
A simpler model is provided by relaxing the assumption of permanent immunity in the SIR model, obtaining the {\bf SIS model} where some infectious diseases are the common cold or influenza. The disease dynamics have the new feature that infected individuals recover and immediately pass to the susceptible compartment at a rate $\alpha$ per unit of time.
Thus, we have removed the recovery compartment, and we add a transition with parameter $\alpha$ connecting the compartments $I$ with $S$. 
Consequently, we obtain the Petri net depicted in Figure \ref{SISmod}.
    \begin{figure} \centering
     \begin{tikzpicture}[scale=0.3,x=1pt,y=-1pt]

\definecolor{BLACK}{RGB}{0,0,0}
\definecolor{YELLOW}{RGB}{255,255,0}
\draw[BLACK, solid, line join=round, line cap=round, line width=1, fill=YELLOW]
	(140,320) ellipse[x radius=35, y radius=35];
\draw[BLACK, solid, line join=round, line cap=round, line width=1, fill=YELLOW]
	(460,320) ellipse[x radius=35, y radius=35];
\definecolor{CYAN}{RGB}{0,255,255}
\draw[BLACK, solid, line join=round, line cap=round, line width=1, fill=CYAN]
	(272,192) rectangle +(56,56);
\draw[BLACK, solid, line join=round, line cap=round, line width=1, fill=CYAN]
	(272,412) rectangle +(56,56);
\draw[BLACK, solid, line join=round, line cap=round, line width=1]
	(170,301) -- (272,238);
\draw[BLACK, solid, line join=round, line cap=round, line width=1, fill=BLACK]
	(272,238) -- (265,246) -- (262,240) -- (272,238) -- cycle;
\draw[BLACK, solid, line join=round, line cap=round, line width=1]
	(430,301) -- (328,238);
\draw[BLACK, solid, line join=round, line cap=round, line width=1, fill=BLACK]
	(328,238) -- (338,240) -- (335,246) -- (328,238) -- cycle;
\draw[BLACK, solid, line join=round, line cap=round, line width=1]
	(272,419) -- (168,341);
\draw[BLACK, solid, line join=round, line cap=round, line width=1, fill=BLACK]
	(168,341) -- (178,344) -- (174,350) -- (168,341) -- cycle;
\draw[BLACK, solid, line join=round, line cap=round, line width=1]
	(432,341) -- (328,419);
\draw[BLACK, solid, line join=round, line cap=round, line width=1, fill=BLACK]
	(328,419) -- (334,410) -- (338,416) -- (328,419) -- cycle;
\draw[BLACK, solid, line join=round, line cap=round, line width=1]
	(328,215) -- (420,200) -- (420,200) -- (449,287);
\draw[BLACK, solid, line join=round, line cap=round, line width=1, fill=BLACK]
	(449,287) -- (443,278) -- (449,276) -- (449,287) -- cycle;

\draw (300,222) node {$\beta$};

\draw (435,182) node {$2$};

\draw (300,440) node {$\alpha$};

\draw (140,322) node {$S$};

\draw (460,320) node {$I$};
 
\end{tikzpicture}
    \caption{The SIS model.}
    \label{SISmod}
\end{figure} 
The SIS model is often used because we can analytically solve the two populations $S(t)$ and $I(t)$ at any point $t$ in time; see \cite[p. 35]{Kuhl21}.

From the Petri net, we can obtain the associated ODE: 
\begin{equation}\begin{array}{l}
S'(t)=-\beta SI+\alpha I\,,\\
I'(t)=\beta SI-\alpha I\,.
\end{array}
\end{equation}
 \end{example}

\begin{example} 
Instead of removing the recovery compartment in the previous example, the disease dynamics consider that infected individuals stay in the recovery compartment and then pass to the susceptible compartment at a rate 
$\gamma$ per unit of time (equivalent to an infected individual spending $1/\gamma$ of time in the recovery compartment to become susceptible). The structure involved is called the {\bf SIRS model}, where some representable diseases are the seasonal influenza, where immunity may wane over time; for example, the reader can consult the book \cite{VW10}. Therefore, the Petri net of this model is illustrated in Figure \ref{SIRSmod}.
        \begin{figure} \centering
     \begin{tikzpicture}[scale=0.3,x=1pt,y=-1pt]

\definecolor{BLACK}{RGB}{0,0,0}
\definecolor{YELLOW}{RGB}{255,255,0}
\draw[BLACK, solid, line join=round, line cap=round, line width=1, fill=YELLOW]
	(160,320) ellipse[x radius=35, y radius=35];
\draw[BLACK, solid, line join=round, line cap=round, line width=1, fill=YELLOW]
	(440,320) ellipse[x radius=35, y radius=35];
\definecolor{CYAN}{RGB}{0,255,255}
\draw[BLACK, solid, line join=round, line cap=round, line width=1, fill=CYAN]
	(272,192) rectangle +(56,56);
\draw[BLACK, solid, line join=round, line cap=round, line width=1]
	(188,300) -- (272,240);
\draw[BLACK, solid, line join=round, line cap=round, line width=1, fill=BLACK]
	(272,240) -- (266,249) -- (262,243) -- (272,240) -- cycle;
\draw[BLACK, solid, line join=round, line cap=round, line width=1]
	(412,300) -- (328,240);
\draw[BLACK, solid, line join=round, line cap=round, line width=1, fill=BLACK]
	(328,240) -- (338,243) -- (334,249) -- (328,240) -- cycle;
\draw[BLACK, solid, line join=round, line cap=round, line width=1, fill=CYAN]
	(492,452) rectangle +(56,56);
\draw[BLACK, solid, line join=round, line cap=round, line width=1, fill=CYAN]
	(592,292) rectangle +(56,56);
\draw[BLACK, solid, line join=round, line cap=round, line width=1, fill=YELLOW]
	(800,320) ellipse[x radius=35, y radius=35];
\draw[BLACK, solid, line join=round, line cap=round, line width=1]
	(475,320) -- (592,320);
\draw[BLACK, solid, line join=round, line cap=round, line width=1, fill=BLACK]
	(592,320) -- (582,323) -- (582,317) -- (592,320) -- cycle;
\draw[BLACK, solid, line join=round, line cap=round, line width=1]
	(648,320) -- (765,320);
\draw[BLACK, solid, line join=round, line cap=round, line width=1, fill=BLACK]
	(765,320) -- (755,323) -- (755,317) -- (765,320) -- cycle;
\draw[BLACK, solid, line join=round, line cap=round, line width=1]
	(800,355) -- (800,480) -- (800,480) -- (548,480);
\draw[BLACK, solid, line join=round, line cap=round, line width=1, fill=BLACK]
	(548,480) -- (558,477) -- (558,483) -- (548,480) -- cycle;
\draw[BLACK, solid, line join=round, line cap=round, line width=1]
	(492,480) -- (160,480) -- (160,480) -- (160,355);
\draw[BLACK, solid, line join=round, line cap=round, line width=1, fill=BLACK]
	(160,355) -- (163,365) -- (157,365) -- (160,355) -- cycle;
\draw[BLACK, solid, line join=round, line cap=round, line width=1]
	(328,220) -- (420,220) -- (420,220) -- (433,286);
\draw[BLACK, solid, line join=round, line cap=round, line width=1, fill=BLACK]
	(433,286) -- (428,277) -- (434,275) -- (433,286) -- cycle;

 \draw (300,222) node {$\beta$};
 
\draw (159,322) node {$S$};

\draw (440,320) node {$I$};

\draw (435,202) node {$2$};

\draw (620,320) node {$\alpha$};

\draw (520,482) node {$\gamma$};

\draw (800,320) node {$R$};

\end{tikzpicture}
    \caption{The SIRS model.}
    \label{SIRSmod}
\end{figure}

The ODE associated with the SIRS model is as follows:
    \begin{equation}\begin{array}{l}
        S'(t)=-\beta SI+\gamma R,\\
        I'(t)=\beta SI-\alpha I,\\
        R'(t)=\alpha I-\gamma R\,.
      \end{array}\end{equation}
\end{example}

\begin{example}\label{maltusiano}
Thomas Malthus \cite{Malth} in the eighteenth-century debates about the population increasing contributing to the first census of England, Scotland, and Wales. The {\bf Malthusian model} assumes that all individuals are identical, the environment is constant in space and time, and, in particular, the resources are unlimited. The population is located in one compartment $N$, and $b$ and $\mu$ denote the birth and death rates, respectively. The Malthusian model becomes $N'(t)=bN-\mu N=rN$, where $r=b-\mu$ is the population growth rate. 
The population is growing exponentially if $r > 0$, decreasing exponentially if $r < 0$, and constant if $r = 0$.
For this work, we use a simplified logistic model assuming a constant birth rate independent of the population size, denoting the total birth rate by $\Lambda$, and where the death rate is $\mu$. The Petri net of the previous model has the form given in Figure \ref{Malthu}.

\begin{figure} \centering
     \begin{tikzpicture}[scale=0.3,x=1pt,y=-1pt]

\definecolor{BLACK}{RGB}{0,0,0}
\definecolor{r255g255b10}{RGB}{255,255,10}
\draw[BLACK, solid, line join=round, line cap=round, line width=1, fill=r255g255b10]
	(200,200) ellipse[x radius=30, y radius=30];
	\draw (200,200) node {$N$};
\definecolor{r33g255b255}{RGB}{33,255,255}
\draw[BLACK, solid, line join=round, line cap=round, line width=1, fill=r33g255b255]
	(15,175) rectangle +(50,50);
	\draw (40,200) node {$\Lambda$};
\draw[BLACK, solid, line join=round, line cap=round, line width=1, fill=r33g255b255]
	(315,175) rectangle +(50,50);
	\draw (340,200) node {$\mu$};
\draw[BLACK, solid, line join=round, line cap=round, line width=1]
	(65,200) -- (170,200);
\draw[BLACK, solid, line join=round, line cap=round, line width=1, fill=BLACK]
	(170,200) -- (160,203) -- (160,197) -- (170,200) -- cycle;
\draw[BLACK, solid, line join=round, line cap=round, line width=1]
	(230,200) -- (315,200);
\draw[BLACK, solid, line join=round, line cap=round, line width=1, fill=BLACK]
	(315,200) -- (305,203) -- (305,197) -- (315,200) -- cycle;
\end{tikzpicture}
    \caption{A simplified logistic Malthusian model.}
    \label{Malthu}
\end{figure} 
There is only one rate equation for this Petri net:
$$N'(t)=\Lambda-\mu N\,.$$ 
The population is asymptotically constant, with $N(t)\To \Lambda/\mu$ as $t\To \infty$.   
It is essential to mention that epidemic models without explicit demography are helpful on a short time scale, as in the cases of childhood disease and influenza. 
On the other hand, there are slow diseases, such as HIV, tuberculosis, and hepatitis C, where the demography has to be included in the models.
\end{example}

\begin{example}
    The {\bf SEIR model} is an immediate generalization of the Kermack-McKendrick SIR model. We incorporate into the SIR model an additional compartment that comprises infected individuals who are not immediately infectious (the infectious individuals are enclosed in the compartment $I$, and the infected individuals that are not immediately infectious are enclosed in the compartment $E$). The disease dynamics have the following interaction: a susceptible individual having contact with an infectious individual becomes exposed. 
We assume that the number of exposed individuals produced in a unit of time by the interaction between the susceptible and infectious populations is given at a transmission rate $\beta$.
The exposed individuals pass to the compartment of infectious individuals with a rate of $\eta$ per unit of time. 
The infected individuals pass to the compartment of recovered individuals with a rate of $\alpha$ per unit of time. 
    Additionally, we include a Malthusian model (the simplified logistic) to the susceptible compartment, and all the individuals in the remaining compartments can disappear at a death rate $\mu$. Consequently, the Petri net associated with the SEIR model has the form in Figure \ref{SEIRfig}.    
    Possibly, the reader asks why there is an arrow from the transition with parameter $\beta$ to the compartment of infectious individuals. This is because an infectious individual remains infectious after contact with a susceptible individual. 
    
\begin{figure}
    \centering
    \begin{tikzpicture}[scale=0.3,x=1pt,y=-1pt]

\definecolor{BLACK}{RGB}{0,0,0}
\definecolor{r255g255b10}{RGB}{255,255,10}
\draw[BLACK, solid, line join=round, line cap=round, line width=1, fill=r255g255b10]
	(200,200) ellipse[x radius=30, y radius=30];
		\draw (200,200) node {$S$};
\draw[BLACK, solid, line join=round, line cap=round, line width=1, fill=r255g255b10]
	(440,200) ellipse[x radius=30, y radius=30];
	    \draw (440,200) node {$E$};
\draw[BLACK, solid, line join=round, line cap=round, line width=1, fill=r255g255b10]
	(700,200) ellipse[x radius=30, y radius=30];
	    \draw (700,200) node {$I$};
\draw[BLACK, solid, line join=round, line cap=round, line width=1, fill=r255g255b10]
	(940,200) ellipse[x radius=30, y radius=30];
	    \draw (940,200) node {$R$};
\definecolor{r33g255b255}{RGB}{33,255,255}
\draw[BLACK, solid, line join=round, line cap=round, line width=1, fill=r33g255b255]
	(75,55) rectangle +(50,50);
	    \draw (100,80) node {$\Lambda$};
\draw[BLACK, solid, line join=round, line cap=round, line width=1, fill=r33g255b255]
	(275,295) rectangle +(50,50);
	  \draw (300,320) node {$\mu$};
\draw[BLACK, solid, line join=round, line cap=round, line width=1, fill=r33g255b255]
	(515,295) rectangle +(50,50);
	\draw (540,320) node {$\mu$};
\draw[BLACK, solid, line join=round, line cap=round, line width=1, fill=r33g255b255]
	(775,295) rectangle +(50,50);
	\draw (800,320) node {$\mu$};
\draw[BLACK, solid, line join=round, line cap=round, line width=1, fill=r33g255b255]
	(1015,295) rectangle +(50,50);
		\draw (1040,320) node {$\mu$};
\draw[BLACK, solid, line join=round, line cap=round, line width=1, fill=r33g255b255]
	(295,175) rectangle +(50,50);
	\draw (320,200) node {$\beta$};
\draw[BLACK, solid, line join=round, line cap=round, line width=1, fill=r33g255b255]
	(555,175) rectangle +(50,50);
	\draw (580,200) node {$\eta$};
\draw[BLACK, solid, line join=round, line cap=round, line width=1, fill=r33g255b255]
	(795,175) rectangle +(50,50);
	\draw (820,200) node {$\alpha$};
\draw[BLACK, solid, line join=round, line cap=round, line width=1]
	(121,105) -- (175,170);
\draw[BLACK, solid, line join=round, line cap=round, line width=1, fill=BLACK]
	(175,170) -- (166,164) -- (171,160) -- (175,170) -- cycle;
\draw[BLACK, solid, line join=round, line cap=round, line width=1]
	(225,230) -- (279,295);
\draw[BLACK, solid, line join=round, line cap=round, line width=1, fill=BLACK]
	(279,295) -- (270,289) -- (275,285) -- (279,295) -- cycle;
\draw[BLACK, solid, line join=round, line cap=round, line width=1]
	(465,230) -- (519,295);
\draw[BLACK, solid, line join=round, line cap=round, line width=1, fill=BLACK]
	(519,295) -- (510,289) -- (515,285) -- (519,295) -- cycle;
\draw[BLACK, solid, line join=round, line cap=round, line width=1]
	(725,230) -- (779,295);
\draw[BLACK, solid, line join=round, line cap=round, line width=1, fill=BLACK]
	(779,295) -- (770,289) -- (775,285) -- (779,295) -- cycle;
\draw[BLACK, solid, line join=round, line cap=round, line width=1]
	(965,230) -- (1019,295);
\draw[BLACK, solid, line join=round, line cap=round, line width=1, fill=BLACK]
	(1019,295) -- (1010,289) -- (1015,285) -- (1019,295) -- cycle;
\draw[BLACK, solid, line join=round, line cap=round, line width=1]
	(230,200) -- (295,200);
\draw[BLACK, solid, line join=round, line cap=round, line width=1, fill=BLACK]
	(295,200) -- (285,203) -- (285,197) -- (295,200) -- cycle;
\draw[BLACK, solid, line join=round, line cap=round, line width=1]
	(345,200) -- (410,200);
\draw[BLACK, solid, line join=round, line cap=round, line width=1, fill=BLACK]
	(410,200) -- (400,203) -- (400,197) -- (410,200) -- cycle;
\draw[BLACK, solid, line join=round, line cap=round, line width=1]
	(470,200) -- (555,200);
\draw[BLACK, solid, line join=round, line cap=round, line width=1, fill=BLACK]
	(555,200) -- (545,203) -- (545,197) -- (555,200) -- cycle;
\draw[BLACK, solid, line join=round, line cap=round, line width=1]
	(605,200) -- (670,200);
\draw[BLACK, solid, line join=round, line cap=round, line width=1, fill=BLACK]
	(670,200) -- (660,203) -- (660,197) -- (670,200) -- cycle;
\draw[BLACK, solid, line join=round, line cap=round, line width=1]
	(730,200) -- (795,200);
\draw[BLACK, solid, line join=round, line cap=round, line width=1, fill=BLACK]
	(795,200) -- (785,203) -- (785,197) -- (795,200) -- cycle;
\draw[BLACK, solid, line join=round, line cap=round, line width=1]
	(845,200) -- (910,200);
\draw[BLACK, solid, line join=round, line cap=round, line width=1, fill=BLACK]
	(910,200) -- (900,203) -- (900,197) -- (910,200) -- cycle;
\draw[BLACK, solid, line join=round, line cap=round, line width=1]
	(700,170) -- (700,120) -- (700,120) -- (320,120) -- (320,175);
\draw[BLACK, solid, line join=round, line cap=round, line width=1, fill=BLACK]
	(320,175) -- (317,165) -- (323,165) -- (320,175) -- cycle;
\draw[BLACK, solid, line join=round, line cap=round, line width=1]
	(313,175) -- (280,60) -- (280,60) -- (740,60) -- (709,170);
\draw[BLACK, solid, line join=round, line cap=round, line width=1, fill=BLACK]
	(709,170) -- (708,159) -- (715,161) -- (709,170) -- cycle;	
\end{tikzpicture}
\caption{The SEIR model.}
    \label{SEIRfig}
\end{figure}
The rate equations associated with this Petri net are:
$$\begin{array}{l}
S'(t)=\Lambda -\beta SI-\mu S,\\
E'(t)=\beta SI-(\eta +\mu )E,\\
I'(t)=\eta E-(\alpha+\mu) I,\\ 
R'(t)=\alpha I-\mu R\,.
\end{array}$$
This model can be found in the book of Martcheva \cite[p. 92]{Ma}.
\end{example}

\begin{example}\label{SEAIR} 
Similarly to the SEIR model, we can add the characterization that a significant group of individuals remain asymptomatic upon infection and can infect others. 
This structure is called the {\bf SEAIR model}, which is also called the SEIIR model in the book of Kuhl \cite[Chap. 8]{Kuhl21}.
A recent case is the COVID-19 pandemic \cite{Kuhl21}. Another example of an asymptomatic disease is HIV, in which an infected individual can live without symptoms for ten years before progressing to symptoms of AIDS \cite{HAF08}.

In the SEAIR model, we distinguish between infection and disease with a new compartment $A$, consisting of individuals whose infection does not show symptoms (the symptomatic infected individuals are enclosed in the compartment $I$, called infectious). 
The disease dynamics have two interactions: first, a susceptible individual having contact with an infectious individual to become exposed, and second, a susceptible individual having contact with an asymptomatic individual to become exposed. We assume that the number of exposed individuals produced in a unit of time by the interaction between the susceptible and infectious populations is given at a transmission rate $\beta$. Similarly, we assume the number of exposed individuals produced in a unit of time by the interaction between the susceptible and asymptomatic populations is given at a transmission rate $q\beta$. 
In this case, there is a bifurcation in the compartment of exposed individuals at a rate $\eta$ in a unit of time to the 
infectious compartment $I$ with probability $p$ and to the asymptomatic compartment $A$ with probability $(1-p)$.
The infected individuals pass to the compartment of recovered individuals with a rate of $\alpha$ per unit of time, and the asymptomatic individuals pass to the compartment of recovered individuals with a rate of $\gamma$ per unit of time. 
 In the same way as in the SEIR model, we include a Malthusian model (the simplified logistic) in the susceptible compartment, and all the individuals in the remaining compartments can disappear at a death rate $\mu$.
The associated Petri net in Figure \ref{SEAIRfig} 
has the particularity that the two arrows from the compartment $E$ to compartments $I$ and $A$ has associated the weights 
$p$ and $(1-p)$, respectively.
\begin{figure}
    \centering
    \begin{tikzpicture}[scale=0.3,x=1pt,y=-1pt]

\definecolor{BLACK}{RGB}{0,0,0}
\definecolor{r255g255b10}{RGB}{255,255,10}
\draw[BLACK, solid, line join=round, line cap=round, line width=1, fill=r255g255b10]
	(200,200) ellipse[x radius=30, y radius=30];
		\draw (200,200) node {$S$};
\draw[BLACK, solid, line join=round, line cap=round, line width=1, fill=r255g255b10]
	(440,200) ellipse[x radius=30, y radius=30];
		\draw (440,200) node {$E$};
\draw[BLACK, solid, line join=round, line cap=round, line width=1, fill=r255g255b10]
	(700,80) ellipse[x radius=30, y radius=30];
	\draw (700,80) node {$A$};
\draw[BLACK, solid, line join=round, line cap=round, line width=1, fill=r255g255b10]
	(940,200) ellipse[x radius=30, y radius=30];
	\draw (940,200) node {$R$};
\definecolor{r33g255b255}{RGB}{33,255,255}
\draw[BLACK, solid, line join=round, line cap=round, line width=1, fill=r33g255b255]
	(180,80) rectangle +(40,40);
		\draw (200,100) node {$\Lambda$};
\draw[BLACK, solid, line join=round, line cap=round, line width=1, fill=r33g255b255]
	(180,280) rectangle +(40,40);
        	\draw (200,300) node {$\mu$};
\draw[BLACK, solid, line join=round, line cap=round, line width=1, fill=r33g255b255]
	(420,280) rectangle +(40,40);
	        	\draw (440,300) node {$\mu$};
\draw[BLACK, solid, line join=round, line cap=round, line width=1, fill=r33g255b255]
	(680,160) rectangle +(40,40);
		\draw (700,180) node {$\mu$};
\draw[BLACK, solid, line join=round, line cap=round, line width=1, fill=r33g255b255]
	(920,280) rectangle +(40,40);
	\draw (940,300) node {$\mu$};
\draw[BLACK, solid, line join=round, line cap=round, line width=1, fill=r33g255b255]
	(295,75) rectangle +(50,50);
	\draw (320,100) node {$q\beta$};
\draw[BLACK, solid, line join=round, line cap=round, line width=1, fill=r33g255b255]
	(555,175) rectangle +(50,50);
	\draw (580,200) node {$\eta$};
\draw[BLACK, solid, line join=round, line cap=round, line width=1, fill=r33g255b255]
	(795,115) rectangle +(50,50);
		\draw (820,140) node {$\gamma$};
\draw[BLACK, solid, line join=round, line cap=round, line width=1]
	(200,120) -- (200,170);
\draw[BLACK, solid, line join=round, line cap=round, line width=1, fill=BLACK]
	(200,170) -- (197,160) -- (203,160) -- (200,170) -- cycle;
\draw[BLACK, solid, line join=round, line cap=round, line width=1]
	(200,230) -- (200,280);
\draw[BLACK, solid, line join=round, line cap=round, line width=1, fill=BLACK]
	(200,280) -- (197,270) -- (203,270) -- (200,280) -- cycle;
\draw[BLACK, solid, line join=round, line cap=round, line width=1]
	(440,230) -- (440,280);
\draw[BLACK, solid, line join=round, line cap=round, line width=1, fill=BLACK]
	(440,280) -- (437,270) -- (443,270) -- (440,280) -- cycle;
\draw[BLACK, solid, line join=round, line cap=round, line width=1]
	(700,110) -- (700,160);
\draw[BLACK, solid, line join=round, line cap=round, line width=1, fill=BLACK]
	(700,160) -- (697,150) -- (703,150) -- (700,160) -- cycle;
\draw[BLACK, solid, line join=round, line cap=round, line width=1]
	(940,230) -- (940,280);
\draw[BLACK, solid, line join=round, line cap=round, line width=1, fill=BLACK]
	(940,280) -- (937,270) -- (943,270) -- (940,280) -- cycle;
\draw[BLACK, solid, line join=round, line cap=round, line width=1]
	(230,175) -- (295,121);
\draw[BLACK, solid, line join=round, line cap=round, line width=1, fill=BLACK]
	(295,121) -- (289,130) -- (285,125) -- (295,121) -- cycle;
\draw[BLACK, solid, line join=round, line cap=round, line width=1]
	(345,121) -- (410,175);
\draw[BLACK, solid, line join=round, line cap=round, line width=1, fill=BLACK]
	(410,175) -- (400,171) -- (404,166) -- (410,175) -- cycle;
\draw[BLACK, solid, line join=round, line cap=round, line width=1]
	(470,200) -- (555,200);
\draw[BLACK, solid, line join=round, line cap=round, line width=1, fill=BLACK]
	(555,200) -- (545,203) -- (545,197) -- (555,200) -- cycle;
\draw[BLACK, solid, line join=round, line cap=round, line width=1]
	(605,175) -- (670,110);
\draw[BLACK, solid, line join=round, line cap=round, line width=1, fill=BLACK]
	(670,110) -- (665,119) -- (661,115) -- (670,110) -- cycle;
\draw[BLACK, solid, line join=round, line cap=round, line width=1]
	(730,95) -- (795,128);
\draw[BLACK, solid, line join=round, line cap=round, line width=1, fill=BLACK]
	(795,128) -- (785,126) -- (788,120) -- (795,128) -- cycle;
\draw[BLACK, solid, line join=round, line cap=round, line width=1]
	(845,153) -- (910,185);
\draw[BLACK, solid, line join=round, line cap=round, line width=1, fill=BLACK]
	(910,185) -- (900,184) -- (903,178) -- (910,185) -- cycle;
\draw[BLACK, solid, line join=round, line cap=round, line width=1, fill=r33g255b255]
	(295,275) rectangle +(50,50);
		\draw (320,300) node {$\beta$};
\draw[BLACK, solid, line join=round, line cap=round, line width=1, fill=r255g255b10]
	(700,300) ellipse[x radius=30, y radius=30];
	\draw (700,300) node {$I$};
\draw[BLACK, solid, line join=round, line cap=round, line width=1, fill=r33g255b255]
	(795,235) rectangle +(50,50);
	\draw (820,260) node {$\alpha$};
\draw[BLACK, solid, line join=round, line cap=round, line width=1]
	(605,221) -- (670,275);
\draw[BLACK, solid, line join=round, line cap=round, line width=1, fill=BLACK]
	(670,275) -- (660,271) -- (664,266) -- (670,275) -- cycle;
\draw[BLACK, solid, line join=round, line cap=round, line width=1]
	(730,290) -- (795,268);
\draw[BLACK, solid, line join=round, line cap=round, line width=1, fill=BLACK]
	(795,268) -- (787,275) -- (784,268) -- (795,268) -- cycle;
\draw[BLACK, solid, line join=round, line cap=round, line width=1]
	(845,248) -- (910,215);
\draw[BLACK, solid, line join=round, line cap=round, line width=1, fill=BLACK]
	(910,215) -- (903,222) -- (900,216) -- (910,215) -- cycle;
\draw[BLACK, solid, line join=round, line cap=round, line width=1]
	(345,279) -- (410,225);
\draw[BLACK, solid, line join=round, line cap=round, line width=1, fill=BLACK]
	(410,225) -- (404,234) -- (400,229) -- (410,225) -- cycle;
\draw[BLACK, solid, line join=round, line cap=round, line width=1]
	(230,225) -- (295,279);
\draw[BLACK, solid, line join=round, line cap=round, line width=1, fill=BLACK]
	(295,279) -- (285,275) -- (289,270) -- (295,279) -- cycle;
\draw[BLACK, solid, line join=round, line cap=round, line width=1, fill=r33g255b255]
	(780,360) rectangle +(40,40);
		\draw (800,380) node {$\mu$};
\draw[BLACK, solid, line join=round, line cap=round, line width=1]
	(730,324) -- (780,364);
\draw[BLACK, solid, line join=round, line cap=round, line width=1, fill=BLACK]
	(780,364) -- (770,360) -- (774,355) -- (780,364) -- cycle;
\draw[BLACK, solid, line join=round, line cap=round, line width=1]
	(310,75) -- (280,0) -- (280,0) -- (760,0) -- (723,50);
\draw[BLACK, solid, line join=round, line cap=round, line width=1, fill=BLACK]
	(723,50) -- (726,40) -- (731,44) -- (723,50) -- cycle;
\draw[BLACK, solid, line join=round, line cap=round, line width=1]
	(700,50) -- (700,20) -- (700,20) -- (320,20) -- (320,75);
\draw[BLACK, solid, line join=round, line cap=round, line width=1, fill=BLACK]
	(320,75) -- (317,65) -- (323,65) -- (320,75) -- cycle;
\draw[BLACK, solid, line join=round, line cap=round, line width=1]
	(700,330) -- (700,360) -- (700,360) -- (320,360) -- (320,325);
\draw[BLACK, solid, line join=round, line cap=round, line width=1, fill=BLACK]
	(320,325) -- (323,335) -- (317,335) -- (320,325) -- cycle;
\draw[BLACK, solid, line join=round, line cap=round, line width=1]
	(308,325) -- (280,380) -- (740,380) -- (715,330);
\draw[BLACK, solid, line join=round, line cap=round, line width=1, fill=BLACK]
	(715,330) -- (722,337) -- (716,340) -- (715,330) -- cycle;

	\draw (580,120) node {$(1-p)$};
	\draw (610,280) node {$p$};

\end{tikzpicture}
    \caption{The SEAIR model.}
    \label{SEAIRfig}
\end{figure}
 Similarly to the Petri net for the SEIR model, we have arrows from the transitions with parameters $\beta$ and $q\beta$ to the compartments of infectious and asymptomatic individuals, respectively. This is because an infectious and asymptomatic individual remains infectious and asymptomatic after contact with a susceptible individual. 

We list the parameters and the variables in Table \ref{SEAIR-T}.
\begin{table}[h!]
    \centering
\begin{tabular}{ l l }
\hline
Notation & Meaning\\
\hline
 $\alpha$ & recovery rate from infected\\ 
 $\beta$ &  transmission rate\\  
 $q\beta$ &  reduced transmission rate\\  
 $\gamma$ & recovery rate from asymptomatic\\
 $\Lambda$ & birth rate\\
 $\mu$ & death rate\\
 $\eta$ & rate of shift from exposed to asymptomatic-infected\\
 $p$ & probability to be infected after being exposed\\
 $1-p$ & probability to be asymptomatic after being exposed\\
 $S(t)$ & susceptible individuals\\
 $E(t)$ & exposed individuals \\ 
 $A(t)$ & asymptomatic individuals \\ 
 $I(t)$ & infected individuals\\
 $R(t)$ & recovered individuals
\end{tabular}
    \caption{List of parameters, variables, and their meanings for the SEAIR model.}
    \label{SEAIR-T}
\end{table}
According to Remark \ref{remarkweight}, the rate equations of the Petri net of the SEAIR model are:
\begin{equation}
\begin{array}{l}
S'(t)=\Lambda -\beta S(I+qA)-\mu S,\\
E'(t)=\beta S(I+qA)-(\eta +\mu )E,\\
A'(t)=(1-p)\eta E-(\gamma+\mu)A,\\
I'(t)=p\eta E-(\alpha+\mu) I,\\
R'(t)=\alpha I+\gamma A-\mu R\,.
\end{array}
\end{equation}
We can find the previous ODE in the book of Martcheva; see \cite[p. 93]{Ma}.
\end{example}

\begin{example}
Something comparable to the SEIR and SEAIR models, now we consider an additional compartment of individuals that do not show 
symptoms or signs of infection but transmit the pathogen from their nose, throat, or feces (we call these individuals carriers, which are enclosed in the compartment $C$). This is called the {\bf SCIR model}, where some representative diseases are viral hepatitis and poliomyelitis, and bacterial diseases, including diphtheria and meningococcal meningitis, see \cite{Ma}.

The disease dynamics have two interactions: first, a susceptible individual having contact with an infected individual to become a carrier individual, and second, a susceptible individual having contact with a carrier individual to become a carrier individual. We assume that the number of carrier individuals produced in a unit of time by the interaction between the susceptible and infected population is given at a transmission rate $\beta$. Similarly, we assume the number  
of carrier individuals produced in a unit of time by the interaction between the susceptible and carrier populations is given at a transmission rate $q\beta$. The individuals in the compartment $C$ can pass to the compartment of recovered individuals $R$ at a rate $\gamma$. Still, they can also pass to the infected individuals' compartment at a rate $\eta$. In addition, the infected individuals in the compartment $I$ can pass to the recovery compartment at a rate $\alpha$, and the recovered individuals in the compartment $R$ can pass to the susceptible compartment $S$ at a rate $\rho$. 
 In the same way as in the SEIR and SEAIR models, we include a Malthusian model (the simplified logistic) in the susceptible compartment, and all the individuals in the remaining compartments can disappear at a death rate $\mu$.
We construct the Petri net of the SCIR model in Figure \ref{SCIRfig}.

\begin{figure}
    \centering
    \begin{tikzpicture}[scale=0.3,x=1pt,y=-1pt]

\definecolor{BLACK}{RGB}{0,0,0}
\definecolor{YELLOW}{RGB}{255,255,0}
\draw[BLACK, solid, line join=round, line cap=round, line width=1, fill=YELLOW]
	(240,320) ellipse[x radius=35, y radius=35];
\draw[BLACK, solid, line join=round, line cap=round, line width=1, fill=YELLOW]
	(440,320) ellipse[x radius=35, y radius=35];
\draw[BLACK, solid, line join=round, line cap=round, line width=1, fill=YELLOW]
	(620,520) ellipse[x radius=35, y radius=35];
\draw[BLACK, solid, line join=round, line cap=round, line width=1, fill=YELLOW]
	(620,120) ellipse[x radius=35, y radius=35];
\definecolor{CYAN}{RGB}{0,255,255}
\draw[BLACK, solid, line join=round, line cap=round, line width=1, fill=CYAN]
	(312,192) rectangle +(56,56);
\draw[BLACK, solid, line join=round, line cap=round, line width=1, fill=CYAN]
	(492,192) rectangle +(56,56);
\draw[BLACK, solid, line join=round, line cap=round, line width=1, fill=CYAN]
	(492,392) rectangle +(56,56);
\draw[BLACK, solid, line join=round, line cap=round, line width=1, fill=CYAN]
	(312,392) rectangle +(56,56);
\draw[BLACK, solid, line join=round, line cap=round, line width=1, fill=CYAN]
	(217,397) rectangle +(46,46);
\draw[BLACK, solid, line join=round, line cap=round, line width=1, fill=CYAN]
	(137,217) rectangle +(46,46);
\draw[BLACK, solid, line join=round, line cap=round, line width=1, fill=CYAN]
	(677,577) rectangle +(46,46);
\draw[BLACK, solid, line join=round, line cap=round, line width=1, fill=CYAN]
	(597,197) rectangle +(46,46);
\draw[BLACK, solid, line join=round, line cap=round, line width=1, fill=CYAN]
	(712,292) rectangle +(56,56);
\draw[BLACK, solid, line join=round, line cap=round, line width=1, fill=CYAN]
	(492,-7) rectangle +(56,56);
\draw[BLACK, solid, line join=round, line cap=round, line width=1]
	(265,295) -- (312,248);
\draw[BLACK, solid, line join=round, line cap=round, line width=1, fill=BLACK]
	(312,248) -- (307,257) -- (303,253) -- (312,248) -- cycle;
\draw[BLACK, solid, line join=round, line cap=round, line width=1]
	(462,293) -- (498,248);
\draw[BLACK, solid, line join=round, line cap=round, line width=1, fill=BLACK]
	(498,248) -- (494,258) -- (489,254) -- (498,248) -- cycle;
\draw[BLACK, solid, line join=round, line cap=round, line width=1]
	(548,192) -- (595,145);
\draw[BLACK, solid, line join=round, line cap=round, line width=1, fill=BLACK]
	(595,145) -- (591,154) -- (586,149) -- (595,145) -- cycle;
\draw[BLACK, solid, line join=round, line cap=round, line width=1]
	(462,347) -- (498,392);
\draw[BLACK, solid, line join=round, line cap=round, line width=1, fill=BLACK]
	(498,392) -- (489,386) -- (494,382) -- (498,392) -- cycle;
\draw[BLACK, solid, line join=round, line cap=round, line width=1]
	(548,448) -- (595,495);
\draw[BLACK, solid, line join=round, line cap=round, line width=1, fill=BLACK]
	(595,495) -- (586,491) -- (591,486) -- (595,495) -- cycle;
\draw[BLACK, solid, line join=round, line cap=round, line width=1]
	(368,248) -- (415,295);
\draw[BLACK, solid, line join=round, line cap=round, line width=1, fill=BLACK]
	(415,295) -- (406,291) -- (411,286) -- (415,295) -- cycle;
\draw[BLACK, solid, line join=round, line cap=round, line width=1]
	(368,392) -- (415,345);
\draw[BLACK, solid, line join=round, line cap=round, line width=1, fill=BLACK]
	(415,345) -- (411,354) -- (406,349) -- (415,345) -- cycle;
\draw[BLACK, solid, line join=round, line cap=round, line width=1]
	(265,345) -- (312,392);
\draw[BLACK, solid, line join=round, line cap=round, line width=1, fill=BLACK]
	(312,392) -- (303,387) -- (307,383) -- (312,392) -- cycle;
\draw[BLACK, solid, line join=round, line cap=round, line width=1]
	(440,285) -- (440,220) -- (440,220) -- (368,220);
\draw[BLACK, solid, line join=round, line cap=round, line width=1, fill=BLACK]
	(368,220) -- (378,217) -- (378,223) -- (368,220) -- cycle;
\draw[BLACK, solid, line join=round, line cap=round, line width=1]
	(240,355) -- (240,397);
\draw[BLACK, solid, line join=round, line cap=round, line width=1, fill=BLACK]
	(240,397) -- (237,387) -- (243,387) -- (240,397) -- cycle;
\draw[BLACK, solid, line join=round, line cap=round, line width=1]
	(183,263) -- (215,295);
\draw[BLACK, solid, line join=round, line cap=round, line width=1, fill=BLACK]
	(215,295) -- (206,291) -- (211,286) -- (215,295) -- cycle;
\draw[BLACK, solid, line join=round, line cap=round, line width=1]
	(645,545) -- (677,577);
\draw[BLACK, solid, line join=round, line cap=round, line width=1, fill=BLACK]
	(677,577) -- (668,572) -- (672,568) -- (677,577) -- cycle;
\draw[BLACK, solid, line join=round, line cap=round, line width=1]
	(620,155) -- (620,197);
\draw[BLACK, solid, line join=round, line cap=round, line width=1, fill=BLACK]
	(620,197) -- (617,187) -- (623,187) -- (620,197) -- cycle;
\draw[BLACK, solid, line join=round, line cap=round, line width=1]
	(351,448) -- (380,520) -- (380,520) -- (585,520);
\draw[BLACK, solid, line join=round, line cap=round, line width=1, fill=BLACK]
	(585,520) -- (575,523) -- (575,517) -- (585,520) -- cycle;
\draw[BLACK, solid, line join=round, line cap=round, line width=1]
	(593,542) -- (520,600) -- (520,600) -- (340,600) -- (340,460) -- (340,448);
\draw[BLACK, solid, line join=round, line cap=round, line width=1, fill=BLACK]
	(340,448) -- (343,458) -- (337,458) -- (340,448) -- cycle;
\draw[BLACK, solid, line join=round, line cap=round, line width=1]
	(655,520) -- (740,520) -- (740,520) -- (740,348);
\draw[BLACK, solid, line join=round, line cap=round, line width=1, fill=BLACK]
	(740,348) -- (743,358) -- (737,358) -- (740,348) -- cycle;
\draw[BLACK, solid, line join=round, line cap=round, line width=1]
	(740,292) -- (740,120) -- (740,120) -- (655,120);
\draw[BLACK, solid, line join=round, line cap=round, line width=1, fill=BLACK]
	(655,120) -- (665,117) -- (665,123) -- (655,120) -- cycle;
\draw[BLACK, solid, line join=round, line cap=round, line width=1]
	(595,95) -- (548,48);
\draw[BLACK, solid, line join=round, line cap=round, line width=1, fill=BLACK]
	(548,48) -- (557,53) -- (553,57) -- (548,48) -- cycle;
\draw[BLACK, solid, line join=round, line cap=round, line width=1]
	(492,20) -- (240,20) -- (240,20) -- (240,285);
\draw[BLACK, solid, line join=round, line cap=round, line width=1, fill=BLACK]
	(240,285) -- (237,275) -- (243,275) -- (240,285) -- cycle;
\draw[BLACK, solid, line join=round, line cap=round, line width=1, fill=CYAN]
	(417,397) rectangle +(46,46);
\draw[BLACK, solid, line join=round, line cap=round, line width=1]
	(440,355) -- (440,397);
\draw[BLACK, solid, line join=round, line cap=round, line width=1, fill=BLACK]
	(440,397) -- (437,387) -- (443,387) -- (440,397) -- cycle;

  \draw (240,320) node {$S$};
  \draw (160,240) node {$\Lambda$};
    \draw (240,425) node {$\mu$};
    \draw (340,225) node {$q\beta$};
    \draw (340,425) node {$\beta$};
    \draw (520,425) node {$\eta$};
    \draw (520,225) node {$\gamma$};
    \draw (520,25) node {$\rho$};
  \draw (440,320) node {$C$};
  \draw (370,280) node {$2$};
  \draw (440,425) node {$\mu$};
  \draw (620,120) node {$R$};
  \draw (620,225) node {$\mu$};
  \draw (620,520) node {$I$};
  \draw (700,605) node {$\mu$};
  
  \draw (740,320) node {$\alpha$};

\end{tikzpicture}
    \caption{The SCIR model.}
    \label{SCIRfig}
\end{figure}
The associated rate equations of the Petri net result in the following ODE:
    $$\begin{array}{l}
S'(t)=\Lambda -\beta S(I+qC)-\mu S+\rho R,\\
C'(t)=\beta S(I+qC)-(\eta +\gamma+\mu )C,\\
I'(t)=\eta C-(\alpha+\mu) I,\\
R'(t)=\alpha I+\gamma C-(\mu+\rho) R\,.
\end{array}$$
This model is found in the book of Martcheva in \cite[p. 93-94]{Ma}.
\end{example}

\begin{example}\label{water}
    Waterborne diseases, such as cholera, hepatitis A and E, norovirus, and rotavirus, are a severe problem for public health today. 
    For instance, Cholera is a disease with a long history that came into prominence in the 19th century, killing thousands of people in India.
    Infection for each disease is typically through pathogen ingestion (e.g., fecal-oral route). For instance, drinking sewage-contaminated water, eating food prepared by an individual with soiled hands, or acquiring an infection during treatment in a hospital. 
    
    The disease dynamics consists of the three compartments $\{S,I,R\}$, denoting susceptible, infected, and recovered, together with an additional compartment $W$ that measures pathogen concentration in a water source. There are two types of interactions: first, a susceptible individual having contact with an infected individual to become infected, and second, a susceptible individual having contact with contaminated water to become infected. We assume that the number of infected individuals produced in a unit of time by the interaction between the susceptible and infected populations is given at the transmission rate $\beta_I$. Similarly, we assume the number of infected individuals produced in a unit of time by the interaction of the susceptible population with contaminated water is given at the transmission rate $\beta_W$. We have the characteristic that infected individuals produce contaminated water at a rate $\alpha$ and remain infected. In addition, the infected individuals can pass to the recovery compartment at a rate $\gamma$. Additionally, we include a Malthusian model in the susceptible compartment where birth is provided at rate $\mu$ from the whole population denoted by $N$ and death rate $\mu$. In all the other compartments we consider death rates $\mu$ except the compartment of contaminated water $W$ where the decay rate of the pathogen in the water is $\xi$. We construct the Petri net of the SIWR model in Figure \ref{SIWRfig}. Similarly to the Petri nets for the SEIR, SEAIR models, we have arrows from the transitions with parameters $\beta_W$ and $\alpha$ to the compartments $W$ and $I$, respectively. This is because the contaminated water remains contaminated, and the infected individual remains infected.

    \begin{figure}
    \centering
    \begin{tikzpicture}[scale=0.3,x=1pt,y=-1pt]

\definecolor{BLACK}{RGB}{0,0,0}
\definecolor{r255g255b10}{RGB}{255,255,10}
\draw[BLACK, solid, line join=round, line cap=round, line width=1, fill=r255g255b10]
	(200,200) ellipse[x radius=30, y radius=30];
\draw[BLACK, solid, line join=round, line cap=round, line width=1, fill=r255g255b10]
	(440,200) ellipse[x radius=30, y radius=30];
\draw[BLACK, solid, line join=round, line cap=round, line width=1, fill=r255g255b10]
	(700,200) ellipse[x radius=30, y radius=30];
\draw[BLACK, solid, line join=round, line cap=round, line width=1, fill=r255g255b10]
	(940,200) ellipse[x radius=30, y radius=30];
\draw[BLACK, solid, line join=round, line cap=round, line width=1, fill=r255g255b10]
	(80,80) ellipse[x radius=30, y radius=30];
\definecolor{r33g255b255}{RGB}{33,255,255}
\draw[BLACK, solid, line join=round, line cap=round, line width=1, fill=r33g255b255]
	(175,55) rectangle +(50,50);
\draw[BLACK, solid, line join=round, line cap=round, line width=1, fill=r33g255b255]
	(182,262) rectangle +(36,36);
\draw[BLACK, solid, line join=round, line cap=round, line width=1, fill=r33g255b255]
	(502,242) rectangle +(36,36);
\draw[BLACK, solid, line join=round, line cap=round, line width=1, fill=r33g255b255]
	(762,242) rectangle +(36,36);
\draw[BLACK, solid, line join=round, line cap=round, line width=1, fill=r33g255b255]
	(922,262) rectangle +(36,36);
\draw[BLACK, solid, line join=round, line cap=round, line width=1, fill=r33g255b255]
	(295,75) rectangle +(50,50);
\draw[BLACK, solid, line join=round, line cap=round, line width=1, fill=r33g255b255]
	(555,175) rectangle +(50,50);
\draw[BLACK, solid, line join=round, line cap=round, line width=1, fill=r33g255b255]
	(795,175) rectangle +(50,50);
\draw[BLACK, solid, line join=round, line cap=round, line width=1, fill=r33g255b255]
	(295,275) rectangle +(50,50);
\draw[BLACK, solid, line join=round, line cap=round, line width=1]
	(200,105) -- (200,170);
\draw[BLACK, solid, line join=round, line cap=round, line width=1, fill=BLACK]
	(200,170) -- (197,160) -- (203,160) -- (200,170) -- cycle;
\draw[BLACK, solid, line join=round, line cap=round, line width=1]
	(200,230) -- (200,262);
\draw[BLACK, solid, line join=round, line cap=round, line width=1, fill=BLACK]
	(200,262) -- (197,252) -- (203,252) -- (200,262) -- cycle;
\draw[BLACK, solid, line join=round, line cap=round, line width=1]
	(470,223) -- (502,247);
\draw[BLACK, solid, line join=round, line cap=round, line width=1, fill=BLACK]
	(502,247) -- (492,243) -- (496,238) -- (502,247) -- cycle;
\draw[BLACK, solid, line join=round, line cap=round, line width=1]
	(730,223) -- (762,247);
\draw[BLACK, solid, line join=round, line cap=round, line width=1, fill=BLACK]
	(762,247) -- (752,243) -- (756,238) -- (762,247) -- cycle;
\draw[BLACK, solid, line join=round, line cap=round, line width=1]
	(940,230) -- (940,262);
\draw[BLACK, solid, line join=round, line cap=round, line width=1, fill=BLACK]
	(940,262) -- (937,252) -- (943,252) -- (940,262) -- cycle;
\draw[BLACK, solid, line join=round, line cap=round, line width=1]
	(230,175) -- (295,121);
\draw[BLACK, solid, line join=round, line cap=round, line width=1, fill=BLACK]
	(295,121) -- (289,130) -- (285,125) -- (295,121) -- cycle;
\draw[BLACK, solid, line join=round, line cap=round, line width=1]
	(470,200) -- (555,200);
\draw[BLACK, solid, line join=round, line cap=round, line width=1, fill=BLACK]
	(555,200) -- (545,203) -- (545,197) -- (555,200) -- cycle;
\draw[BLACK, solid, line join=round, line cap=round, line width=1]
	(605,200) -- (670,200);
\draw[BLACK, solid, line join=round, line cap=round, line width=1, fill=BLACK]
	(670,200) -- (660,203) -- (660,197) -- (670,200) -- cycle;
\draw[BLACK, solid, line join=round, line cap=round, line width=1]
	(845,200) -- (910,200);
\draw[BLACK, solid, line join=round, line cap=round, line width=1, fill=BLACK]
	(910,200) -- (900,203) -- (900,197) -- (910,200) -- cycle;
\draw[BLACK, solid, line join=round, line cap=round, line width=1]
	(110,80) -- (175,80);
\draw[BLACK, solid, line join=round, line cap=round, line width=1, fill=BLACK]
	(175,80) -- (165,83) -- (165,77) -- (175,80) -- cycle;
\draw[BLACK, solid, line join=round, line cap=round, line width=1]
	(230,225) -- (295,279);
\draw[BLACK, solid, line join=round, line cap=round, line width=1, fill=BLACK]
	(295,279) -- (285,275) -- (289,270) -- (295,279) -- cycle;
\draw[BLACK, solid, line join=round, line cap=round, line width=1]
	(345,279) -- (410,225);
\draw[BLACK, solid, line join=round, line cap=round, line width=1, fill=BLACK]
	(410,225) -- (404,234) -- (400,229) -- (410,225) -- cycle;
\draw[BLACK, solid, line join=round, line cap=round, line width=1]
	(700,230) -- (700,300) -- (700,300) -- (345,300);
\draw[BLACK, solid, line join=round, line cap=round, line width=1, fill=BLACK]
	(345,300) -- (355,297) -- (355,303) -- (345,300) -- cycle;
\draw[BLACK, solid, line join=round, line cap=round, line width=1]
	(320,325) -- (320,340) -- (320,340) -- (760,340) -- (713,230);
\draw[BLACK, solid, line join=round, line cap=round, line width=1, fill=BLACK]
	(713,230) -- (720,238) -- (714,241) -- (713,230) -- cycle;
\draw[BLACK, solid, line join=round, line cap=round, line width=1]
	(452,170) -- (480,100) -- (820,100) -- (820,175);
\draw[BLACK, solid, line join=round, line cap=round, line width=1, fill=BLACK]
	(820,175) -- (817,165) -- (823,165) -- (820,175) -- cycle;
\draw[BLACK, solid, line join=round, line cap=round, line width=1]
	(410,175) -- (345,121);
\draw[BLACK, solid, line join=round, line cap=round, line width=1, fill=BLACK]
	(345,121) -- (355,125) -- (351,130) -- (345,121) -- cycle;
\draw[BLACK, solid, line join=round, line cap=round, line width=1]
	(345,100) -- (440,100) -- (440,100) -- (440,170);
\draw[BLACK, solid, line join=round, line cap=round, line width=1, fill=BLACK]
	(440,170) -- (437,160) -- (443,160) -- (440,170) -- cycle;
\draw[BLACK, solid, line join=round, line cap=round, line width=1]
	(580,175) -- (580,120) -- (580,120) -- (520,120) -- (470,170);
\draw[BLACK, solid, line join=round, line cap=round, line width=1, fill=BLACK]
	(470,170) -- (475,161) -- (479,165) -- (470,170) -- cycle;

\draw (80,80) node {$N$};
\draw (200,80) node {$\mu$};
 \draw (200,200) node {$S$};

 \draw (200,285) node {$\mu$};
  \draw (322,100) node {$\beta_I$};

\draw (390,80) node {$2$};

 \draw (322,300) node {$\beta_W$};
 \draw (440,200) node {$I$};  
  \draw (520,265) node {$\mu$}; 
    \draw (780,263) node {$\xi$}; 
 \draw (580,200) node {$\alpha$};
 \draw (700,200) node {$W$};  
  \draw (820,200) node {$\gamma$};
  \draw (940,200) node {$R$};
   \draw (940,285) node {$\mu$};
\end{tikzpicture}
    \caption{The SIWR model.}
    \label{SIWRfig}
\end{figure}

Table \ref{SIWR} lists the parameters and the variables.
\begin{table}[h!]
    \centering
\begin{tabular}{ l l }
\hline
Notation & Meaning\\
\hline
 $\alpha$ & person-reservoir contact rate ("shedding") \\ 
 $\beta_W$ &  transmission rate for water-to-person\\  
 $\beta_I$ &  transmission rate for person-to-person\\  
 $1/\gamma$ & infectious period \\
 $1/\xi$ & pathogen lifetime in water reservoir \\
 $\mu$ & birth/death rate\\
 $S(t)$ & susceptible individuals\\
 $I(t)$ & infected individuals\\
 $W(t)$ &  pathogen concentration in water reservoir\\
 $R(t)$ & recovered individuals\\
 $N(t)$ & total population 
\end{tabular}
    \caption{List of parameters, variables, and their meanings for the SIWR model.}
    \label{SIWR}
\end{table}
The rate equations of this Petri net are the following system of ODEs:
$$\begin{array}{l}
S'(t)=\mu N-\beta_W SW -\beta_I SI -\mu S\,,\\
I'(t)=\beta_W SW+\beta_I SI -(\gamma+\mu)I\,,\\
W'(t)=\alpha I-\xi W\,,\\
R'(t)=\gamma I-\mu R\,.
\end{array}$$
This coincides with the system of ODEs studied in \cite[Sec. 2]{TE}.
\end{example}

\begin{example}
Malaria is one of the diseases that are constantly present in the human population. It is caused by
the entry of the malaria parasite called Plasmodium into the bloodstream due to the bite of an infected female
Anopheles mosquito. A single bite by a malaria-carrying mosquito can lead to extreme sickness or death. 

The disease dynamics of Malaria consist of two populations (humans/mosquitoes) with a common disease. There are three human compartments $\{S_H,I_H,R_H\}$, denoting susceptible, infected, and recovered. Meanwhile, there are two mosquito compartments $\{S_M,I_M\}$, denoting susceptible and infected. 
We have two types of interactions: first, a susceptible human is bitten by an infected mosquito to become infected, and second, a susceptible mosquito bites an infected human to become an infected mosquito. 
We assume the number of infected humans produced in a unit of time by the interaction of an infected mosquito with a susceptible human is given at a transmission rate $\beta_{HM}$. Similarly, we assume the number of infected mosquitoes produced in a unit of time by the interaction of an infected human with a susceptible mosquito is given at a transmission rate $\beta_{MH}$. 
Humans enter the infected class through the immigration rate $\delta$. All human individuals are subject to a natural death, which occurs at a rate $\mu_H$ and a disease-induced death rate $\alpha$. The infected humans can pass to the recovered compartment at a rate $\sigma$. We include Malthusian models (the simplified logistic) for the two susceptible populations of humans and mosquitoes. We construct the Petri net of the model of Malaria in Figure \ref{hig2}.
\begin{figure}
    \centering
    \begin{tikzpicture}[scale=0.3,x=1pt,y=-1pt]

\definecolor{BLACK}{RGB}{0,0,0}
\definecolor{YELLOW}{RGB}{255,255,0}
\draw[BLACK, solid, line join=round, line cap=round, line width=1, fill=YELLOW]
	(100,180) ellipse[x radius=35, y radius=35];
\draw[BLACK, solid, line join=round, line cap=round, line width=1, fill=YELLOW]
	(540,180) ellipse[x radius=35, y radius=35];
\draw[BLACK, solid, line join=round, line cap=round, line width=1, fill=YELLOW]
	(700,320) ellipse[x radius=35, y radius=35];
\draw[BLACK, solid, line join=round, line cap=round, line width=1, fill=YELLOW]
	(100,580) ellipse[x radius=30, y radius=30];
\draw[BLACK, solid, line join=round, line cap=round, line width=1, fill=YELLOW]
	(540,580) ellipse[x radius=30, y radius=30];
\definecolor{CYAN}{RGB}{0,255,255}
\draw[BLACK, solid, line join=round, line cap=round, line width=1, fill=CYAN]
	(77,37) rectangle +(46,46);
\draw[BLACK, solid, line join=round, line cap=round, line width=1, fill=CYAN]
	(77,277) rectangle +(46,46);
\draw[BLACK, solid, line join=round, line cap=round, line width=1, fill=CYAN]
	(272,152) rectangle +(56,56);
\draw[BLACK, solid, line join=round, line cap=round, line width=1, fill=CYAN]
	(517,257) rectangle +(46,46);
\draw[BLACK, solid, line join=round, line cap=round, line width=1, fill=CYAN]
	(517,677) rectangle +(46,46);
\draw[BLACK, solid, line join=round, line cap=round, line width=1, fill=CYAN]
	(677,157) rectangle +(46,46);
\draw[BLACK, solid, line join=round, line cap=round, line width=1, fill=CYAN]
	(82,462) rectangle +(36,36);
\draw[BLACK, solid, line join=round, line cap=round, line width=1, fill=CYAN]
	(82,662) rectangle +(36,36);
\draw[BLACK, solid, line join=round, line cap=round, line width=1, fill=CYAN]
	(272,552) rectangle +(56,56);
\draw[BLACK, solid, line join=round, line cap=round, line width=1, fill=CYAN]
	(677,417) rectangle +(46,46);
\draw[BLACK, solid, line join=round, line cap=round, line width=1, fill=CYAN]
	(505,17) rectangle +(70,46);
\draw[BLACK, solid, line join=round, line cap=round, line width=1]
	(100,83) -- (100,145);
\draw[BLACK, solid, line join=round, line cap=round, line width=1, fill=BLACK]
	(100,145) -- (97,135) -- (103,135) -- (100,145) -- cycle;
\draw[BLACK, solid, line join=round, line cap=round, line width=1]
	(100,215) -- (100,277);
\draw[BLACK, solid, line join=round, line cap=round, line width=1, fill=BLACK]
	(100,277) -- (97,267) -- (103,267) -- (100,277) -- cycle;
\draw[BLACK, solid, line join=round, line cap=round, line width=1]
	(135,180) -- (272,180);
\draw[BLACK, solid, line join=round, line cap=round, line width=1, fill=BLACK]
	(272,180) -- (262,183) -- (262,177) -- (272,180) -- cycle;
\draw[BLACK, solid, line join=round, line cap=round, line width=1]
	(540,215) -- (540,257);
\draw[BLACK, solid, line join=round, line cap=round, line width=1, fill=BLACK]
	(540,257) -- (537,247) -- (543,247) -- (540,257) -- cycle;
\draw[BLACK, solid, line join=round, line cap=round, line width=1]
	(700,203) -- (700,285);
\draw[BLACK, solid, line join=round, line cap=round, line width=1, fill=BLACK]
	(700,285) -- (697,275) -- (703,275) -- (700,285) -- cycle;
\draw[BLACK, solid, line join=round, line cap=round, line width=1]
	(100,498) -- (100,550);
\draw[BLACK, solid, line join=round, line cap=round, line width=1, fill=BLACK]
	(100,550) -- (97,540) -- (103,540) -- (100,550) -- cycle;
\draw[BLACK, solid, line join=round, line cap=round, line width=1]
	(100,610) -- (100,662);
\draw[BLACK, solid, line join=round, line cap=round, line width=1, fill=BLACK]
	(100,662) -- (97,652) -- (103,652) -- (100,662) -- cycle;
\draw[BLACK, solid, line join=round, line cap=round, line width=1]
	(540,145) -- (540,63);
\draw[BLACK, solid, line join=round, line cap=round, line width=1, fill=BLACK]
	(540,63) -- (543,73) -- (537,73) -- (540,63) -- cycle;
\draw[BLACK, solid, line join=round, line cap=round, line width=1]
	(575,180) -- (677,180);
\draw[BLACK, solid, line join=round, line cap=round, line width=1, fill=BLACK]
	(677,180) -- (667,183) -- (667,177) -- (677,180) -- cycle;
\draw[BLACK, solid, line join=round, line cap=round, line width=1]
	(700,355) -- (700,417);
\draw[BLACK, solid, line join=round, line cap=round, line width=1, fill=BLACK]
	(700,417) -- (697,407) -- (703,407) -- (700,417) -- cycle;
\draw[BLACK, solid, line join=round, line cap=round, line width=1]
	(540,610) -- (540,677);
\draw[BLACK, solid, line join=round, line cap=round, line width=1, fill=BLACK]
	(540,677) -- (537,667) -- (543,667) -- (540,677) -- cycle;
\draw[BLACK, solid, line join=round, line cap=round, line width=1]
	(328,580) -- (510,580);
\draw[BLACK, solid, line join=round, line cap=round, line width=1, fill=BLACK]
	(510,580) -- (500,583) -- (500,577) -- (510,580) -- cycle;
\draw[BLACK, solid, line join=round, line cap=round, line width=1]
	(130,580) -- (272,580);
\draw[BLACK, solid, line join=round, line cap=round, line width=1, fill=BLACK]
	(272,580) -- (262,583) -- (262,577) -- (272,580) -- cycle;
\draw[BLACK, solid, line join=round, line cap=round, line width=1]
	(525,554) -- (317,208);
\draw[BLACK, solid, line join=round, line cap=round, line width=1, fill=BLACK]
	(317,208) -- (325,215) -- (319,218) -- (317,208) -- cycle;
\draw[BLACK, solid, line join=round, line cap=round, line width=1]
	(522,210) -- (317,552);
\draw[BLACK, solid, line join=round, line cap=round, line width=1, fill=BLACK]
	(317,552) -- (319,542) -- (325,545) -- (317,552) -- cycle;
\draw[BLACK, solid, line join=round, line cap=round, line width=1]
	(328,180) -- (505,180);
\draw[BLACK, solid, line join=round, line cap=round, line width=1, fill=BLACK]
	(505,180) -- (495,183) -- (495,177) -- (505,180) -- cycle;

\draw (100,61) node {$\Pi$};
\draw (100,180) node {$S_H$};
\draw (102,305) node {\scalebox{0.9}{$\mu_H$}};

\draw (703,180) node {$\sigma$};

\draw (540,40) node {\scalebox{0.8}{$\alpha-\delta$}};

\draw (542,180) node {$I_H$};
\draw (542,284) node {\scalebox{0.9}{$\mu_H$}};
\draw (303,180) node {\scalebox{0.7}{$\beta_{HM}$}};
\draw (703,323) node {$R_H$};
\draw (703,444) node {\scalebox{0.9}{$\mu_H$}};

\draw (100,480) node {$\Lambda$};

\draw (100,580) node {$S_M$};
\draw (542,580) node {$I_M$};

\draw (303,580) node {\scalebox{0.7}{$\beta_{MH}$}};

\draw (102,680) node {\scalebox{0.7}{$\mu_M$}};
\draw (542,700) node {\scalebox{0.85}{$\mu_M$}};

\end{tikzpicture}
    \caption{A SIR model for Malaria.}
    \label{hig2}
\end{figure}
We list the parameters and the variables in the Table \ref{Malarion}.
\begin{table}[h!]
    \centering
\begin{tabular}{ l l }
\hline
Notation & Meaning\\
\hline
 $\beta_{HM}$ &  human-mosquito transmission rate\\  
 $\beta_{MH}$ &  mosquito-human transmission rate\\  
 $\Pi$ & birth human rate\\
 $\Lambda$ & birth mosquito rate\\
 $\mu_H$ & death human rate\\
 $\mu_M$ & death mosquito rate\\
 $\alpha$ & disease-induced rate for humans\\
 $\delta$ & infected migration rate for humans\\
 $S_H(t)$ & susceptible human individuals\\
 $I_H(t)$ & infected human individuals\\
 $R_H(t)$ & recovered human individuals\\
  $S_M(t)$ & susceptible mosquito individuals\\
 $I_M(t)$ & infected mosquito individuals\\
\end{tabular}
    \caption{List of parameters, variables, and their meanings for the model of Malaria.}
    \label{Malarion}
\end{table}

The rate equations of the Petri net of the model of Malaria are:
\begin{equation}\label{Malaria}
\begin{array}{l}
S'_H(t)=\Pi-\beta_{HM}S_HI_M-\mu_HS_H\,,\\
I'_H(t)=\beta_{HM}S_HI_M-(\mu_H+\sigma)I_H- (\alpha-\delta) I_H\,,\\
R'_H(t)=\sigma I_H-\mu_HR_H\,,\\
S'_M(t)=\Lambda-\beta_{MH}S_MI_H-\mu_MS_M\,,\\
I'_M(t)=\beta_{MH}S_MI_H-\mu_MI_M\,.
\end{array}
\end{equation}
This system of ODEs is studied in \cite[Sec. II]{WBK}.

\end{example}

Now, we study the Petri nets associated with control strategies implemented in epidemiological models. We base our systems of ODE on the book \cite{Ma}.
\begin{example}\label{1stVac}
    Vaccination is among the most outstanding public health achievements. For instance, vaccination has led to the complete eradication of smallpox worldwide and a near eradication of polio. There are two points at which vaccination models can differ from one another. The first model treats vaccination as equivalent to going through the disease, so the vaccinated individuals resist the disease. 
    Here, we are assuming a perfect vaccine where everybody is wholly protected.

    The disease dynamics consider the usual SIR model with the three compartments ${S,I,R}$, denoting susceptible, infected, and recovered, together with the addition of a compartment $N$, meaning the whole population. 
    The process of vaccination can be understood as follows:
    individuals in the compartment $N$ has a bifurcation and can pass  
    at a rate $\mu$ where a fraction $p$ are vaccinated and pass to the recovered compartment, meanwhile a fraction $q=1-p$ pass to the susceptible compartment. 
    Now, there is only one interaction from a susceptible individual having contact with an infected individual to become infected. We assume the number of infected individuals produced by this interaction in a unit of time is given at rate $\beta$.
    The infected individuals can pass to the recovered compartment at a rate $\alpha$.
    Finally, all the individuals in the compartments $S$, $I$, and $R$ can disappear at a death rate $\mu$. 
    The Petri net resulting from this epidemic process is given in Figure \ref{Vac1}. The rate equations of this Petri net recover the system of ODEs in \cite[p. 217]{Ma}, given as follows:
    \begin{figure}
    \centering
   \begin{tikzpicture}[scale=0.3,x=1pt,y=-1pt]

\definecolor{BLACK}{RGB}{0,0,0}
\definecolor{r255g255b10}{RGB}{255,255,10}
\draw[BLACK, solid, line join=round, line cap=round, line width=1, fill=r255g255b10]
	(200,120) ellipse[x radius=30, y radius=30];
\draw[BLACK, solid, line join=round, line cap=round, line width=1, fill=r255g255b10]
	(400,120) ellipse[x radius=30, y radius=30];
\draw[BLACK, solid, line join=round, line cap=round, line width=1, fill=r255g255b10]
	(200,280) ellipse[x radius=30, y radius=30];
\draw[BLACK, solid, line join=round, line cap=round, line width=1, fill=r255g255b10]
	(100,40) ellipse[x radius=30, y radius=30];
\definecolor{r33g255b255}{RGB}{33,255,255}
\draw[BLACK, solid, line join=round, line cap=round, line width=1, fill=r33g255b255]
	(75,175) rectangle +(50,50);
\draw[BLACK, solid, line join=round, line cap=round, line width=1, fill=r33g255b255]
	(182,182) rectangle +(36,36);
\draw[BLACK, solid, line join=round, line cap=round, line width=1, fill=r33g255b255]
	(275,175) rectangle +(50,50);
\draw[BLACK, solid, line join=round, line cap=round, line width=1, fill=r33g255b255]
	(275,15) rectangle +(50,50);
\draw[BLACK, solid, line join=round, line cap=round, line width=1, fill=r33g255b255]
	(382,182) rectangle +(36,36);
\draw[BLACK, solid, line join=round, line cap=round, line width=1, fill=r33g255b255]
	(182,342) rectangle +(36,36);
\draw[BLACK, solid, line join=round, line cap=round, line width=1]
	(125,180) -- (170,144);
\draw[BLACK, solid, line join=round, line cap=round, line width=1, fill=BLACK]
	(170,144) -- (164,153) -- (160,148) -- (170,144) -- cycle;
\draw[BLACK, solid, line join=round, line cap=round, line width=1]
	(200,150) -- (200,182);
\draw[BLACK, solid, line join=round, line cap=round, line width=1, fill=BLACK]
	(200,182) -- (197,172) -- (203,172) -- (200,182) -- cycle;
\draw[BLACK, solid, line join=round, line cap=round, line width=1]
	(370,144) -- (325,180);
\draw[BLACK, solid, line join=round, line cap=round, line width=1, fill=BLACK]
	(325,180) -- (331,171) -- (335,176) -- (325,180) -- cycle;
\draw[BLACK, solid, line join=round, line cap=round, line width=1]
	(230,96) -- (275,60);
\draw[BLACK, solid, line join=round, line cap=round, line width=1, fill=BLACK]
	(275,60) -- (269,69) -- (265,64) -- (275,60) -- cycle;
\draw[BLACK, solid, line join=round, line cap=round, line width=1]
	(400,150) -- (400,182);
\draw[BLACK, solid, line join=round, line cap=round, line width=1, fill=BLACK]
	(400,182) -- (397,172) -- (403,172) -- (400,182) -- cycle;
\draw[BLACK, solid, line join=round, line cap=round, line width=1]
	(275,220) -- (230,256);
\draw[BLACK, solid, line join=round, line cap=round, line width=1, fill=BLACK]
	(230,256) -- (236,247) -- (240,252) -- (230,256) -- cycle;
\draw[BLACK, solid, line join=round, line cap=round, line width=1]
	(125,220) -- (170,256);
\draw[BLACK, solid, line join=round, line cap=round, line width=1, fill=BLACK]
	(170,256) -- (160,252) -- (164,247) -- (170,256) -- cycle;
\draw[BLACK, solid, line join=round, line cap=round, line width=1]
	(200,310) -- (200,342);
\draw[BLACK, solid, line join=round, line cap=round, line width=1, fill=BLACK]
	(200,342) -- (197,332) -- (203,332) -- (200,342) -- cycle;
\draw[BLACK, solid, line join=round, line cap=round, line width=1]
	(100,70) -- (100,175);
\draw[BLACK, solid, line join=round, line cap=round, line width=1, fill=BLACK]
	(100,175) -- (97,165) -- (103,165) -- (100,175) -- cycle;
\draw[BLACK, solid, line join=round, line cap=round, line width=1]
	(325,40) -- (400,40) -- (400,40) -- (400,90);
\draw[BLACK, solid, line join=round, line cap=round, line width=1, fill=BLACK]
	(400,90) -- (397,80) -- (403,80) -- (400,90) -- cycle;
\draw[BLACK, solid, line join=round, line cap=round, line width=1]
	(370,96) -- (325,60);
\draw[BLACK, solid, line join=round, line cap=round, line width=1, fill=BLACK]
	(325,60) -- (335,64) -- (331,69) -- (325,60) -- cycle;

\draw (100,40) node {$N$};
\draw (100,205) node {$\mu$};
\draw (200,120) node {$S$};
 \draw (200,205) node {$\mu$};
\draw (140,125) node {$q$};
\draw (300,200) node {$\alpha$};
\draw (300,40) node {$\beta$};

\draw (365,20) node {$2$};

\draw (400,120) node {$I$};
\draw (400,205) node {$\mu$};
\draw (200,280) node {$R$};
\draw (200,365) node {$\mu$};
\draw (140,275) node {$p$}; 
\end{tikzpicture}
    \caption{First model of vaccination.}
    \label{Vac1}
\end{figure}

    $$\begin{array}{l}
S'(t)=q\mu N-\beta SI-\mu S,\\
I'(t)=\beta SI-(\mu+\alpha) I,\\
R'(t)=p\mu N+\alpha I-\mu R\,.
\end{array}$$
\end{example}

\begin{example}\label{2ndVac}
The second model for vaccination is when we consider that vaccines are rarely perfect. 
Now, the disease dynamics consider the SIS model with the two compartments $\{S, I\}$ denoting
susceptible and infected. 
We add a third compartment $V$ of vaccinated individuals. 
The susceptible population is vaccinated at a rate $\psi$ and passes to the compartment $V$.
Infected individuals has a bifurcation and can pass at a rate $\gamma$ where a fraction $\chi$ pass to the compartment of susceptible individuals and a fraction $1-\chi$ pass to the compartment of vaccinated individuals.
There are two types of interactions: first, a susceptible individual having contact with an infected individual to become infected, and second, a vaccinated individual having contact with an infected individual to become infected. 
We assume a proportion of $1/N$ (where $N$ is the total population) of the susceptible population interacts with the infected population and produces infected individuals in a unit of time at a rate $\beta$.
Similarly, we assume a proportion of $1/N$ of the vaccinated population interacts with the infected population and produces infected individuals in a unit of time at a rate $\beta\delta$.
We include a Malthusian model (the simplified logistic) in the susceptible compartment with birth rate $\Lambda$ and death rate $\mu$, and all the individuals in the remaining compartments can disappear at a death rate $\mu$. We construct the associated Petri net of this epidemic process in Figure \ref{Vac2}. 
\begin{figure}
    \centering
   \begin{tikzpicture}[scale=0.3,x=1pt,y=-1pt]

\definecolor{BLACK}{RGB}{0,0,0}
\definecolor{r255g255b10}{RGB}{255,255,10}
\draw[BLACK, solid, line join=round, line cap=round, line width=1, fill=r255g255b10]
	(400,200) ellipse[x radius=30, y radius=30];
\definecolor{r33g255b255}{RGB}{33,255,255}
\draw[BLACK, solid, line join=round, line cap=round, line width=1, fill=r33g255b255]
	(482,182) rectangle +(36,36);
\draw[BLACK, solid, line join=round, line cap=round, line width=1, fill=r33g255b255]
	(275,95) rectangle +(50,50);
\draw[BLACK, solid, line join=round, line cap=round, line width=1]
	(430,200) -- (482,200);
\draw[BLACK, solid, line join=round, line cap=round, line width=1, fill=BLACK]
	(482,200) -- (472,203) -- (472,197) -- (482,200) -- cycle;
\draw[BLACK, solid, line join=round, line cap=round, line width=1, fill=r255g255b10]
	(300,400) ellipse[x radius=30, y radius=30];
\draw[BLACK, solid, line join=round, line cap=round, line width=1, fill=r33g255b255]
	(282,482) rectangle +(36,36);
\draw[BLACK, solid, line join=round, line cap=round, line width=1]
	(300,430) -- (300,482);
\draw[BLACK, solid, line join=round, line cap=round, line width=1, fill=BLACK]
	(300,482) -- (297,472) -- (303,472) -- (300,482) -- cycle;
\draw[BLACK, solid, line join=round, line cap=round, line width=1, fill=r255g255b10]
	(180,200) ellipse[x radius=30, y radius=30];
\draw[BLACK, solid, line join=round, line cap=round, line width=1, fill=r33g255b255]
	(15,175) rectangle +(50,50);
\draw[BLACK, solid, line join=round, line cap=round, line width=1]
	(65,200) -- (150,200);
\draw[BLACK, solid, line join=round, line cap=round, line width=1, fill=BLACK]
	(150,200) -- (140,203) -- (140,197) -- (150,200) -- cycle;
\draw[BLACK, solid, line join=round, line cap=round, line width=1]
	(210,180) -- (275,137);
\draw[BLACK, solid, line join=round, line cap=round, line width=1, fill=BLACK]
	(275,137) -- (269,145) -- (265,139) -- (275,137) -- cycle;
\draw[BLACK, solid, line join=round, line cap=round, line width=1]
	(370,176) -- (325,140);
\draw[BLACK, solid, line join=round, line cap=round, line width=1, fill=BLACK]
	(325,140) -- (335,144) -- (331,149) -- (325,140) -- cycle;
\draw[BLACK, solid, line join=round, line cap=round, line width=1, fill=r33g255b255]
	(102,262) rectangle +(36,36);
\draw[BLACK, solid, line join=round, line cap=round, line width=1]
	(158,230) -- (134,262);
\draw[BLACK, solid, line join=round, line cap=round, line width=1, fill=BLACK]
	(134,262) -- (137,252) -- (142,256) -- (134,262) -- cycle;
\draw[BLACK, solid, line join=round, line cap=round, line width=1, fill=r33g255b255]
	(275,255) rectangle +(50,50);
\draw[BLACK, solid, line join=round, line cap=round, line width=1]
	(370,224) -- (325,260);
\draw[BLACK, solid, line join=round, line cap=round, line width=1, fill=BLACK]
	(325,260) -- (331,251) -- (335,256) -- (325,260) -- cycle;
\draw[BLACK, solid, line join=round, line cap=round, line width=1, fill=r33g255b255]
	(375,375) rectangle +(50,50);
\draw[BLACK, solid, line join=round, line cap=round, line width=1]
	(400,230) -- (400,375);
\draw[BLACK, solid, line join=round, line cap=round, line width=1, fill=BLACK]
	(400,375) -- (397,365) -- (403,365) -- (400,375) -- cycle;
\draw[BLACK, solid, line join=round, line cap=round, line width=1, fill=r33g255b255]
	(155,375) rectangle +(50,50);
\draw[BLACK, solid, line join=round, line cap=round, line width=1]
	(180,230) -- (180,375);
\draw[BLACK, solid, line join=round, line cap=round, line width=1, fill=BLACK]
	(180,375) -- (177,365) -- (183,365) -- (180,375) -- cycle;
\draw[BLACK, solid, line join=round, line cap=round, line width=1]
	(205,400) -- (270,400);
\draw[BLACK, solid, line join=round, line cap=round, line width=1, fill=BLACK]
	(270,400) -- (260,403) -- (260,397) -- (270,400) -- cycle;
\draw[BLACK, solid, line join=round, line cap=round, line width=1]
	(325,120) -- (400,120) -- (400,120) -- (400,170);
\draw[BLACK, solid, line join=round, line cap=round, line width=1, fill=BLACK]
	(400,170) -- (397,160) -- (403,160) -- (400,170) -- cycle;
\draw[BLACK, solid, line join=round, line cap=round, line width=1]
	(300,305) -- (300,370);
\draw[BLACK, solid, line join=round, line cap=round, line width=1, fill=BLACK]
	(300,370) -- (297,360) -- (303,360) -- (300,370) -- cycle;
\draw[BLACK, solid, line join=round, line cap=round, line width=1]
	(275,263) -- (210,220);
\draw[BLACK, solid, line join=round, line cap=round, line width=1, fill=BLACK]
	(210,220) -- (220,223) -- (216,228) -- (210,220) -- cycle;
\draw[BLACK, solid, line join=round, line cap=round, line width=1]
	(330,400) -- (375,400);
\draw[BLACK, solid, line join=round, line cap=round, line width=1, fill=BLACK]
	(375,400) -- (365,403) -- (365,397) -- (375,400) -- cycle;
\draw[BLACK, solid, line join=round, line cap=round, line width=1]
	(425,400) -- (480,400) -- (480,400) -- (412,230);
\draw[BLACK, solid, line join=round, line cap=round, line width=1, fill=BLACK]
	(412,230) -- (419,238) -- (413,241) -- (412,230) -- cycle;

 \draw (40,200) node {$\Lambda$};
 \draw (180,200) node {$S$};
 \draw (120,285) node {$\mu$};
 \draw (180,400) node {$\psi$};
  \draw (400,200) node {$I$};

 \draw (415,105) node {$2$};

 \draw (215,135) node {\scalebox{0.8}{$1/N$}};
  
  \draw (500,205) node {$\mu$};
  \draw (400,400) node {$\beta\delta$};
   \draw (300,400) node {$V$};

 \draw (345,440) node {\scalebox{0.8}{$1/N$}};
   
   \draw (300,505) node {$\mu$};
    \draw (300,120) node {$\beta$};
      \draw (300,280) node {$\gamma$};
      \draw (260,220) node {\scalebox{0.8}{$\chi$}};
        \draw (335,340) node {\scalebox{0.8}{$1-\chi$}};
         \draw (478,340) node {2};
 
\end{tikzpicture}
    \caption{Second model of vaccination.}
    \label{Vac2}
\end{figure}
In Table \ref{vacuna}, we detail all the parameters used in this model and the description of the compartments.
\begin{table}[h!]
    \centering
\begin{tabular}{ l l }
\hline
Notation & Meaning\\
\hline
 $\Lambda$ & birth rate\\
 $\mu$ & death rate\\
 $\beta$ & transmission rate\\
 $\gamma$ & recovery rate\\
 $\chi$ & proportion of individuals who recover from the vaccination class\\
 $1-\chi$ & proportion of individuals who recover to the susceptible class\\
 $\psi$ & vaccination rate \\
 $\epsilon =1-\delta$ & vaccine efficacy\\
 $S(t)$ & susceptible individuals\\
 $I(t)$ & infected individuals\\
 $V(t)$ & vaccinated individuals 
\end{tabular}
    \caption{List of parameters, variables, and meanings for the second vaccination model.}
    \label{vacuna}
\end{table}
The rate equations of this Petri net are as follows:
\begin{equation}
\label{SIRVAC}
\begin{array}{l}
S'(t)=\Lambda-\beta SI/N-(\mu+\psi)S+\chi \gamma I,\\
I'(t)= \beta SI/N+\beta\delta VI/N-(\mu+\gamma) I,\\
V'(t)=\psi S-\beta\delta VI/N+(1-\chi)\gamma I-\mu V\,.
\end{array}
\end{equation}
This model coincides with the system of ODEs in the book \cite[p. 219]{Ma}, which is a minor modification of the model in \cite{XVelasco}.
\end{example}

\begin{example} Quarantine is one of the more antique control strategies to prevent the spread of disease. It is essential when vaccination or treatment is not possible.

The disease dynamics consider the SIR model with the compartments $\{S,I,R\}$, denoting susceptible, infected, and recovered.
We add a compartment $Q$ which comprises the isolated individuals in quarantine. 
Now, there is only one interaction from a susceptible individual having contact with an infected individual to become infected. 
We assume a proportion of $1/A$ of the susceptible population interacts with the infected population and produces
infected individuals in a unit of time at a rate $\beta$. The class $A$ is given by the sum $A(t)=S(t) + I (t) + R (t)$ of all the mixing individuals.
The infected individuals can pass to the compartment $R$ at a rate $\alpha$ or can pass to the compartment $Q$ at a rate $\gamma$. The individuals in the compartment $Q$ can pass to the compartment $R$ at a rate $\eta$.  
We include a Malthusian model (the simplified logistic) in the susceptible compartment $S$ with birth rate $\Lambda$ and death rate $\mu$, and all the individuals in the remaining compartments can disappear at a death rate $\mu$. We construct the associated Petri net of this epidemic process in Figure \ref{Qua}. 
\begin{figure}
    \centering
   \begin{tikzpicture}[scale=0.3,x=1pt,y=-1pt]

\definecolor{BLACK}{RGB}{0,0,0}
\definecolor{YELLOW}{RGB}{255,255,0}
\draw[BLACK, solid, line join=round, line cap=round, line width=1, fill=YELLOW]
	(140,320) ellipse[x radius=35, y radius=35];
\draw[BLACK, solid, line join=round, line cap=round, line width=1, fill=YELLOW]
	(460,320) ellipse[x radius=35, y radius=35];
\definecolor{CYAN}{RGB}{0,255,255}
\draw[BLACK, solid, line join=round, line cap=round, line width=1, fill=CYAN]
	(272,192) rectangle +(56,56);
\draw[BLACK, solid, line join=round, line cap=round, line width=1]
	(170,301) -- (272,238);
\draw[BLACK, solid, line join=round, line cap=round, line width=1, fill=BLACK]
	(272,238) -- (265,246) -- (262,240) -- (272,238) -- cycle;
\draw[BLACK, solid, line join=round, line cap=round, line width=1]
	(430,301) -- (328,238);
\draw[BLACK, solid, line join=round, line cap=round, line width=1, fill=BLACK]
	(328,238) -- (338,240) -- (335,246) -- (328,238) -- cycle;
\draw[BLACK, solid, line join=round, line cap=round, line width=1, fill=CYAN]
	(117,177) rectangle +(46,46);
\draw[BLACK, solid, line join=round, line cap=round, line width=1, fill=CYAN]
	(117,417) rectangle +(46,46);
\draw[BLACK, solid, line join=round, line cap=round, line width=1, fill=CYAN]
	(437,417) rectangle +(46,46);
\draw[BLACK, solid, line join=round, line cap=round, line width=1]
	(460,355) -- (460,417);
\draw[BLACK, solid, line join=round, line cap=round, line width=1, fill=BLACK]
	(460,417) -- (457,407) -- (463,407) -- (460,417) -- cycle;
\draw[BLACK, solid, line join=round, line cap=round, line width=1]
	(140,355) -- (140,417);
\draw[BLACK, solid, line join=round, line cap=round, line width=1, fill=BLACK]
	(140,417) -- (137,407) -- (143,407) -- (140,417) -- cycle;
\draw[BLACK, solid, line join=round, line cap=round, line width=1]
	(140,223) -- (140,285);
\draw[BLACK, solid, line join=round, line cap=round, line width=1, fill=BLACK]
	(140,285) -- (137,275) -- (143,275) -- (140,285) -- cycle;
\draw[BLACK, solid, line join=round, line cap=round, line width=1]
	(328,220) -- (460,220) -- (460,220) -- (460,285);
\draw[BLACK, solid, line join=round, line cap=round, line width=1, fill=BLACK]
	(460,285) -- (457,275) -- (463,275) -- (460,285) -- cycle;
\draw[BLACK, solid, line join=round, line cap=round, line width=1, fill=YELLOW]
	(740,120) ellipse[x radius=35, y radius=35];
\draw[BLACK, solid, line join=round, line cap=round, line width=1, fill=YELLOW]
	(740,520) ellipse[x radius=35, y radius=35];
\draw[BLACK, solid, line join=round, line cap=round, line width=1, fill=CYAN]
	(572,192) rectangle +(56,56);
\draw[BLACK, solid, line join=round, line cap=round, line width=1, fill=CYAN]
	(572,392) rectangle +(56,56);
\draw[BLACK, solid, line join=round, line cap=round, line width=1]
	(488,340) -- (572,400);
\draw[BLACK, solid, line join=round, line cap=round, line width=1, fill=BLACK]
	(572,400) -- (562,397) -- (566,391) -- (572,400) -- cycle;
\draw[BLACK, solid, line join=round, line cap=round, line width=1]
	(628,440) -- (712,500);
\draw[BLACK, solid, line join=round, line cap=round, line width=1, fill=BLACK]
	(712,500) -- (701,497) -- (705,491) -- (712,500) -- cycle;
\draw[BLACK, solid, line join=round, line cap=round, line width=1]
	(488,300) -- (572,240);
\draw[BLACK, solid, line join=round, line cap=round, line width=1, fill=BLACK]
	(572,240) -- (566,249) -- (562,243) -- (572,240) -- cycle;
\draw[BLACK, solid, line join=round, line cap=round, line width=1]
	(628,200) -- (712,140);
\draw[BLACK, solid, line join=round, line cap=round, line width=1, fill=BLACK]
	(712,140) -- (705,149) -- (701,143) -- (712,140) -- cycle;
\draw[BLACK, solid, line join=round, line cap=round, line width=1, fill=CYAN]
	(952,292) rectangle +(56,56);
\draw[BLACK, solid, line join=round, line cap=round, line width=1, fill=CYAN]
	(717,217) rectangle +(46,46);
\draw[BLACK, solid, line join=round, line cap=round, line width=1, fill=CYAN]
	(717,617) rectangle +(46,46);
\draw[BLACK, solid, line join=round, line cap=round, line width=1]
	(740,155) -- (740,217);
\draw[BLACK, solid, line join=round, line cap=round, line width=1, fill=BLACK]
	(740,217) -- (737,207) -- (743,207) -- (740,217) -- cycle;
\draw[BLACK, solid, line join=round, line cap=round, line width=1]
	(740,555) -- (740,617);
\draw[BLACK, solid, line join=round, line cap=round, line width=1, fill=BLACK]
	(740,617) -- (737,607) -- (743,607) -- (740,617) -- cycle;
\draw[BLACK, solid, line join=round, line cap=round, line width=1]
	(767,142) -- (952,297);
\draw[BLACK, solid, line join=round, line cap=round, line width=1, fill=BLACK]
	(952,297) -- (942,293) -- (946,288) -- (952,297) -- cycle;
\draw[BLACK, solid, line join=round, line cap=round, line width=1]
	(952,343) -- (767,498);
\draw[BLACK, solid, line join=round, line cap=round, line width=1, fill=BLACK]
	(767,498) -- (772,489) -- (777,494) -- (767,498) -- cycle;

\draw (215,240) node {$1/A$};

\draw (140,205) node {$\Lambda$};

 \draw (140,320) node {$S$};

   \draw (140,445) node {$\mu$};

 \draw (460,320) node {$I$}; 

 \draw (480,198) node {$2$};

 \draw (980,320) node {$\eta$};

  \draw (300,220) node {$\beta$};
 
 \draw (460,445) node {$\mu$};

 \draw (740,120) node {$Q$}; 

\draw (740,245) node {$\mu$};

\draw (600,220) node {$\gamma$};

\draw (600,420) node {$\alpha$};

  \draw (740,520) node {$R$};
 \draw (740,645) node {$\mu$};

\end{tikzpicture}
    \caption{The SIR model with quarantine.}
    \label{Qua}
\end{figure}
The rate equations of this Petri net are as follows:
 $$\begin{array}{l}
S'(t)=\Lambda-\beta SI/A-\mu S,\\
I'(t)= \beta SI/A-(\alpha+\gamma+\mu)I,\\
Q'(t)=\gamma I-(\eta+\mu)Q,\\
R'(t)=\alpha I+\eta Q-\mu R\,.
\end{array}$$
This system of ODEs is studied in \cite[p. 95]{Ma}.
\end{example}

\subsection{Correspondence between ODEs and Petri nets}
\label{inverseODE}
One of the aims of our work is to propose Petri nets as a way of thinking about an epidemiological model of a disease.
We have studied in Section \ref{secPN} several different models of diseases (together with the Malthusian model) to 
characterize the patterns observed in their associated Petri net diagrams. We assemble the different parts or components of a model to obtain the whole picture of the Petri net that models our phenomenon. The theory of ordinary differential equations is one of the basic tools of mathematical science; see \cite{arnold}. Therefore, it is of great interest to consider the geometric approach to thinking about evolutionary processes. Petri nets provide a way of thinking about models for designing a playing board to simulate the evolution of our phenomenon. However, for Petri nets to be an acceptable alternative to the theory of ODEs, we need to establish a certain correspondence between ODEs and Petri nets. 
From Section \ref{secPN}, we know that Petri nets, together with a rate function, uniquely define a system of 
ODE. We accomplished this section by solving the inverse problem of finding, for any ordinary differential equation (ODE), a Petri net whose rate equations recover the original ODE. 

\subsubsection{The inverse problem between ODEs and Petri nets}
The general form of a system of ODEs is composed of $k$ equations for $x_1'(t)$, $\cdots$, $x_k'(t)$.
Suppose an order of all the exponents in the monomials $x_1^{m_{1,j}}\cdots x_k^{m_{k,j}}$ by the relation $<$ starting from left to right (e.g., for $k=1$, we can take the order $0,1,\cdots$). Thus, we are ordering the monomials by the index $1\leq j\leq l$, where $l$ is the number of tuples $(m_{1,j},\cdots,m_{k,j})$ not equal to the zero vector.
Any system of ODEs has the following form:
\begin{equation}\label{epi2} x_i'(t)=\sum_{j= 1}^{l_i}f_{i,j}(m_{1,j},\cdots,m_{k,j})\alpha_{i,j}(m_{1,j},\cdots,m_{k,j})x_1^{m_{1,j}}\dots x_k^{m_{k,j}}\,,\end{equation}
for $i\in \{1,\cdots,k\}$. The description of the components of \eqref{epi2} are as follows:
\begin{itemize}
\item $f_{i,j}(m_{1,j},\cdots,m_{k,j})$ is an integer function depending on $m_{1,j},\cdots,m_{k,j}$; and 
\item $\alpha_{i,j}(m_{1,j},\cdots,m_{k,j})$ is a parameter function\footnote{We refer to a parameter function that associates to each tuple an indeterminate variable.} depending on $m_{1,j},\cdots,m_{k,j}$. 
\end{itemize}
For the base case $k=1$. The ODE has the form:
\begin{equation}\label{re1}x_1'(t)=\sum_{j=1}^l f_j(m_j)\alpha_j(m_j)x^{m_j}\,.\end{equation}
Consider the integer values 
$$n_j:=f_j(m_j)+m_j\,,$$for $1\leq j\leq l$.
The Petri net is constructed in Figure \ref{onecomp}. 
\begin{figure}
    \centering
\begin{tikzpicture}[scale=0.3,x=1pt,y=-1pt]

\definecolor{BLACK}{RGB}{0,0,0}
\definecolor{YELLOW}{RGB}{255,255,0}
\draw[BLACK, solid, line join=round, line cap=round, line width=1, fill=YELLOW]
(340,340) ellipse[x radius=45, y radius=45];
\definecolor{CYAN}{RGB}{0,255,255}
\draw[BLACK, solid, line join=round, line cap=round, line width=1, fill=CYAN]
(550,170) rectangle +(140,60);
\draw[BLACK, solid, line join=round, line cap=round, line width=1, fill=CYAN]
(550,450) rectangle +(140,60);
\draw[BLACK, solid, line join=round, line cap=round, line width=1]
(340,295) -- (340,200) -- (340,200) -- (550,200);
\draw[BLACK, solid, line join=round, line cap=round, line width=1, fill=BLACK]
(550,200) -- (540,203) -- (540,197) -- (550,200) -- cycle;
\draw[BLACK, solid, line join=round, line cap=round, line width=1]
(560,230) -- (380,320);
\draw[BLACK, solid, line join=round, line cap=round, line width=1, fill=BLACK]
(380,320) -- (388,312) -- (391,318) -- (380,320) -- cycle;
\draw[BLACK, solid, line join=round, line cap=round, line width=1]
(340,385) -- (340,480) -- (550,480);
\draw[BLACK, solid, line join=round, line cap=round, line width=1, fill=BLACK]
(550,480) -- (540,483) -- (540,477) -- (550,480) -- cycle;
\draw[BLACK, solid, line join=round, line cap=round, line width=1]
(560,450) -- (380,360);
\draw[BLACK, solid, line join=round, line cap=round, line width=1, fill=BLACK]
(380,360) -- (391,362) -- (388,368) -- (380,360) -- cycle;

\draw (620,200) node {\scalebox{1.0}{$\alpha_1(m_1)$}};
\draw (620,320) node {\scalebox{1.7}{\vdots}};
\draw (620,480) node {\scalebox{1.0}{$\alpha_l(m_l)$}};
\draw (340,340) node {\scalebox{1.4}{$x_1$}};

\draw (490,310) node {\scalebox{1.4}{$n_1$}};
\draw (300,440) node {\scalebox{1.4}{$m_l$}};

\draw (300,240) node {\scalebox{1.4}{$m_1$}};
\draw (490,380) node {\scalebox{1.4}{$n_l$}};

\end{tikzpicture}
    \caption{Petri net associated with an ODE with one compartment.}
    \label{onecomp}
\end{figure}
In this Petri net we have $l$ transitions $z_1,\cdots, z_l$ with respectively associated parameters $\alpha_1(m_1)$, $\cdots$, $\alpha_l(m_l)$. For each transition $z_j$, with $1\leq j\leq l$, the arrows in the Petri net have the following description:
\begin{itemize}
    \item there are $m_j$ arrows from $x_1$ to $z_j$; and
    \item there are $n_j$ arrows from $z_j$ to $x_1$.
\end{itemize}
It is straightforward to check that the rate equation of this Petri net gives the equation \eqref{re1}.

It is illustrative to present the case $k=2$. We have the following ODE:
\begin{flalign}\label{re23}
     x_1'(t)=\sum_{j=1}^{l_1}f_{1,j}(m_{1,j},m_{2,j})\alpha_{1,j}(m_{1,j},m_{2,j})x_1^{m_{1,j}}x_2^{m_{2,j}} \\\label{re24}
      x_2'(t)=\sum_{j=1}^{l_2}f_{2,j}(m_{1,j},m_{2,j})\alpha_{2,j}(m_{1,j},m_{2,j})x_1^{m_{1,j}}x_2^{m_{2,j}}\,.& 
\end{flalign}
Consider $l$ the number of pairs $(m_{1,j},m_{2,j})$ not equal to the zero vector in both equations. 
    
Now, the associated Petri net has two set of transitions $z_{1,1},\cdots, z_{1,l}$, and $z_{2,1},\cdots, z_{2,l}$ associated with the parameters $\alpha_{1,1}(m_{1,1},m_{2,1})$, $\cdots$, $\alpha_{1,l}(m_{1,l},m_{2,l})$, and $\alpha_{2,1}(m_{1,1},m_{2,1})$, $\cdots$, $\alpha_{2,l}(m_{1,l},m_{2,l})$, where some of them can be zero. Consider the integer values 
$$n_{1,j}:=f_{1,j}(m_{1,j},m_{2,j})+m_{1,j}\,,\textrm{ and }n_{2,j}:=f_{2,j}(m_{1,j},m_{2,j})+m_{2,j}\,,$$ 
for $1\leq j\leq l$. 
We represent the associated Petri net in Figure \ref{twocomp}.
\begin{figure}
    \centering
\begin{tikzpicture}[scale=0.3,x=1pt,y=-1pt]

\definecolor{BLACK}{RGB}{0,0,0}
\definecolor{YELLOW}{RGB}{255,255,0}
\draw[BLACK, solid, line join=round, line cap=round, line width=1, fill=YELLOW]
(340,340) ellipse[x radius=45, y radius=45];
\definecolor{CYAN}{RGB}{0,255,255}
\draw[BLACK, solid, line join=round, line cap=round, line width=1, fill=CYAN]
(540,170) rectangle +(200,60);
\draw[BLACK, solid, line join=round, line cap=round, line width=1, fill=CYAN]
(540,450) rectangle +(200,60);
\draw[BLACK, solid, line join=round, line cap=round, line width=1]
(340,295) -- (340,200) -- (340,200) -- (540,200);
\draw[BLACK, solid, line join=round, line cap=round, line width=1, fill=BLACK]
(540,200) -- (530,203) -- (530,197) -- (540,200) -- cycle;
\draw[BLACK, solid, line join=round, line cap=round, line width=1]
(576,230) -- (381,321);
\draw[BLACK, solid, line join=round, line cap=round, line width=1, fill=BLACK]
(381,321) -- (388,314) -- (391,320) -- (381,321) -- cycle;
\draw[BLACK, solid, line join=round, line cap=round, line width=1]
(340,385) -- (340,480) -- (540,480);
\draw[BLACK, solid, line join=round, line cap=round, line width=1, fill=BLACK]
(540,480) -- (530,483) -- (530,477) -- (540,480) -- cycle;
\draw[BLACK, solid, line join=round, line cap=round, line width=1]
(576,450) -- (381,359);
\draw[BLACK, solid, line join=round, line cap=round, line width=1, fill=BLACK]
(381,359) -- (391,360) -- (388,366) -- (381,359) -- cycle;
\draw[BLACK, solid, line join=round, line cap=round, line width=1, fill=YELLOW]
(940,340) ellipse[x radius=45, y radius=45];
\draw[BLACK, solid, line join=round, line cap=round, line width=1]
(940,295) -- (940,200) -- (940,200) -- (740,200);
\draw[BLACK, solid, line join=round, line cap=round, line width=1, fill=BLACK]
(740,200) -- (750,197) -- (750,203) -- (740,200) -- cycle;
\draw[BLACK, solid, line join=round, line cap=round, line width=1]
(940,385) -- (940,480) -- (940,480) -- (740,480);
\draw[BLACK, solid, line join=round, line cap=round, line width=1, fill=BLACK]
(740,480) -- (750,477) -- (750,483) -- (740,480) -- cycle;
\draw[BLACK, solid, line join=round, line cap=round, line width=1]
(704,450) -- (899,359);
\draw[BLACK, solid, line join=round, line cap=round, line width=1, fill=BLACK]
(899,359) -- (892,366) -- (889,360) -- (899,359) -- cycle;
\draw[BLACK, solid, line join=round, line cap=round, line width=1]
(704,230) -- (899,321);
\draw[BLACK, solid, line join=round, line cap=round, line width=1, fill=BLACK]
(899,321) -- (889,320) -- (892,314) -- (899,321) -- cycle;

\draw (340,340) node {\scalebox{1.4}{$x_1$}};
\draw (940,340) node {\scalebox{1.4}{$x_2$}};

\draw (640,200) node {\scalebox{0.80}{$\alpha_{1,j}(m_{1,j},m_{2,j})$}};
\draw (640,480) node {\scalebox{0.80}{$\alpha_{2,j}(m_{1,j},m_{2,j})$}};

\draw (495,315) node {\scalebox{1.2}{$n_{1,j}$}};
\draw (280,440) node {\scalebox{1.2}{$m_{1,j}$}};

\draw (780,315) node {\scalebox{1.2}{$m_{2,j}$}};
\draw (780,380) node {\scalebox{1.2}{$n_{2,j}$}};

\draw (1000,440) node {\scalebox{1.2}{$m_{2,j}$}};
\draw (1000,240) node {\scalebox{1.2}{$m_{2,j}$}};

\draw (280,240) node {\scalebox{1.2}{$m_{1,j}$}};
\draw (495,380) node {\scalebox{1.2}{$m_{1,j}$}};

\end{tikzpicture}
    \caption{Petri net associated with an ODE with two compartments.}
    \label{twocomp}
\end{figure}
Consider the transposition $\tau:\{1,2\}\rightarrow\{1,2\}$.
For each transition $z_{i,j}$, with $i=1,2$ and $1\leq j\leq l$, the arrows in the Petri net have the following description:
\begin{itemize}
    \item there are $m_{i,j}$ arrows from $x_i$ to $z_{i,j}$; 
    \item there are $n_{i,j}$ arrows from $z_{i,j}$ to $x_i$;
    \item there are $m_{i,j}$ arrows from $x_i$ to $z_{\tau(i),j}$; and 
    \item there are $m_{i,j}$ arrow $z_{\tau(i),j}$ to $x_{i}$.
\end{itemize}
It is straightforward to check that the rate equations of this Petri net give the equations \eqref{re23} and \eqref{re24}.

Notice that the generic case is analogous to $k=2$. We have set of transitions $z_{1,j}$, $z_{2,j}$,$\cdots$,$z_{k,j}$, with $1\leq j\leq l$, where $l$ is the number of tuples $(m_{1,j},\cdots,m_{k,j})$ not equal to the zero vector.
Like $k=2$, we assume the functions $f_{i,j}$ and $\alpha_{i,j}$ are zero when required. 
Since we are taking an order of all the exponents of the variables $x_1$, $\cdots$, $x_k$ by integers $m_{1,j}$, $\cdots$, $m_{k,j}$, then each rate equation has the form 
\begin{equation}\label{epi222} x_i'(t)=\sum_{j=1}^lf_{i,j}(m_{1,j},\cdots,m_{k,j})\alpha_{i,j}(m_{1,j},\cdots,m_{k,j})x_1^{m_{1,j}}\dots x_k^{m_{k,j}}\,,\end{equation}
Consider the integer values 
$$n_{i,j}:=f_{i,j}(m_{1,j},\cdots,m_{k,j})+m_{i,j}\,,$$
for $1\leq j\leq l$. For each transition $z_{i,j}$, with $1\leq i\leq k$ and $1\leq j\leq l$, the arrows in the Petri net have the following description:
\begin{itemize}
    \item there are $m_{i,j}$ arrows from $x_i$ to $z_{i,j}$;
    \item there are $n_{i,j}$ arrows from $z_{i,j}$ to $x_i$;
    \item there are $m_{i,j}$ arrows from $x_i$ to $z_{s,j}$ for $s\neq i$; and 
    \item there are $m_{i,j}$ arrows from $z_{s,j}$ to $x_i$ for $s\neq i$.
\end{itemize}
It is straightforward to check that the rate equations of this Petri net give the equations \eqref{epi222}.

The algorithm presented in this section provides a procedure to obtain a Petri net for any system of ODEs. Nevertheless, the efficiency of this procedure remains an open question. As the reader can notice, our Petri net will present some meaningless transitions and arrows that could be removed from the Petri net without affecting the representability of the original system of ODEs. An essential project for the future is the simplification of this Petri net. 
\begin{example}
    We use the algorithm to create the Petri net associated with the ODEs of Examples \ref{SIR1} and \ref{2ndVac}; respectively, the SIR model and the second vaccination model. 

    The SIR model will have compartments $x_1=S$, $x_2=I$, and $x_3=R$. Starting with the ODE in \eqref{SIR-model}, we obtain transitions $z_{1,2}$, $z_{2,1}$, $z_{2,2}$, and $z_{3,1}$ associated to the parameters $\beta$, $\alpha$, $\beta$, and $\alpha$; respectively. Thus, $1\leq i\leq 3$ and $1\leq j\leq 2$, and we obtain the values for tuples, denoted by $m$ and $n$, and the functions $\alpha$ and $f$:
    \begin{equation*}
    \begin{tabular}{ c|c|c } 
j & m & n\\
\hline
1 & (0,1,0) & (0,0,1) \\ 
2& (1,1,0)  &(0,2,0)\\ 
\end{tabular}
\hspace{1cm}
\begin{tabular}{ c|cc } 
$\alpha_{ij}$ & 1 & 2\\
\hline
1 & 0 & $\beta$ \\ 
2&  $\alpha$ & $\beta$\\
3 & $\alpha$ & 0
\end{tabular}
\hspace{1cm}
\begin{tabular}{ c|cc} 
$f_{ij}$ & 1 & 2\\
\hline
1 & 0 & -1 \\ 
2& -1  & 1\\
3 & 1 & 0
\end{tabular}
    \end{equation*}
In Figure \ref{algSIR}, we include the associated Petri net. Notice that after the identification of transitions with the same parameter, we obtain the original Petri net in Figure \ref{figu2}.

The ODEs in \eqref{SIRVAC} of the second model of vaccination have compartments $x_1=S$, $x_2=I$, and $x_3=V$.  We obtain transitions $z_{1,1}$, $z_{1,6}$, $z_{1,5}$, $z_{2,6}$, $z_{2,4}$, $z_{2,3}$, $z_{3,5}$, $z_{3,4}$, $z_{3,3}$ and $z_{3,2}$. Thus, $1\leq i\leq 3$ and $1\leq j\leq 6$, and we obtain the functions $\alpha$ and $f$:
 \begin{equation*}
\begin{tabular}{ c|c|c|c|c|c|cc } 
$\alpha_{ij}$ & 1 & 2 & 3 & 4 & 5 & 6\\
\hline
1 & $\Lambda$ & 0 & 0 & 0 & $\mu+\psi$ & $\beta/N$\\ 
2& 0  & 0 & $\mu+\gamma$ & $\beta\delta/N$ &0 & $\beta/N$ \\
3 & 0 & $\mu$ & $(1-\chi)\gamma$ & $\beta\delta/N$ & $\psi$ &0
\end{tabular}
\hspace{0.5cm}
\begin{tabular}{ c|c|c|c|c|c|cc} 
$f_{ij}$ & 1 & 2 & 3 & 4 & 5 & 6\\
\hline
1 & 1 & 0 & 0 & 0 & -1 & -1\\ 
2& 0  & 0 & -1 & 1& 0& 1\\
3 & 0 & -1 & 1 & -1& 1& 0
\end{tabular}
    \end{equation*}
We use the value of $f$ to obtain the tuple $n$, where $m+f=n$:
\begin{equation*}
     \begin{tabular}{ c|c|c|c } 
j & m & f& n\\
\hline
1 & (0,0,0)& (1,0,0) & (1,0,0)\\
2 & (0,0,1) & (0,0,-1) & (0,0,0) \\ 
3& (0,1,0)  & (0,-1,1) & (0,0,1)\\ 
4& (0,1,1)  & (0,1,-1) & (0,2,0)\\
5& (1,0,0)  & (-1,0,1) & (0,0,1)\\
6& (1,1,0)  & (-1,1,0) & (0,2,0) 
\end{tabular}
\end{equation*}

\begin{figure}
    \centering
\begin{tikzpicture}[scale=0.3,x=1pt,y=-1pt]

\definecolor{BLACK}{RGB}{0,0,0}
\definecolor{YELLOW}{RGB}{255,255,0}
\draw[BLACK, solid, line join=round, line cap=round, line width=1, fill=YELLOW]
	(160,320) ellipse[x radius=35, y radius=35];
\draw[BLACK, solid, line join=round, line cap=round, line width=1, fill=YELLOW]
	(440,320) ellipse[x radius=35, y radius=35];
\draw[BLACK, solid, line join=round, line cap=round, line width=1, fill=YELLOW]
	(660,320) ellipse[x radius=35, y radius=35];
\definecolor{CYAN}{RGB}{0,255,255}
\draw[BLACK, solid, line join=round, line cap=round, line width=1, fill=CYAN]
	(272,292) rectangle +(56,56);
\draw[BLACK, solid, line join=round, line cap=round, line width=1, fill=CYAN]
	(412,452) rectangle +(56,56);
\draw[BLACK, solid, line join=round, line cap=round, line width=1]
	(195,320) -- (272,320);
\draw[BLACK, solid, line join=round, line cap=round, line width=1, fill=BLACK]
	(272,320) -- (262,323) -- (262,317) -- (272,320) -- cycle;
\draw[BLACK, solid, line join=round, line cap=round, line width=1]
	(405,320) -- (328,320);
\draw[BLACK, solid, line join=round, line cap=round, line width=1, fill=BLACK]
	(328,320) -- (338,317) -- (338,323) -- (328,320) -- cycle;
\draw[BLACK, solid, line join=round, line cap=round, line width=1, fill=CYAN]
	(272,452) rectangle +(56,56);
\draw[BLACK, solid, line join=round, line cap=round, line width=1, fill=CYAN]
	(632,552) rectangle +(56,56);
\draw[BLACK, solid, line join=round, line cap=round, line width=1]
	(183,346) -- (276,452);
\draw[BLACK, solid, line join=round, line cap=round, line width=1, fill=BLACK]
	(276,452) -- (266,447) -- (271,442) -- (276,452) -- cycle;
\draw[BLACK, solid, line join=round, line cap=round, line width=1]
	(417,346) -- (325,452);
\draw[BLACK, solid, line join=round, line cap=round, line width=1, fill=BLACK]
	(325,452) -- (329,442) -- (334,447) -- (325,452) -- cycle;
\draw[BLACK, solid, line join=round, line cap=round, line width=1]
	(440,355) -- (440,452);
\draw[BLACK, solid, line join=round, line cap=round, line width=1, fill=BLACK]
	(440,452) -- (437,442) -- (443,442) -- (440,452) -- cycle;
\draw[BLACK, solid, line join=round, line cap=round, line width=1]
	(660,552) -- (660,355);
\draw[BLACK, solid, line join=round, line cap=round, line width=1, fill=BLACK]
	(660,355) -- (663,365) -- (657,365) -- (660,355) -- cycle;
\draw[BLACK, solid, line join=round, line cap=round, line width=1]
	(458,350) -- (529,465) .. controls (600,580) .. (600,580) .. controls (600,580) .. (616,580) -- (632,580);
\draw[BLACK, solid, line join=round, line cap=round, line width=1, fill=BLACK]
	(632,580) -- (622,583) -- (622,577) -- (632,580) -- cycle;
\draw[BLACK, solid, line join=round, line cap=round, line width=1]
	(636,552) -- (463,347);
\draw[BLACK, solid, line join=round, line cap=round, line width=1, fill=BLACK]
	(463,347) -- (472,352) -- (467,357) -- (463,347) -- cycle;
\draw[BLACK, solid, line join=round, line cap=round, line width=1]
	(328,309) -- (364,294.5) .. controls (400,280) .. (400,280) .. controls (400,280) .. (407.5,287.5) -- (415,295);
\draw[BLACK, solid, line join=round, line cap=round, line width=1, fill=BLACK]
	(415,295) -- (406,291) -- (411,286) -- (415,295) -- cycle;
\draw[BLACK, solid, line join=round, line cap=round, line width=1]
	(300,452) -- (300,426) .. controls (300,400) .. (300,400) .. controls (300,400) .. (355,368.5) -- (410,337);
\draw[BLACK, solid, line join=round, line cap=round, line width=1, fill=BLACK]
	(410,337) -- (403,345) -- (399,339) -- (410,337) -- cycle;

\draw (160,320) node {\scalebox{1.2}{$S$}};

\draw (300,320) node {\scalebox{1}{$\beta$}};

\draw (300,480) node {\scalebox{1}{$\beta$}};

\draw (440,480) node {\scalebox{1}{$\alpha$}};

\draw (280,390) node {\scalebox{1}{$2$}};

\draw (440,320) node {\scalebox{1.2}{$I$}};
\draw (660,320) node {\scalebox{1.2}{$R$}};

\draw (660,580) node {\scalebox{1}{$\alpha$}};
\end{tikzpicture}

    \caption{The algorithm applied to the ODEs of the SIR model.}
    \label{algSIR}
\end{figure}

\end{example}

\section{The basic reproduction number}
\label{secNGM}

The basic reproduction number $R_0$ is one of the most essential concepts in mathematical epidemiology. Historical references \cite{He02,Pe18} assert this number $R_0$ started in demography and ecology, but it gained popularity in epidemiology, where it
is a threshold parameter that controls the spread of an infectious or contagious disease.

There are several definitions of $R_0$ depending on the field of study. In epidemiology, this quantity is defined as the expected number of new cases of an infection caused by a typical infected individual in a population consisting of susceptibles only. In mathematics, there is a method proposed by Diekmann--Heesterbeek--Metz \cite{DHM90} and elaborated by van den Driessche--Watmough \cite{DW02,DW08}, which defines $R_0$ as the dominant eigenvalue of the so-called next-generation matrix.

\subsection{The next generation matrix method}
\label{NGMM}
We start with an ordinary differential equation (ODE) where the individuals are sorted into compartments $\{x_1,\cdots,x_n\}$. Assume the first $m<n$ compartments contain the infected individuals. 
Denote by $(x,y)$ the pair where $x$ consists of the first coordinates $x_1,\cdots, x_m$ and $y$ consists of $x_{m+1},\cdots,x_n$. The rate equations are as follows:
\begin{flalign}\label{ratequ}
          x'_i(t)=\ma{F}_i(x,y)-\ma{V}_i(x,y)\,, \hspace{0.5cm}i=1,\cdots,n\,,\\
      y'_j=g_j(x,y),\hspace{0.5cm}j=1,\cdots,m\,.
\end{flalign}

The function $\ma{F}_i$ denotes the rate secondary infections increase the $i$th disease compartment, and $\ma{V}_i$ denotes the rate disease progression, death, and recovery decrease the $i$th compartment.

A disease-free equilibrium (DFE) is a state where the epidemiological model remains in the absence of disease. We need some assumptions to ensure the existence of this equilibrium and to ensure that the model is well-posed. In what follows, we enumerate these assumptions established by van den Driessche--Watmough \cite{DW08}, which are equivalent to those in \cite{DW02}.
\begin{itemize}
    \item[(A1)] Assume $\ma{F}_i(0,y)=0$ and $\ma{V}_i(0,y)=0$ for all $y\geq 0$ and $i=1,\cdots n$. All new infections are secondary infections arising from infected hosts; there is no immigration of individuals into the disease compartments.
    \item[(A2)] Assume $\ma{F}_i(x,y)\geq 0$ for all nonnegative $x$ and $y$ and $i=1,\cdots, n$. The function $\mathcal{F}$ represents new infections and cannot be negative.
    \item[(A3)] Assume $\ma{V}_i(x,y)\leq 0$ whenever $x_i=0$, $i=1,\cdots, n$. Each component, $\mathcal{V}_i$, represents a net outflow from compartment $i$ and must be negative (inflow only) whenever the compartment is empty. 
    \item[(A4)] Assume $\sum_{i=1}^n\ma{V}_i(x,y)\geq 0$ for all nonnegative $x$ and $y$. This sum represents the total outflow from all infected compartments. Terms in the model leading to increases in $\sum_{i=1}^nx_i$ are assumed to represent secondary infections and therefore belong in $\mathcal{F}$.
    \item[(A5)] Assume the disease-free system $y'=g(0,y)$ has a unique equilibrium that is {\bf asymptotically stable}. That is, all solutions with initial condition of the form $(0,y)$ approach a point $(0,y_0)$ as $t\rightarrow\infty$. We refer to this point as the disease-free equilibrium.
    
\end{itemize}
As a consequence, we obtain the disease compartments are decoupled from the remaining equations $(x_1'(t),\cdots,x_n'(t))^{tr}=
(F-V) (x_1,\cdots,x_n)^{tr}$,
where the $n\times n$ matrices $F$ and $V$ are given by 
$$F=\left(\frac{\partial \ma{F}_i}{\partial x_j}(0,y_0)\right)\textrm{ and }V=\left(\frac{\partial \ma{V}_i}{\partial x_j}(0,y_0)\right)\,,$$
for $i,j=1,\dots, n$. The basic reproduction number is defined by the dominant eigenvalue of the nonnegative generation matrix $FV^{-1}$ as follows $$R_0=\rho(FV^{-1})\,.$$

The interpretation of the entries in the next generation matrix $FV^{-1}$ is explained by van den Driessche--Wartmough \cite{DW02}.
Assuming the population remains near the DFE and excepting reinfection, the $(j,k)$ entry of $V^{-1}$ is the average time this individual spends in compartment $j$ during its lifetime. The $(i,j)$ entry of $F$ is the rate at which infected individuals in compartment $j$ produce new infections in compartment $i$. Hence, {\bf the $(i,k)$ entry of the product $FV^{-1}$ is the expected number of new infections in compartment $i$ produced by the infected individual originally introduced into compartment $k$}. 

In what follows, we apply the next-generation matrix method to the SEAIR model, the model of Malaria, and the two vaccination models. The reader can find a list of this type of examples for a variety of models in epidemiology in Table \ref{Tabla1} at the end of the paper.

 \begin{example}
We consider the SEAIR model from the book \cite{Ma}, which was studied in Example \ref{SEAIR}. The infection compartments contain exposed $E$, asymptomatic $A$, and infected $I$. The associated function $\ma{F}$ and $\ma{V}$ are:
       \begin{center}
            $\begin{array}{cl}
                 \ma{F}_E =& \beta S(I+q A)  \\
                 \ma{F}_A =& 0 \\
                 \ma{F}_I =& 0
            \end{array}$, \hspace{1cm}   $\begin{array}{cl}
                 \ma{V}_E =& (\eta + \mu) E \\
                 \ma{V}_A =& \eta(p-1)E +(\gamma+\mu)A\\
                 \ma{V}_I =& -\eta p E+(\alpha+\mu)I
            \end{array}$
 \end{center}
Consequently, the $F$ and $V$ matrices are given as follows:
\begin{center}
$
F=\begin{pmatrix}
0 & \beta q S & \beta S\\
0 & 0 & 0 \\
0 & 0 & 0
\end{pmatrix}$, \hspace{1cm}
$V=\begin{pmatrix}
\eta+\mu & 0 & 0\\
\eta (p-1) & \gamma+\mu & 0 \\
-\eta p & 0 & \alpha+\mu
\end{pmatrix}$
 \end{center}
with the next-generation matrix 
$$FV^{-1}=\begin{pmatrix}
\frac{\beta \eta (1-p)qS}{(\eta+\mu)(\gamma+\mu)}+\frac{\beta\eta p S}{(\eta+\mu)(\alpha+\mu)} & \frac{\beta q S}{\gamma+\mu} & \frac{\beta S}{\alpha+\mu}\\
0 & 0 & 0 \\
0 & 0 & 0
\end{pmatrix}\,,$$
and the basic reproduction number is 
$$R_0=\frac{\beta \eta (1-p)qS}{(\eta+\mu)(\gamma+\mu)}+\frac{\beta\eta p S}{(\eta+\mu)(\alpha+\mu)}\,.$$
 which is included in Table \ref{Tabla1} at the end of the paper. 
 \end{example}

\begin{example} The model of Malaria in \cite{WBK} has the infection compartments $I_H$ and $I_M$. The functions $\ma{F}$ and $\ma{V}$ are as follows:
 \begin{center}
            $\begin{array}{cl}
                 \ma{F}_{I_H} =& \beta_{HM} S_H I_M  \\
                 \ma{F}_{I_M} =& \beta_{MH} S_M I_H 
            \end{array}$, \hspace{1cm}   $\begin{array}{cl}
                 \ma{V}_{I_H} =& (\mu_H+\alpha+\sigma)I_H-\delta I_H \\
                 \ma{V}_{I_M} =& \mu_M I_M
            \end{array}$
 \end{center}
 Consequently, the $F$ and $V$ matrices are given as follows:
 \begin{center}
     $F=\begin{pmatrix}
0 & \beta_{HM} S_H\\
\beta_{MH} S_M & 0
\end{pmatrix}$, \hspace{1cm}$V=\begin{pmatrix}
\alpha-\delta+\mu_H+\sigma&0\\
0 & \mu_M
\end{pmatrix}$
 \end{center}
 with the next-generation matrix 
 $$FV^{-1}=\begin{pmatrix}
0 & \frac{\beta_{HM} S_H}{\mu_M}\\
\frac{\beta_{MH} S_M}{\alpha-\delta+\mu_H+\sigma} & 0
\end{pmatrix}$$
and the basic reproduction number is 
$$R_0=\sqrt{\frac{\beta_{HM}\beta_{MH}S_HS_M}{(\alpha-\delta+\mu_H+\sigma)\mu_M}}\,.$$
This case is special since the basic reproduction number is not realized as an element of the diagonal of $FV^{-1}$. We obtain a square root since the matrix $F$ has rank two, whereas the other examples have $F$
with rank one. 
\end{example}

\begin{remark}
   There are different reproduction numbers in the literature \cite{muchos}, depending on the conditions of the susceptible population. The basic reproduction number $R_0$ describes the expected number of secondary infections
caused by a typical infected individual in a completely susceptible host population under standard behavioral patterns.
The effective reproduction number $R_e$, also denoted by $R_t$, describes the expected number of secondary infections under the current conditions of population mixing, transmission, and immunity. 
The control reproduction number $R_c$ describes the expected number of secondary infections under the current contact and transmission patterns in an otherwise fully susceptible population. This is the case of treatment and vaccination. 
   \end{remark}

\begin{example}\label{ejVac}
   The reproduction number for the two models of vaccinations is as follows:   
   \begin{itemize}
       \item For the first model in Example \ref{1stVac}, the disease-free equilibrium with $I=0$ and $S'(t)=0$ implies that $S=qN$, which is the flow from the susceptible compartment to the transition of $\beta$. The $F$-matrix and $V$-matrix are $(\beta S)$ and $(\mu+\alpha)$, which implies that the basic reproduction number is $R_0=\beta S/(\mu+\alpha)$. 
       Without vaccination at the disease-free equilibrium ($p=0$), we obtain $ R_0=\beta N/(\mu+\alpha)$. In the presence of vaccination, we obtain the reproduction number \begin{equation}qR_0=\frac{\beta qN}{(\mu+\alpha)}\,.\end{equation} 
       \item For the second model of vaccination in Example \ref{2ndVac}, the disease-free equilibrium with $I=0$, $S'(t)$ and $V'(t)=0$ implies that $S=\Lambda \mu/(\mu+\psi)$ and $V=\Lambda\psi/\mu(\mu+\psi)$. They are the flows from the susceptible and vaccination compartment to the infection transitions associated with $\beta$ and $q\beta$, respectively.
       Since the equation of the total population is $N'(t)=\Lambda -\mu N$, the equilibrium total population size is $N=\Lambda/\mu$.
       The $F$-matrix and $V$-matrix are $(\beta S/N+\beta\delta V/N)$ and $(\mu+\gamma)$, which implies that the basic reproduction number is $R_0=(\beta S/N+\beta\delta V/N)/(\mu+\gamma)$. In the disease-free equilibrium at the equilibrium total population, we obtain 
       \begin{equation}
           \frac{S}{N}=\frac{\mu}{\mu+\psi}\,,\textrm{ and }\frac{V}{N}=\frac{\psi}{\mu+\psi}\,.
       \end{equation}
       and the reproduction number is 
       \begin{equation}
           R_0=\frac{\beta(\mu+\delta\psi)}{(\mu+\gamma)(\mu+\psi)}\,.
       \end{equation}
        \end{itemize}
\end{example}

\section{The basic reproduction number of a Petri net}

\label{R0PN}
The basic reproduction number is not only a concept that pertains to epidemiology. For instance, there are applications outside epidemiology in phenomena that are different from a pandemic or the growth of a population: In informatics, we can use a particular type of SEIR model to give a mathematical explanation of computer virus propagation \cite{CHHLN20,Mu88}; also, the standard SIR model is used in \cite{BLT12} to simulate key properties of real spreading cascades of file sharing. The SIR model is used in social networks to study rumor spreading; see \cite{WW15}. A summary of interesting applications is found in \cite{Ro16}.

The next-generation matrix method has assumptions (A1)-(A5) from Section \ref{secNGM}, which assure that the basic reproduction number is the dominant eigenvalue of the next-generation matrix.
We now present a special case of the next-generation matrix method for the rate equations of Petri nets. The assumptions (A1)-(A5) has a geometric interpretation with assumptions (G1)-(G5) which we will explained in what follows.
In the next section we explain the first four assumptions and we delay the fifth assumption for Section \ref{threshold}.

\subsection{Geometric next-generation matrix method}
\label{GeoNGM}
In order to provide a geometric interpretation of the assumptions (A1)-(A5), that we called assumptions (G1)-(G5), we studied in deep a plenty of examples of epidemiological models of diseases; see Section \ref{secPN}. 
We observe that the calculation of the basic reproduction number for a highly sophisticated epidemiological model depends in these three basic substructures which we called {\it modules}, as a component in the construction of our entire model. We mean with these modules as nested-structures comprising the susceptible population, the infection process and the infection population. When we have that the model of a disease can be described by a Petri net these structures consists of three disjoint Petri nets inside which are defined below.
\begin{itemize}
    \item {\bf The susceptible module}: consider the Petri net composed of the susceptible compartments and all the transitions that model the behavior of the susceptible population (birth, death, migration, emigration, etc.).
   
    \item {\bf The infection-process module}: consider the Petri net composed of the infection transitions and some compartments that connect them (for example, when it occurs, reinfection of two different diseases, communication between two other infection transitions, etc.). 
    
    \item {\bf The infection module}:  consider the Petri net with all the different disease compartments of the study population (infected, exposed, asymptomatic, carrier, etc.). The transitions included are those relating to disease compartments and all transitions that do not belong to the susceptible or infection-process modules involving the input of a disease compartment.

\end{itemize}

We will refer to these structures as the {\bf Kermack--McKendrick modules} or, more briefly, the KM modules. Assume that all sets of compartments and transitions in the different KM modules are disjoint from one another.

Notice that the union of the compartments and transitions of the KM modules could not sum to the whole set of compartments and transitions of the original Petri net.

\begin{remark}
    We want to clarify more about the construction of the infection module: from the total compartments $S=\{x_1,\cdots,x_k\}$, the Petri net of the infection module has the first $s<k$ compartments that contain the infected individuals as in the next generation matrix method from Section \ref{NGMM}. Assume $T=\{z_1,\cdots,z_l\}$ are the whole transitions, the Petri net of the infection module has transitions $z\in T$, with $z\not\in T_{S}\cup T_{IP}$, such that there is an arrow $x\To z$ with $x\in \{x_1,\cdots,x_s\}$. In other words, all transitions that do not belong to the susceptible or infection process modules with an input of a disease compartment. 
    
\end{remark}

\begin{example}
The SEAIR model has the compartments $\{S,E,A,I,R\}$ denoting Susceptibles, Exposed, Asymptomatic, Infected, and Recovered. The KM modules are described as follows: The susceptible module is a Malthusian model (see Example \ref{maltusiano}) associated with the susceptible population. The Petri net has one species $S$ and two transitions given by birth and decease or death. The infection-process module is the disjoint union of the infection transitions with parameters $q\beta$ and $\beta$. 
The infection module is a Petri net with compartments $\{E,A,I\}$ and transitions with parameters $\{\eta,\gamma,\alpha\}$ and three transitions of death denoted by $\mu$.
These modules are illustrated in Figure \ref{SEAIR2}.
\begin{figure}
    \centering
     \begin{tikzpicture}[scale=0.3,x=1pt,y=-1pt]

\definecolor{BLACK}{RGB}{0,0,0}
\definecolor{r255g255b10}{RGB}{255,255,10}
\draw[BLACK, solid, line join=round, line cap=round, line width=1, fill=r255g255b10]
	(120,260) ellipse[x radius=30, y radius=30];
\draw[BLACK, solid, line join=round, line cap=round, line width=1, fill=r255g255b10]
	(700,260) ellipse[x radius=30, y radius=30];
\draw[BLACK, solid, line join=round, line cap=round, line width=1, fill=r255g255b10]
	(1000,140) ellipse[x radius=30, y radius=30];
\draw[BLACK, solid, line join=round, line cap=round, line width=1, fill=r255g255b10]
	(1000,380) ellipse[x radius=30, y radius=30];
\definecolor{r33g255b255}{RGB}{33,255,255}
\draw[BLACK, solid, line join=round, line cap=round, line width=1, fill=r33g255b255]
	(95,135) rectangle +(50,50);
\draw[BLACK, solid, line join=round, line cap=round, line width=1, fill=r33g255b255]
	(95,355) rectangle +(50,50);
\draw[BLACK, solid, line join=round, line cap=round, line width=1, fill=r33g255b255]
	(675,355) rectangle +(50,50);
\draw[BLACK, solid, line join=round, line cap=round, line width=1, fill=r33g255b255]
	(1135,195) rectangle +(50,50);
\draw[BLACK, solid, line join=round, line cap=round, line width=1, fill=r33g255b255]
	(395,135) rectangle +(50,50);
\draw[BLACK, solid, line join=round, line cap=round, line width=1, fill=r33g255b255]
	(835,235) rectangle +(50,50);
\draw[BLACK, solid, line join=round, line cap=round, line width=1, fill=r33g255b255]
	(1135,15) rectangle +(50,50);
\draw[BLACK, solid, line join=round, line cap=round, line width=1, fill=r33g255b255]
	(395,355) rectangle +(50,50);
\draw[BLACK, solid, line join=round, line cap=round, line width=1, fill=r33g255b255]
	(1135,295) rectangle +(50,50);
\draw[BLACK, solid, line join=round, line cap=round, line width=1, fill=r33g255b255]
	(1135,455) rectangle +(50,50);
\draw[BLACK, solid, line join=round, line cap=round, line width=1]
	(120,185) -- (120,230);
\draw[BLACK, solid, line join=round, line cap=round, line width=1, fill=BLACK]
	(120,230) -- (117,220) -- (123,220) -- (120,230) -- cycle;
\draw[BLACK, solid, line join=round, line cap=round, line width=1]
	(120,290) -- (120,355);
\draw[BLACK, solid, line join=round, line cap=round, line width=1, fill=BLACK]
	(120,355) -- (117,345) -- (123,345) -- (120,355) -- cycle;
\draw[BLACK, solid, line join=round, line cap=round, line width=1]
	(700,290) -- (700,355);
\draw[BLACK, solid, line join=round, line cap=round, line width=1, fill=BLACK]
	(700,355) -- (697,345) -- (703,345) -- (700,355) -- cycle;
\draw[BLACK, solid, line join=round, line cap=round, line width=1]
	(1030,155) -- (1135,208);
\draw[BLACK, solid, line join=round, line cap=round, line width=1, fill=BLACK]
	(1135,208) -- (1125,206) -- (1128,200) -- (1135,208) -- cycle;
\draw[BLACK, solid, line join=round, line cap=round, line width=1]
	(885,239) -- (970,166);
\draw[BLACK, solid, line join=round, line cap=round, line width=1, fill=BLACK]
	(970,166) -- (965,175) -- (960,170) -- (970,166) -- cycle;
\draw[BLACK, solid, line join=round, line cap=round, line width=1]
	(1030,121) -- (1135,56);
\draw[BLACK, solid, line join=round, line cap=round, line width=1, fill=BLACK]
	(1135,56) -- (1128,64) -- (1125,58) -- (1135,56) -- cycle;
\draw[BLACK, solid, line join=round, line cap=round, line width=1]
	(885,281) -- (970,354);
\draw[BLACK, solid, line join=round, line cap=round, line width=1, fill=BLACK]
	(970,354) -- (960,350) -- (965,345) -- (970,354) -- cycle;
\draw[BLACK, solid, line join=round, line cap=round, line width=1]
	(1030,369) -- (1135,329);
\draw[BLACK, solid, line join=round, line cap=round, line width=1, fill=BLACK]
	(1135,329) -- (1127,336) -- (1124,330) -- (1135,329) -- cycle;
\draw[BLACK, solid, line join=round, line cap=round, line width=1]
	(1030,399) -- (1135,464);
\draw[BLACK, solid, line join=round, line cap=round, line width=1, fill=BLACK]
	(1135,464) -- (1125,462) -- (1128,456) -- (1135,464) -- cycle;
\draw[BLACK, solid, line join=round, line cap=round, line width=1]
	(730,260) -- (835,260);
\draw[BLACK, solid, line join=round, line cap=round, line width=1, fill=BLACK]
	(835,260) -- (825,263) -- (825,257) -- (835,260) -- cycle;

\draw (80,0) node {Susceptible module};


\draw (120,258) node {$S$};
\draw (120,158) node {$\Lambda$};
\draw (120,380) node {$\mu$};

\draw (480,0) node {Infection-process module};


\draw (418,158) node {$q\beta$};
\draw (418,380) node {$\beta$};

\draw (880,0) node {Infection module};


\draw (700,258) node {$E$};
\draw (700,380) node {$\mu$};
\draw (860,261) node {$\eta$};

\draw (1000,137) node {$A$};

\draw (1000,378) node {$I$};
\draw (1158,45) node {$\gamma$};
\draw (1158,225) node {$\mu$};
\draw (1158,325) node {$\alpha$};
\draw (1158,485) node {$\mu$};

\end{tikzpicture}
    \caption{Susceptible, infection-process and infection modules for the SEAIR model.}
    \label{SEAIR2}
\end{figure}
\end{example}

\begin{example}\label{malaroso}
The malaria model has compartments $ S_H, I_H, R_H, S_M, I_M$ where two different types of populations interact. The KM modules are described as follows: The susceptible module is a disjoint union of two Malthusian models associated with the human and mosquito populations. 
The infection-process module comprises only the two transitions with parameters $\{\beta_{HM},\beta_{MH}\}$. The infection module has two disjoint components, each associated with one of the compartments $\{I_H, I_M\}$ with transitions to those with input from one of these compartments.  
These modules are illustrated in Figure \ref{Malach}.
\begin{figure}
    \centering
\begin{tikzpicture}[scale=0.3,x=1pt,y=-1pt]

\definecolor{BLACK}{RGB}{0,0,0}
\definecolor{r255g255b10}{RGB}{255,255,10}
\draw[BLACK, solid, line join=round, line cap=round, line width=1, fill=r255g255b10]
	(40,260) ellipse[x radius=30, y radius=30];
\draw[BLACK, solid, line join=round, line cap=round, line width=1, fill=r255g255b10]
	(780,160) ellipse[x radius=30, y radius=30];
\draw[BLACK, solid, line join=round, line cap=round, line width=1, fill=r255g255b10]
	(780,400) ellipse[x radius=30, y radius=30];
\definecolor{r33g255b255}{RGB}{33,255,255}
\draw[BLACK, solid, line join=round, line cap=round, line width=1, fill=r33g255b255]
	(15,135) rectangle +(50,50);
\draw[BLACK, solid, line join=round, line cap=round, line width=1, fill=r33g255b255]
	(15,355) rectangle +(50,50);
\draw[BLACK, solid, line join=round, line cap=round, line width=1, fill=r33g255b255]
	(975,255) rectangle +(50,50);
\draw[BLACK, solid, line join=round, line cap=round, line width=1, fill=r33g255b255]
	(390,130) rectangle +(60,60);
\draw[BLACK, solid, line join=round, line cap=round, line width=1, fill=r33g255b255]
	(960,15) rectangle +(80,50);
\draw[BLACK, solid, line join=round, line cap=round, line width=1, fill=r33g255b255]
	(390,350) rectangle +(60,60);
\draw[BLACK, solid, line join=round, line cap=round, line width=1, fill=r33g255b255]
	(975,375) rectangle +(50,50);
\draw[BLACK, solid, line join=round, line cap=round, line width=1]
	(40,185) -- (40,230);
\draw[BLACK, solid, line join=round, line cap=round, line width=1, fill=BLACK]
	(40,230) -- (37,220) -- (43,220) -- (40,230) -- cycle;
\draw[BLACK, solid, line join=round, line cap=round, line width=1]
	(40,290) -- (40,355);
\draw[BLACK, solid, line join=round, line cap=round, line width=1, fill=BLACK]
	(40,355) -- (37,345) -- (43,345) -- (40,355) -- cycle;
\draw[BLACK, solid, line join=round, line cap=round, line width=1]
	(810,176) -- (975,266);
\draw[BLACK, solid, line join=round, line cap=round, line width=1, fill=BLACK]
	(975,266) -- (965,265) -- (968,259) -- (975,266) -- cycle;
\draw[BLACK, solid, line join=round, line cap=round, line width=1]
	(810,144) -- (960,62);
\draw[BLACK, solid, line join=round, line cap=round, line width=1, fill=BLACK]
	(960,62) -- (953,70) -- (950,64) -- (960,62) -- cycle;
\draw[BLACK, solid, line join=round, line cap=round, line width=1]
	(810,400) -- (975,400);
\draw[BLACK, solid, line join=round, line cap=round, line width=1, fill=BLACK]
	(975,400) -- (965,403) -- (965,397) -- (975,400) -- cycle;
\draw[BLACK, solid, line join=round, line cap=round, line width=1, fill=r33g255b255]
	(135,355) rectangle +(50,50);
\draw[BLACK, solid, line join=round, line cap=round, line width=1, fill=r33g255b255]
	(135,135) rectangle +(50,50);
\draw[BLACK, solid, line join=round, line cap=round, line width=1, fill=r255g255b10]
	(160,260) ellipse[x radius=25, y radius=25];
\draw[BLACK, solid, line join=round, line cap=round, line width=1]
	(160,185) -- (160,235);
\draw[BLACK, solid, line join=round, line cap=round, line width=1, fill=BLACK]
	(160,235) -- (157,225) -- (163,225) -- (160,235) -- cycle;
\draw[BLACK, solid, line join=round, line cap=round, line width=1]
	(160,285) -- (160,355);
\draw[BLACK, solid, line join=round, line cap=round, line width=1, fill=BLACK]
	(160,355) -- (157,345) -- (163,345) -- (160,355) -- cycle;
\draw[BLACK, solid, line join=round, line cap=round, line width=1, fill=r33g255b255]
	(975,135) rectangle +(50,50);
\draw[BLACK, solid, line join=round, line cap=round, line width=1]
	(810,160) -- (975,160);
\draw[BLACK, solid, line join=round, line cap=round, line width=1, fill=BLACK]
	(975,160) -- (965,163) -- (965,157) -- (975,160) -- cycle;

\draw (60,-10) node {Susceptible module};


\draw (45,258) node {$S_H$};
\draw (42,158) node {$\Pi$};
\draw (42,380) node {$\mu_H$};

\draw (162,258) node {\scalebox{0.95}{$S_M$}};
\draw (162,158) node {\scalebox{1}{$\Lambda$}};
\draw (162,380) node {\scalebox{1}{$\mu_M$}};


\draw (460,-10) node {Infection-process module};

\draw (422,160) node {\scalebox{0.8}{$\beta_{HM}$}};
\draw (422,382) node {\scalebox{0.8}{$\beta_{MH}$}};


\draw (860,-10) node {Infection module};

\draw (782,160) node {\scalebox{0.85}{$I_H$}};
\draw (782,400) node {\scalebox{0.85}{$I_M$}};

\draw (1000,40) node {$\alpha-\delta$};
\draw (1000,165) node {$\mu_H$};
\draw (1000,280) node {$\sigma$};

\draw (1000,405) node {$\mu_M$};

\end{tikzpicture}
    \caption{Susceptible, infection-process and infection modules for the model of Malaria.}
    \label{Malach}
\end{figure}
\end{example}

The mathematical definition of the KM modules consists of three disjoint Petri nets inside the whole Petri net. 
\begin{definition}\label{SIRmod}
Assume a Petri net with a set of compartments $S$, a set of transitions $T$, a set of arrows $\ma{A}$, and a matching function $f$. The {\bf KM modules} are defined by three disjoint subsets of compartments $S_S,S_{IP},S_{I}\subset S$, three disjoint subsets of transitions $T_S,T_{IP},T_{I}\subset T$, three disjoint subsets of arrows $\ma{A}_S,\ma{A}_{I},\ma{A}_{IP}\subset \ma{A}$, and three functions $f_i:\ma{A}_i\longrightarrow S_i\times T_i\sqcup T_i\times S_i$, for $i\in \{S,IP,I\}$, such that $f_i=f|_{\ma{A}_i}$.
\end{definition}
The next-generation matrix method has assumptions (A1)-(A5) to guarantee that the basic reproduction numbers result as the dominant eigenvalue of the next-generation matrix. 

Denote the compartments of the whole Petri net by $S=\{x_1,\cdots,x_k\}$ with the disease compartments by $S_I=\{x_1,\cdots, x_s\}$. The geometric version of the assumptions (A1)-(A4) are as follows:
\begin{enumerate}
    \item[(G1)] For a disease compartment $x_i\in S_I$ and a transition in the susceptible compartment $z_j\in T_S$, we have $m_{ij}=n_{ij}=0$. Additionally, if $x_i\not\in S_I$ and a transition in the infection compartment $z_j\in T_I$, we have $m_{ij}=0$. In other words, we are prohibiting arrows between susceptible transitions and disease compartments; and arrows from compartments outside the infection module to an infection transition.


    \item[(G2)]
    Assume an infection transition $ z_j\in T_{IP}$ such that for all disease compartment $ x_i\in S_I$, we have $ m_{ij}=0$; therefore, we obtain $ n_{ij}=0$ for all disease compartment $ x_i\in S_I$. 
    
    This condition means all new infections are secondary infections arising from infected hosts; there is no immigration of individuals into the disease compartments.
    
    \item[(G3)] If $x_i\in S_I$ and $z_j\in T_{IP}\sqcup T_I$, then we have the inequality $n_{ij}\geq m_{ij}$.
    
    This means that there is no disappearance of infected individuals.

    \item[(G4)] We require a positive outflow of all the disease compartments. Thus, for compartments and transitions inside the infection module, i.e., $x_i\in S_I$ and $z_j\in T_I$, we have the inequality 
    $$\sum_{i=1}^s(m_{ij}-n_{ij})\geq 0\,.$$   
    \end{enumerate}

Consider a Petri net with compartments $S=\{x_1,\cdots,x_k\}$ and transitions $T=\{z_1,\cdots,z_l\}$, with disease compartments $S_I=\{x_1,\cdots,x_s\}$ with $s\leq k$.

We derive an explicit form for the functions $\ma{F}_i$ and $\ma{V}_i$ that define the rate equations of the disease compartments; see equation \ref{ratequ}.

\begin{lemma}\label{lem13}
For a Petri net satisfying the assumption (G1), the rate equation of a disease compartment $x_i\in S_I$ can be written as follows:
\begin{equation}\label{eq13}x_i'(t)=\sum_{z_j\in T_{IP}} r_j(n_{ij}-m_{ij})x_1^{m_{1j}}\cdots x_k^{m_{kj}}-\sum_{z_j\in T_I}r_j(m_{ij}-n_{ij})x_1^{m_{1j}}\cdots x_s^{m_{sj}}\,.\end{equation}   
\end{lemma}
\begin{proof}     
    The rate equation associated with a disease compartment $x_i\in S_I$ sums along all the transitions connected to this compartment. 
    Also recall that for $z_j\in T$ some transition, the integer $m_{ij}$ is the number of arrows with source $x_i$ and target $z_j$, and the integer $n_{ij}$ 
    is the number of arrows with source $z_j$ and target $x_i$.    

Recall that the infection module is a substructure that encompasses all disease compartments of the study population, as well as all transitions that do not belong to the susceptible or infection process modules, with an input of a disease compartment. As a consequence, the rate equation of a disease compartment only sums along transitions in the susceptible, infection process, and infection modules. 
    
    Assumption (G1) implies that we do not allow arrows between a disease compartment and a transition in the susceptible module, established as $m_{ij}=n_{ij}=0$ for $x_i\in S_I$ and $z_j\in T_S$.   
    Thus, in the expression \eqref{eq13}, we do not have a sum along transitions in the susceptible module. 
    Additionally, again by assumption (G1) for a transition in the infection compartment $z_j\in T_I$, we have $m_{ij}=0$ for $x_i\not\in S_I$ which explain that $m_{ij}=0$ for $i>s$. This explains the form of the expression \eqref{eq13}.
\end{proof}
    

The functions $\ma{F}_i$ and $\ma{V}_i$ from the next-generation matrix method are now given by the following expressions:
\begin{equation}\label{equacion1}\ma{F}_i(x_1,\cdots,x_k):=\sum_{z_j\in T_{IP}} r_j(n_{ij}-m_{ij})x_1^{m_{1j}}\cdots x_k^{m_{kj}}\end{equation}and\begin{equation}\label{equacion2}
\ma{V}_i(x_1,\cdots x_k):=\sum_{z_j\in T_I}r_j(m_{ij}-n_{ij})x_1^{m_{1j}}\cdots x_s^{m_{sj}}
\end{equation}

\begin{theorem}\label{kma}
The KM modules of a Petri net with the assumptions (G1)-(G4) satisfy the assumptions (A1)-(A4) of the next-generation matrix method.    
\end{theorem}
\begin{proof}
Assume $S=\{x_1,\cdots,x_k\}$ are the whole compartments and $S_I=\{x_1,\cdots, x_s\}$ are the compartments in the infection module, with $s\leq k$.
In order to short the notation, set $x=\{x_1,\cdots, x_s\}$, $y=\{x_{s+1},\cdots, x_k\}$, and denote $x^{m_j}:=x_1^{m_{1j}}\cdots x_s^{m_{sj}} $ and $y^{m_j}:=x_{s+1}^{m_{s+1j}}\cdots x_k^{m_{kj}}$. The functions $\ma{F}_i$ and $\ma{V}_i$ have the following expressions: 
$$\ma{F}_i(x,y)=\sum_{z_j\in T_{IP}}r_j(n_{ij}-m_{ij})x^{m_j}y^{m_j}\textrm{ and }\ma{V}_i(x,y)=\sum_{z_j\in T_I}r_j(m_{ij}-n_{ij})x^{m_j}\,.$$
Using the expression \eqref{eq13} deduced in Lemma \ref{lem13}, the rate equation of a disease compartment $x_i\in S_I$ has the form 
$$x_i'(t)=\ma{F}_i(x,y)-\ma{V}_i(x,y)\,.$$

For (A1), assume $x=0$ and $y\geq 0$, hence $\ma{V}_i(0,y)=0$ and $\ma{F}_i(0,y)=\sum_{z_j\in T_{IP}}r_jn_{ij}y^{m_j}$ which is also zero by the assumption (G2). More precisely, $\ma{F}_i(0,y)$ sums along all transitions in the infection module $z_j\in T_{IP}$
where all the monomials $x_1^{m_{1j}}\cdots x_k^{m_{kj}}$ have $m_{ij}=0$ for $1\leq i\leq s$, hence assumption (G2) implies 
that $n_{ij}=0$ for $1\leq i\leq s$. Consequently, we obtain $\ma{F}_i(0,y)=0$.

For (A2), take $x,y\geq 0$ and from assumption (G3) it follows that 
for disease compartment $x_i\in S_I$ and a transition in the infection process module $z_j\in T_{IP}$, we obtain $n_{ij}\geq m_{ij}$, which implies that $\ma{F}_i(x,y)\geq 0$.

For (A3), fix an integer $1\leq i\leq s$. We can write $\ma{V}_i(x,y)$ by the sum 
$$\ma{V}_i(x,y)=\sum_{z_j\in \ma{T}_I,m_{ij}\neq 0}r_j(m_{ij}-n_{ij})x^{m_j}+\sum_{z_j\in \ma{T}_I,m_{ij}= 0}
r_j(-n_{ij})x^{m_j}\,.$$ 
The first summand vanishes whenever $ x_i = 0$. As a consequence, we obtain that $\ma{V}_i(x,y)$ with $x_i=0$, reduces to the summand $\sum_{z_j\in \ma{T}_I}r_j(-n_{ij})x^{m_j}$ which is less or equal than zero.

Finally, for (A4), we use the assumption (G4) and the sum has the form $$\sum_{i=1}^s\ma{V}_i(x,y)=\sum_{i=1}^s\sum_{z_j\in T_I}r_j(m_{ij}-n_{ij})x^{m_j}=\sum_{z_j\in T_I}r_j\left(\sum_{i=1}^s m_{ij}-n_{ij}\right)x^{m_j}\geq 0 \,.$$

\end{proof}

\subsubsection{Threshold for an outbreak in the Petri net models}
\label{threshold}
The previous section provides a special case where the assumptions (A1)-(A4) have a geometric interpretation in terms of Petri nets. The last assumption (A5), in terms of Petri nets, states that the structure obtained by removing the infection module from the entire Petri net has rate equations with a disease-free equilibrium that is locally asymptotically stable. 

Locally asymptotically stable means an equilibrium point of a dynamical system is stable in a neighborhood of the equilibrium point, and that solutions starting close to it converge to the equilibrium point as time goes to infinity.
This is equivalent to the eigenvalues of the Jacobian matrix have negative real part; see \cite[Thm. 2, p. 56]{Perko}.

It is reasonable to expect that removing the infection module, we obtain a demographic system that is locally asymptotically stable. 

In terms of Petri nets, the assumption (A5) has the following description:

\begin{itemize}
    \item[(G5)]  The rate equations of the Petri net, which result from removing the infection module, have a disease-free equilibrium that is locally asymptotically stable. In other words, we are assuming a unique equilibrium point where the eigenvalues of the Jacobian have negative real part.
\end{itemize}

The advantage of our approach is a precise expression of the functions $\mathcal{F}$ and $\mathcal{V}$. Thus, we can  define the $F$ and $V$ matrices used in the next-generation matrix by
$$F=\left( \frac{\partial \ma{F}_i}{\partial x_j}(0,y)\right)\textrm{ and }V=\left( \frac{\partial \ma{V}_i}{\partial x_j}(0,y)\right)\,,$$

\begin{definition}
    The basic number for a Petri net satisfying the assumptions (G1)-(G5) is defined by the spectral radius $R_0:=\rho(FV^{-1})$.
\end{definition}

In what follows, we demonstrate that the reproduction number computed for Petri nets serves as a threshold for an outbreak. We need specific definitions and results from \cite{DW08}.

\begin{definition}
If each entry of a matrix $T$ is nonnegative, we write $T \geq 0$ and refer to $T$ as a {\it nonnegative matrix}.     
A matrix of the form $A = sI -B$, with $B \geq  0$, is said to have the {\it $Z$ sign pattern}. These are matrices whose offdiagonal entries are negative or zero. If in addition, the spectral radius of $B$ satisfies the inequality $s\geq \rho(B)$, then A is called an {\it $M$-matrix}.
\end{definition}


An important application of the Perron-Frobenius theorem \cite[Thm. 2.1.1, p. 133]{BP70} is the following lemma.

\begin{lemma}[\cite{DW08}]\label{lemita}
    If $F$ is nonnegative and $V$ is a nonsingular $M$-matrix, then $R_0 = \rho(FV^{-1}) < 1$ if and only if all eigenvalues of $(F-V)$ have negative real parts.
\end{lemma}

Now, we use the explicit expression of the function $\mathcal{F}$ and $\mathcal{V}$ to show properties for the matrices $F$ and $V$.

\begin{proposition}\label{pipi1}
   The matrix $F$ is nonnegative, $F\geq 0$, and the matrix $V$ is a (possibly singular) $M$-matrix.
\end{proposition}

\begin{proof}
For $x_j\in S_I$, we differentiate $\ma{F}_i(x,y)$ and we obtain
\begin{equation}\label{eqFV1}
    \frac{\partial \ma{F}_i}{\partial x_j}(x,y)=\sum_{z_l\in T_{IP}}r_l(n_{il}-m_{il})m_{jl}x_1^{m_{1l}}\cdots x_j^{m_{jl}-1}\cdots x_s^{m_{sl}}x_{s+1}^{m_{s+1l}}\cdots x_k^{m_{kl}}\,.
\end{equation}
From assumption (G3) we have $n_{il}\geq m_{il}$ for $x_i\in S_I$ and $z_l\in T_{IP}$. It follows that each entry of $F$ is nonnegative. 

For $x_i\in S_I$, we differentiate $\ma{V}_i(x,y)$ and we obtain
\begin{equation}\label{eqFV2}\frac{\partial \ma{V}_i}{\partial x_j}(x,y)  =\sum_{z_l\in T_I, m_{jl}\neq 0}r_l(m_{il}-n_{il})m_{jl}x_1^{m_{1l}}\cdots x_j^{m_{jl}-1}\cdots x_s^{m_{sl}}\,.
\end{equation}
From assumption (G3) we have $n_{il}\geq m_{il}$ for $x_i\in S_I$ and $z_l\in T_{I}$. It follows that the off-diagonal entries of $V$ are negative or zero.

The $M$-matrices are deeply studied in the book of Berman--Plemmons \cite{BP70}. In fact, it is so interesting the 
50 conditions equivalent to the statement {\it $A$ is a nonsingular $M$-matrix}, where $A$ has the $Z$ sign pattern; see \cite[p. 134-138]{BP70}. One of this characterization of a non-singular $M$-matrix, is the existence of a positive diagonal matrix $D$ such that $AD$ has all positive row sums (this is condition $I_{29}$ in \cite[p.136]{BP70}). From assumption (G4), we ensure that the row sums of $V$ are positive or zero, hence we obtain that $V$ is a (possibly singular) $M$-matrix. 
\end{proof}

\begin{assum}
Similar to the work of van den Driessche--Watmough \cite{DW02,DW08}, we assume that $V$ is a non-singular matrix.
\end{assum}


Now we state that the basic reproduction number of a Petri net provide a threshold for an outbreak.

\begin{theorem}\label{teoremon} The disease free equilibrium of the system conformed by the rate equations of a Petri net satisfying the assumptions (G1)-(G5) is locally asymptotically stable  if $\rho(FV^{-1})<1$, but unstable if $\rho(FV^{-1})$.
\end{theorem}

\begin{proof} This is similar as Theorem 1 in \cite{DW08} adapted for Petri nets. More precisely, a Petri net satisfying the assumptions (G1)-(G5) have a Jacobian matrix for the rate equations:
 $$J=\begin{bmatrix} 
    F-V & 0\\J_{21}& J_{22}
    \end{bmatrix}$$
We showed in Proposition \ref{pipi1} that $F$ is nonnegative and $V$ is a (possibly singular) $M$-matrix. We assumed $V$
is a singular matrix, then $\rho(FV^{-1})<1$ if and only if all eigenvalues of $(F-V)$ have negative real parts. The submatrix $J_{22}$ has all eigenvalues with negative real part by assumption (G5). It follows that the disease free equilibrium is locally asymptotically if $\rho(FV^{-1})<1$. 

The instability of $\rho(FV^{-1})$ is equal as in the proof of Theorem 1 in \cite{DW08}.
\end{proof}

\subsection{Geometric interpretation of the next-generation matrix}

The present section is motivated by the following assertion of van den Driessche--Watmough \cite[p. 33]{DW02} about the next-generation matrix: the $(i,k)$ entry of the product $FV^{-1}$ is the expected number of new infections in compartment $i$ produced by the infected individual originally introduced into compartment $k$.

Consider a Petri net with compartments $S$ and transitions $T$ with KM modules satisfying the assumptions (G1)-(G5).
A {\it path} between two compartment $x,x'\in S$ consists of a sequence of compartments  $x_{i_1},\cdots, x_{i_k}$, with $x_{i_1}=x$ and $x_{i_k}=x'$, and a sequence of transitions $z_{i_1},\cdots,z_{i_{k-1}}$, with arrows $x_{i_l}\rightarrow z_{i_l}$ and $z_{i_l}\rightarrow x_{i_{l+1}}$, for $1\leq l\leq k-1$. 
We are interested in paths between compartments with the following properties:
\begin{itemize}
    \item each path does not have repeating compartments and repeating transitions;
    \item all the compartments $x_{i_1},\cdots, x_{i_k}$ belong to the infection compartments $S_I$; and 
    \item all the transitions $z_{i_1},\cdots,z_{i_{k-1}}$ belong to the infection module except once which belongs to the infection-process module. 
\end{itemize}
For $x,x'$ disease compartments, we denote by $\op{Path}(x,x')$ all the paths from $x$ to $x'$
that traverse at some point some transition $\beta$ in the infection-process module.
Consider a path in $\op{Path}(x,x')$ and denote by $x_{i_t}$ the compartment just before the associated transition $\beta$; see Figure \ref{fefo} for an illustration. For each compartment $x_{i_l}$, with $1\leq l \leq k-1$, we have 
a subset of transitions $Z_{i_l}$at length one from the compartment. Thus all the transitions $z_{i_l,j}$, for $1\leq l \leq k-1$ and $1\leq j\leq r_i$ with only one arrow $x_{i_l}\rightarrow z_{i_l,j}$ and no return arrow. 
We assume the compartment $Z_{i_t}$ does not include the infection transition $\beta$. We denote these sets by $Z_i=\{z_{i,j}:  1\leq j\leq r_i\}$. 

\begin{assum} Notice that one assumption in our models is that there is only one arrow from each $x_{i_l}$ to $z_{i_l,j}$, for $1\leq j\leq r_i$, and not return arrows. For instance, the Example \ref{water} is a counterexample since there are arrows $I\To\alpha$ and $\alpha\To I$. Another assumption is that there are no arrows from a susceptible compartment $x$ to a transition $z_{i_l,j}$, for $1\leq l \leq k-1$ and $1\leq j\leq r_i$. For instance, a counterexample is presented by an SIS model in ecoepidemiology \cite{venturino}, where a Lotka-Volterra model is combined with competition and mutualism.
Nevertheless, these issues can be implemented in our geometric next-generation matrix model. The proper expectation values of secondary infections can be worked out to generalize our construction to other contexts, which is an interesting project in the future. 
\end{assum}

    \begin{figure}
        \centering
        \begin{tikzpicture}[scale=0.3,x=1pt,y=-1pt]

\definecolor{BLACK}{RGB}{0,0,0}
\definecolor{YELLOW}{RGB}{255,255,0}
\draw[BLACK, solid, line join=round, line cap=round, line width=1, fill=YELLOW]
	(100,180) ellipse[x radius=45, y radius=45];
\draw[BLACK, solid, line join=round, line cap=round, line width=1, fill=YELLOW]
	(400,180) ellipse[x radius=45, y radius=45];
\draw[BLACK, solid, line join=round, line cap=round, line width=1, fill=YELLOW]
	(920,180) ellipse[x radius=45, y radius=45];
\definecolor{CYAN}{RGB}{0,255,255}
\draw[BLACK, solid, line join=round, line cap=round, line width=1, fill=CYAN]
	(200,140) rectangle +(80,80);
\draw[BLACK, solid, line join=round, line cap=round, line width=1, fill=CYAN]
	(200,280) rectangle +(80,80);
\draw[BLACK, solid, line join=round, line cap=round, line width=1, fill=CYAN]
	(520,140) rectangle +(80,80);
\draw[BLACK, solid, line join=round, line cap=round, line width=1, fill=CYAN]
	(520,280) rectangle +(80,80);
\draw[BLACK, solid, line join=round, line cap=round, line width=1, fill=CYAN]
	(880,280) rectangle +(80,80);
\draw[BLACK, solid, line join=round, line cap=round, line width=1, fill=CYAN]
	(1020,280) rectangle +(80,80);
\draw[BLACK, solid, line join=round, line cap=round, line width=1]
	(145,180) -- (200,180);
\draw[BLACK, solid, line join=round, line cap=round, line width=1, fill=BLACK]
	(200,180) -- (190,183) -- (190,177) -- (200,180) -- cycle;
\draw[BLACK, solid, line join=round, line cap=round, line width=1]
	(132,212) -- (200,280);
\draw[BLACK, solid, line join=round, line cap=round, line width=1, fill=BLACK]
	(200,280) -- (191,275) -- (195,271) -- (200,280) -- cycle;
\draw[BLACK, solid, line join=round, line cap=round, line width=1]
	(280,180) -- (355,180);
\draw[BLACK, solid, line join=round, line cap=round, line width=1, fill=BLACK]
	(355,180) -- (345,183) -- (345,177) -- (355,180) -- cycle;
\draw[BLACK, solid, line join=round, line cap=round, line width=1]
	(445,180) -- (520,180);
\draw[BLACK, solid, line join=round, line cap=round, line width=1, fill=BLACK]
	(520,180) -- (510,183) -- (510,177) -- (520,180) -- cycle;
\draw[BLACK, solid, line join=round, line cap=round, line width=1]
	(434,210) -- (520,285);
\draw[BLACK, solid, line join=round, line cap=round, line width=1, fill=BLACK]
	(520,285) -- (510,281) -- (515,276) -- (520,285) -- cycle;
\draw[BLACK, solid, line join=round, line cap=round, line width=1]
	(600,180) -- (645,180);
\draw[BLACK, solid, line join=round, line cap=round, line width=1, fill=BLACK]
	(645,180) -- (635,183) -- (635,177) -- (645,180) -- cycle;
\draw[BLACK, solid, line join=round, line cap=round, line width=1]
	(788,180) -- (875,180);
\draw[BLACK, solid, line join=round, line cap=round, line width=1, fill=BLACK]
	(875,180) -- (865,183) -- (865,177) -- (875,180) -- cycle;
\draw[BLACK, solid, line join=round, line cap=round, line width=1]
	(920,225) -- (920,280);
\draw[BLACK, solid, line join=round, line cap=round, line width=1, fill=BLACK]
	(920,280) -- (917,270) -- (923,270) -- (920,280) -- cycle;
\draw[BLACK, solid, line join=round, line cap=round, line width=1]
	(952,212) -- (1020,280);
\draw[BLACK, solid, line join=round, line cap=round, line width=1, fill=BLACK]
	(1020,280) -- (1011,275) -- (1015,271) -- (1020,280) -- cycle;
\draw[BLACK, solid, line join=round, line cap=round, line width=1, fill=CYAN]
	(1020,140) rectangle +(80,80);
\draw[BLACK, solid, line join=round, line cap=round, line width=1]
	(965,180) -- (1020,180);
\draw[BLACK, solid, line join=round, line cap=round, line width=1, fill=BLACK]
	(1020,180) -- (1010,183) -- (1010,177) -- (1020,180) -- cycle;
\draw[BLACK, solid, line join=round, line cap=round, line width=1, fill=CYAN]
	(1030,820) rectangle +(140,80);
\draw[BLACK, solid, line join=round, line cap=round, line width=1, fill=YELLOW]
	(920,700) ellipse[x radius=45, y radius=45];
\draw[BLACK, solid, line join=round, line cap=round, line width=1]
	(920,588) -- (920,655);
\draw[BLACK, solid, line join=round, line cap=round, line width=1, fill=BLACK]
	(920,655) -- (917,645) -- (923,645) -- (920,655) -- cycle;
\draw[BLACK, solid, line join=round, line cap=round, line width=1]
	(920,360) -- (920,445);
\draw[BLACK, solid, line join=round, line cap=round, line width=1, fill=BLACK]
	(920,445) -- (917,435) -- (923,435) -- (920,445) -- cycle;
\draw[BLACK, solid, line join=round, line cap=round, line width=1, fill=CYAN]
	(873,820) rectangle +(95,80);
\draw[BLACK, solid, line join=round, line cap=round, line width=1]
	(920,745) -- (920,820);
\draw[BLACK, solid, line join=round, line cap=round, line width=1, fill=BLACK]
	(920,820) -- (917,810) -- (923,810) -- (920,820) -- cycle;
\draw[BLACK, solid, line join=round, line cap=round, line width=1]
	(954,730) -- (1055,820);
\draw[BLACK, solid, line join=round, line cap=round, line width=1, fill=BLACK]
	(1055,820) -- (1045,816) -- (1050,811) -- (1055,820) -- cycle;
\draw[BLACK, solid, line join=round, line cap=round, line width=1, fill=YELLOW]
	(920,1000) ellipse[x radius=40, y radius=40];
\draw[BLACK, solid, line join=round, line cap=round, line width=1]
	(920,900) -- (920,960);
\draw[BLACK, solid, line join=round, line cap=round, line width=1, fill=BLACK]
	(920,960) -- (917,950) -- (923,950) -- (920,960) -- cycle;
\draw[BLACK, solid, line join=round, line cap=round, line width=1, fill=YELLOW]
	(1260,60) ellipse[x radius=55, y radius=55];
\draw[BLACK, solid, line join=round, line cap=round, line width=1]
	(1213,88) -- (1100,156);
\draw[BLACK, solid, line join=round, line cap=round, line width=1, fill=BLACK]
	(1100,156) -- (1107,148) -- (1110,154) -- (1100,156) -- cycle;

\draw (105,185) node {$x_{i_1}$};
\draw (240.5,185) node {$y_{i_11}$};
\draw (240.5,185) node {$y_{i_11}$};
\draw (242,245) node {$\vdots$};
\draw (242,325) node {$y_{i_1r_{i_1}}$};

\draw (560.5,185) node {$y_{i_21}$};
\draw (562,245) node {$\vdots$};
\draw (562,325) node {$y_{i_2r_{i_2}}$};

\draw (1060.5,185) node {$\beta$};

\draw (1260,60) node {$\scalebox{1.8}{S}$};

\draw (990,325) node {$\hdots$};

\draw (924,325) node {$y_{i_t1}$};

\draw (1062,325) node {$y_{i_tr_{i_t}}$};

\draw (405,185) node {$x_{i_2}$};
\draw (924,185) node {$x_{i_{t}}$};

\draw (995,860) node {$\hdots$};

\draw (925,860) node {$y_{i_{k-1}1}$};
\draw (1102,860) node {$y_{i_{k-1}r_{i_{k-1}}}$};

\draw (922,505) node {\scalebox{1.8}{$\vdots$}};

\draw (925,705) node {$x_{i_{k-1}}$};

\draw (925,1005) node {$x_{i_k}$};

\draw (730,180) node {\scalebox{1.8}{$\cdots$}};
 
\end{tikzpicture}
        \caption{A path between the compartments $x=x_{i_1}$ and $x'=x_{i_k}$ passing through the transition $\beta$.}
        \label{fefo}
    \end{figure}

\begin{proposition}\label{exo}
    Assume a path $\gamma\in \op{Path}(x,x')$ of the form $(x_{i_1},\cdots,x_{i_k};z_{i_1},\cdots,z_{i_{k-1}};\beta)$ passing through the infection transition $\beta$. Additionally, we assume $z_{i_l}=z_{i_l1}$ for $1\leq l \leq k-1$ and denote by $\alpha_{i_lj}$ the parameter associated with $z_{i_lj}\in Z_{i_l}$ for $1\leq j\leq r_{i_l}$ and $1\leq l \leq k-1$. Therefore, the expected time an individual spends to cross the path $\gamma$ is obtained by the formula
    \begin{equation}
        \mathcal{E}(\gamma)=\left(\prod_{l=1}^{k-1}\alpha_{i_l1}\right)\left(\prod_{l=1}^{k-1}\sum_{t=1}^{r_{i_l}}\alpha_{i_lt}\right)^{-1}\,.
    \end{equation}
\end{proposition}
\begin{proof}
Whenever we pass through the compartment $x_{i_l}$, $1\leq l\leq k-1$, we divide by the sum of the parameters corresponding to the transitions in $Z_{i_l}$. 
The transition associated with $\beta$ does not contribute to the time we spend to pass the compartment $x_{i_t}$ because there is, in a certain sense, a return arrow canceling the arrow $x_{i_t}\To \beta$.
Finally, whenever we pass through the transition $z_{i_l}$, we must multiply by $\alpha_{i_l1}$ where recall we assumed $z_{i_l}=z_{i_l1}$ for $1\leq l \leq k-1$.
\end{proof}

Each path $\gamma\in \op{Path}(x,x')$ passes to the unique infection transition $\beta$, associated with a flow of susceptible individuals that linearly depends on $\beta$. We denote this flow of susceptible individuals by $\mathcal{S}(\gamma)$. 

The {\bf matrix of flows} of a Petri net $(S,T)$ with $S=\{x_1,\cdots, x_k\}$, $T=\{z_1,\cdots, z_l\}$ and disease compartments $S_I=\{x_1,\cdots, x_s\}$, with $s\leq k$, is a $(s\times s)$-matrix which we denoted by $M(S,T,T_I)$ with entries given as follows
\begin{equation}\label{matflow}
    M(S,T,T_I)_{ij}:=\sum_{\gamma \in\op{Path}(x_i,x_j)}\mathcal{S}(\gamma)\mathcal{E}(\gamma)\,.
\end{equation}
We have the following immediate consequence as an application of Theorem \ref{kma} and the discussion from the last section.

\begin{cor}
    The next-generation matrix associated with the rate equation of a Petri net $(S,T)$ satisfying the assumptions (G1)-(G5) is equal to the matrix of flows.
\end{cor}

\begin{example}
Now, we use this geometric procedure to calculate the basic reproduction of the SEAIR model (see Example \ref{SEAIR}). We have already found the next-generation matrix in Table \ref{Tabla1} at the end of the paper, but now we check that this coincides with the matrix of flows. The KM modules were also described in Figure \ref{SEAIR2}. It is easy to verify that the arrows between the KM modules satisfy the (G1)-(G5) conditions of the geometric next-generation method. The matrix of flows is as follows:
\begin{equation}\label{matr1}
\begin{blockarray}{cccc}
& E & A & I \\
\begin{block}{c(ccc)}
 E & \frac{\beta\eta(1-p)qS}{(\eta+\mu)(\gamma+\mu)}+\frac{\beta\eta p S}{(\eta+\mu)(\alpha+\mu)} & \frac{\beta qS}{\gamma+\mu}  & \frac{\beta S}{\alpha+\mu} \\
 A & 0 & 0 & 0 \\
 I & 0 & 0 & 0 \\
\end{block}
\end{blockarray}\end{equation}
The term associated with the $E$-$E$ coordinate corresponds to the two paths of flows in the following pictures:

\begin{minipage}[t]{0.5\textwidth} 
        \begin{center}
        \begin{tikzpicture}[scale=0.2,x=1pt,y=-1pt]

\definecolor{BLACK}{RGB}{0,0,0}
\definecolor{YELLOW}{RGB}{255,255,0}
\draw[BLACK, solid, line join=round, line cap=round, line width=1, fill=YELLOW]
	(100,280) ellipse[x radius=45, y radius=45];
\draw[BLACK, solid, line join=round, line cap=round, line width=1, fill=YELLOW]
	(400,280) ellipse[x radius=45, y radius=45];
\draw[BLACK, solid, line join=round, line cap=round, line width=1, fill=YELLOW]
	(580,100) ellipse[x radius=55, y radius=55];
\definecolor{CYAN}{RGB}{0,255,255}
\draw[BLACK, solid, line join=round, line cap=round, line width=1, fill=CYAN]
	(210,250) rectangle +(60,60);
\draw[BLACK, solid, line join=round, line cap=round, line width=1, fill=CYAN]
	(70,390) rectangle +(60,60);
\draw[BLACK, solid, line join=round, line cap=round, line width=1]
	(145,280) -- (210,280);
\draw[BLACK, solid, line join=round, line cap=round, line width=1, fill=BLACK]
	(210,280) -- (200,283) -- (200,277) -- (210,280) -- cycle;
\draw[BLACK, solid, line join=round, line cap=round, line width=1]
	(100,325) -- (100,390);
\draw[BLACK, solid, line join=round, line cap=round, line width=1, fill=BLACK]
	(100,390) -- (97,380) -- (103,380) -- (100,390) -- cycle;
\draw[BLACK, solid, line join=round, line cap=round, line width=1]
	(270,280) -- (355,280);
\draw[BLACK, solid, line join=round, line cap=round, line width=1, fill=BLACK]
	(355,280) -- (345,283) -- (345,277) -- (355,280) -- cycle;
\draw[BLACK, solid, line join=round, line cap=round, line width=1, fill=CYAN]
	(550,250) rectangle +(60,60);
\draw[BLACK, solid, line join=round, line cap=round, line width=1, fill=CYAN]
	(550,370) rectangle +(60,60);
\draw[BLACK, solid, line join=round, line cap=round, line width=1, fill=CYAN]
	(550,490) rectangle +(60,60);
\draw[BLACK, solid, line join=round, line cap=round, line width=1, fill=YELLOW]
	(760,280) ellipse[x radius=45, y radius=45];
\draw[BLACK, solid, line join=round, line cap=round, line width=1]
	(445,280) -- (550,280);
\draw[BLACK, solid, line join=round, line cap=round, line width=1, fill=BLACK]
	(550,280) -- (540,283) -- (540,277) -- (550,280) -- cycle;
\draw[BLACK, solid, line join=round, line cap=round, line width=1]
	(437,305) -- (550,380);
\draw[BLACK, solid, line join=round, line cap=round, line width=1, fill=BLACK]
	(550,380) -- (540,377) -- (544,372) -- (550,380) -- cycle;
\draw[BLACK, solid, line join=round, line cap=round, line width=1]
	(427,316) -- (558,490);
\draw[BLACK, solid, line join=round, line cap=round, line width=1, fill=BLACK]
	(558,490) -- (549,484) -- (554,480) -- (558,490) -- cycle;
\draw[BLACK, solid, line join=round, line cap=round, line width=1]
	(610,280) -- (715,280);
\draw[BLACK, solid, line join=round, line cap=round, line width=1, fill=BLACK]
	(715,280) -- (705,283) -- (705,277) -- (715,280) -- cycle;
\draw[BLACK, solid, line join=round, line cap=round, line width=1]
	(580,155) -- (580,250);
\draw[BLACK, solid, line join=round, line cap=round, line width=1, fill=BLACK]
	(580,250) -- (577,240) -- (583,240) -- (580,250) -- cycle;

 \draw (100,280) node {$E$};

\draw (245,280) node {$\eta$};

\draw (320,220) node {$1-p$};
 
\draw (100,420) node {$\mu$};

  \draw (400,280) node {$A$};
  \draw (580,280) node {$q\beta$};

\draw (580,400) node {$\gamma$};

\draw (580,520) node {$\mu$};
  
   \draw (760,280) node {$E$};
   \draw (580,100) node {$S$};
 
\end{tikzpicture}     
        \end{center}
\end{minipage}
\begin{minipage}[t]{0.5\textwidth} 
\begin{center}
 \begin{tikzpicture}[scale=0.2,x=1pt,y=-1pt]

\definecolor{BLACK}{RGB}{0,0,0}
\definecolor{YELLOW}{RGB}{255,255,0}
\draw[BLACK, solid, line join=round, line cap=round, line width=1, fill=YELLOW]
	(100,280) ellipse[x radius=45, y radius=45];
\draw[BLACK, solid, line join=round, line cap=round, line width=1, fill=YELLOW]
	(400,280) ellipse[x radius=45, y radius=45];
\draw[BLACK, solid, line join=round, line cap=round, line width=1, fill=YELLOW]
	(580,100) ellipse[x radius=55, y radius=55];
\definecolor{CYAN}{RGB}{0,255,255}
\draw[BLACK, solid, line join=round, line cap=round, line width=1, fill=CYAN]
	(210,250) rectangle +(60,60);
\draw[BLACK, solid, line join=round, line cap=round, line width=1, fill=CYAN]
	(70,390) rectangle +(60,60);
\draw[BLACK, solid, line join=round, line cap=round, line width=1]
	(145,280) -- (210,280);
\draw[BLACK, solid, line join=round, line cap=round, line width=1, fill=BLACK]
	(210,280) -- (200,283) -- (200,277) -- (210,280) -- cycle;
\draw[BLACK, solid, line join=round, line cap=round, line width=1]
	(100,325) -- (100,390);
\draw[BLACK, solid, line join=round, line cap=round, line width=1, fill=BLACK]
	(100,390) -- (97,380) -- (103,380) -- (100,390) -- cycle;
\draw[BLACK, solid, line join=round, line cap=round, line width=1]
	(270,280) -- (355,280);
\draw[BLACK, solid, line join=round, line cap=round, line width=1, fill=BLACK]
	(355,280) -- (345,283) -- (345,277) -- (355,280) -- cycle;
\draw[BLACK, solid, line join=round, line cap=round, line width=1, fill=CYAN]
	(550,250) rectangle +(60,60);
\draw[BLACK, solid, line join=round, line cap=round, line width=1, fill=CYAN]
	(550,370) rectangle +(60,60);
\draw[BLACK, solid, line join=round, line cap=round, line width=1, fill=CYAN]
	(550,490) rectangle +(60,60);
\draw[BLACK, solid, line join=round, line cap=round, line width=1, fill=YELLOW]
	(760,280) ellipse[x radius=45, y radius=45];
\draw[BLACK, solid, line join=round, line cap=round, line width=1]
	(445,280) -- (550,280);
\draw[BLACK, solid, line join=round, line cap=round, line width=1, fill=BLACK]
	(550,280) -- (540,283) -- (540,277) -- (550,280) -- cycle;
\draw[BLACK, solid, line join=round, line cap=round, line width=1]
	(437,305) -- (550,380);
\draw[BLACK, solid, line join=round, line cap=round, line width=1, fill=BLACK]
	(550,380) -- (540,377) -- (544,372) -- (550,380) -- cycle;
\draw[BLACK, solid, line join=round, line cap=round, line width=1]
	(427,316) -- (558,490);
\draw[BLACK, solid, line join=round, line cap=round, line width=1, fill=BLACK]
	(558,490) -- (549,484) -- (554,480) -- (558,490) -- cycle;
\draw[BLACK, solid, line join=round, line cap=round, line width=1]
	(610,280) -- (715,280);
\draw[BLACK, solid, line join=round, line cap=round, line width=1, fill=BLACK]
	(715,280) -- (705,283) -- (705,277) -- (715,280) -- cycle;
\draw[BLACK, solid, line join=round, line cap=round, line width=1]
	(580,155) -- (580,250);
\draw[BLACK, solid, line join=round, line cap=round, line width=1, fill=BLACK]
	(580,250) -- (577,240) -- (583,240) -- (580,250) -- cycle;

 \draw (100,280) node {$E$};

\draw (245,280) node {$\eta$};

\draw (320,220) node {$p$};
 
\draw (100,420) node {$\mu$};

  \draw (400,280) node {$I$};
  \draw (580,280) node {$\beta$};

\draw (580,400) node {$\alpha$};

\draw (580,520) node {$\mu$};
  
   \draw (760,280) node {$E$};
   \draw (580,100) node {$S$};

\end{tikzpicture}
 \end{center}
\end{minipage}
The terms associated with the $E$-$A$ and $E$-$I$ correspond to the following paths:

\begin{minipage}[t]{0.5\textwidth} 
    \begin{center}
        \begin{tikzpicture}[scale=0.2,x=1pt,y=-1pt]

\definecolor{BLACK}{RGB}{0,0,0}
\definecolor{YELLOW}{RGB}{255,255,0}
\draw[BLACK, solid, line join=round, line cap=round, line width=1, fill=YELLOW]
	(400,280) ellipse[x radius=45, y radius=45];
\draw[BLACK, solid, line join=round, line cap=round, line width=1, fill=YELLOW]
	(580,100) ellipse[x radius=55, y radius=55];
\definecolor{CYAN}{RGB}{0,255,255}
\draw[BLACK, solid, line join=round, line cap=round, line width=1, fill=CYAN]
	(550,250) rectangle +(60,60);
\draw[BLACK, solid, line join=round, line cap=round, line width=1, fill=CYAN]
	(550,370) rectangle +(60,60);
\draw[BLACK, solid, line join=round, line cap=round, line width=1, fill=CYAN]
	(550,490) rectangle +(60,60);
\draw[BLACK, solid, line join=round, line cap=round, line width=1, fill=YELLOW]
	(760,280) ellipse[x radius=45, y radius=45];
\draw[BLACK, solid, line join=round, line cap=round, line width=1]
	(445,280) -- (550,280);
\draw[BLACK, solid, line join=round, line cap=round, line width=1, fill=BLACK]
	(550,280) -- (540,283) -- (540,277) -- (550,280) -- cycle;
\draw[BLACK, solid, line join=round, line cap=round, line width=1]
	(437,305) -- (550,380);
\draw[BLACK, solid, line join=round, line cap=round, line width=1, fill=BLACK]
	(550,380) -- (540,377) -- (544,372) -- (550,380) -- cycle;
\draw[BLACK, solid, line join=round, line cap=round, line width=1]
	(427,316) -- (558,490);
\draw[BLACK, solid, line join=round, line cap=round, line width=1, fill=BLACK]
	(558,490) -- (549,484) -- (554,480) -- (558,490) -- cycle;
\draw[BLACK, solid, line join=round, line cap=round, line width=1]
	(610,280) -- (715,280);
\draw[BLACK, solid, line join=round, line cap=round, line width=1, fill=BLACK]
	(715,280) -- (705,283) -- (705,277) -- (715,280) -- cycle;
\draw[BLACK, solid, line join=round, line cap=round, line width=1]
	(580,155) -- (580,250);
\draw[BLACK, solid, line join=round, line cap=round, line width=1, fill=BLACK]
	(580,250) -- (577,240) -- (583,240) -- (580,250) -- cycle;

  \draw (400,280) node {$A$};
  \draw (580,280) node {$q\beta$};

\draw (580,400) node {$\gamma$};

\draw (580,520) node {$\mu$};
  
   \draw (760,280) node {$E$};
   \draw (580,100) node {$S$};
\end{tikzpicture}
    \end{center}
\end{minipage}
\begin{minipage}[t]{0.5\textwidth} 
\begin{center}
 \begin{tikzpicture}[scale=0.2,x=1pt,y=-1pt]

\definecolor{BLACK}{RGB}{0,0,0}
\definecolor{YELLOW}{RGB}{255,255,0}
\draw[BLACK, solid, line join=round, line cap=round, line width=1, fill=YELLOW]
	(400,280) ellipse[x radius=45, y radius=45];
\draw[BLACK, solid, line join=round, line cap=round, line width=1, fill=YELLOW]
	(580,100) ellipse[x radius=55, y radius=55];
\definecolor{CYAN}{RGB}{0,255,255}
\draw[BLACK, solid, line join=round, line cap=round, line width=1, fill=CYAN]
	(550,250) rectangle +(60,60);
\draw[BLACK, solid, line join=round, line cap=round, line width=1, fill=CYAN]
	(550,370) rectangle +(60,60);
\draw[BLACK, solid, line join=round, line cap=round, line width=1, fill=CYAN]
	(550,490) rectangle +(60,60);
\draw[BLACK, solid, line join=round, line cap=round, line width=1, fill=YELLOW]
	(760,280) ellipse[x radius=45, y radius=45];
\draw[BLACK, solid, line join=round, line cap=round, line width=1]
	(445,280) -- (550,280);
\draw[BLACK, solid, line join=round, line cap=round, line width=1, fill=BLACK]
	(550,280) -- (540,283) -- (540,277) -- (550,280) -- cycle;
\draw[BLACK, solid, line join=round, line cap=round, line width=1]
	(437,305) -- (550,380);
\draw[BLACK, solid, line join=round, line cap=round, line width=1, fill=BLACK]
	(550,380) -- (540,377) -- (544,372) -- (550,380) -- cycle;
\draw[BLACK, solid, line join=round, line cap=round, line width=1]
	(427,316) -- (558,490);
\draw[BLACK, solid, line join=round, line cap=round, line width=1, fill=BLACK]
	(558,490) -- (549,484) -- (554,480) -- (558,490) -- cycle;
\draw[BLACK, solid, line join=round, line cap=round, line width=1]
	(610,280) -- (715,280);
\draw[BLACK, solid, line join=round, line cap=round, line width=1, fill=BLACK]
	(715,280) -- (705,283) -- (705,277) -- (715,280) -- cycle;
\draw[BLACK, solid, line join=round, line cap=round, line width=1]
	(580,155) -- (580,250);
\draw[BLACK, solid, line join=round, line cap=round, line width=1, fill=BLACK]
	(580,250) -- (577,240) -- (583,240) -- (580,250) -- cycle;

  \draw (400,280) node {$I$};
  \draw (580,280) node {$\beta$};

\draw (580,400) node {$\alpha$};

\draw (580,520) node {$\mu$};
  
   \draw (760,280) node {$E$};
   \draw (580,100) node {$S$};
   
\end{tikzpicture}  
\end{center}
\end{minipage}
Notice that the other terms in the matrix \eqref{matr1} are zero because the possible paths do not pass through an infection transition. The basic reproduction number is the dominant eigenvalue of the matrix \eqref{matr1} as follows:
\begin{equation}R_0=\frac{\beta\eta(1-p)qS}{(\eta+\mu)(\gamma+\mu)}+\frac{\beta\eta p S}{(\eta+\mu)(\alpha+\mu)}\,.\end{equation}
\end{example}

\begin{example}  The KM modules of the model of Malaria are found in Example \ref{malaroso} where we have the condition (G1)-(G5).  
The next-generation matrix of the model of Malaria was given in Table \ref{Tabla1} at the end of the paper. We want to notice the particularity of the form of this matrix, which is as follows:
\begin{equation}\label{mat22}
\begin{blockarray}{ccc}
 & I_H & I_M \\
\begin{block}{c(cc)}
 I_H & 0 & \frac{\beta_{HM} S_H}{\mu_M} \\
 I_M & \frac{\beta_{MH} S_M}{\alpha-\delta+\mu_H+\sigma} & 0 \\
\end{block}
\end{blockarray}\end{equation}
Notice that the diagonal entries are zero, so the eigenvalues are calculated by a square root. This case is special since no path realizes the basic reproduction number. The two nontrivial components of the matrix of flows \eqref{mat22} are realized by the following two paths:

\begin{minipage}[t]{0.5\textwidth} 
    \begin{center}
        \begin{tikzpicture}[scale=0.2,x=1pt,y=-1pt]

\definecolor{BLACK}{RGB}{0,0,0}
\definecolor{YELLOW}{RGB}{255,255,0}
\draw[BLACK, solid, line join=round, line cap=round, line width=1, fill=YELLOW]
	(400,280) ellipse[x radius=45, y radius=45];
\draw[BLACK, solid, line join=round, line cap=round, line width=1, fill=YELLOW]
	(580,100) ellipse[x radius=55, y radius=55];
\draw[BLACK, solid, line join=round, line cap=round, line width=1, fill=YELLOW]
	(760,280) ellipse[x radius=45, y radius=45];
\definecolor{CYAN}{RGB}{0,255,255}
\draw[BLACK, solid, line join=round, line cap=round, line width=1, fill=CYAN]
	(520,240) rectangle +(120,80);
\draw[BLACK, solid, line join=round, line cap=round, line width=1, fill=CYAN]
	(520,370) rectangle +(120,60);
\draw[BLACK, solid, line join=round, line cap=round, line width=1, fill=CYAN]
	(550,490) rectangle +(60,60);
\draw[BLACK, solid, line join=round, line cap=round, line width=1, fill=CYAN]
	(540,610) rectangle +(80,60);
\draw[BLACK, solid, line join=round, line cap=round, line width=1]
	(445,280) -- (520,280);
\draw[BLACK, solid, line join=round, line cap=round, line width=1, fill=BLACK]
	(520,280) -- (510,283) -- (510,277) -- (520,280) -- cycle;
\draw[BLACK, solid, line join=round, line cap=round, line width=1]
	(437,305) -- (535,370);
\draw[BLACK, solid, line join=round, line cap=round, line width=1, fill=BLACK]
	(535,370) -- (525,367) -- (529,362) -- (535,370) -- cycle;
\draw[BLACK, solid, line join=round, line cap=round, line width=1]
	(427,316) -- (558,490);
\draw[BLACK, solid, line join=round, line cap=round, line width=1, fill=BLACK]
	(558,490) -- (549,484) -- (554,480) -- (558,490) -- cycle;
\draw[BLACK, solid, line join=round, line cap=round, line width=1]
	(640,280) -- (715,280);
\draw[BLACK, solid, line join=round, line cap=round, line width=1, fill=BLACK]
	(715,280) -- (705,283) -- (705,277) -- (715,280) -- cycle;
\draw[BLACK, solid, line join=round, line cap=round, line width=1]
	(580,155) -- (580,240);
\draw[BLACK, solid, line join=round, line cap=round, line width=1, fill=BLACK]
	(580,240) -- (577,230) -- (583,230) -- (580,240) -- cycle;
\draw[BLACK, solid, line join=round, line cap=round, line width=1]
	(420,320) -- (565,610);
\draw[BLACK, solid, line join=round, line cap=round, line width=1, fill=BLACK]
	(565,610) -- (558,603) -- (564,600) -- (565,610) -- cycle;

  \draw (400,280) node {$I_H$};
  \draw (580,283) node {$\beta_{MH}$};

\draw (580,400) node {$\alpha-\delta$};

\draw (580,520) node {$\sigma$};

\draw (580,640) node {$\mu_H$};
  
   \draw (760,280) node {$I_M$};
   \draw (580,103) node {$S_M$}; 
\end{tikzpicture}
          \end{center}
\end{minipage}
\begin{minipage}[t]{0.5\textwidth} 
\begin{center}
 \begin{tikzpicture}[scale=0.2,x=1pt,y=-1pt]

\definecolor{BLACK}{RGB}{0,0,0}
\definecolor{YELLOW}{RGB}{255,255,0}
\draw[BLACK, solid, line join=round, line cap=round, line width=1, fill=YELLOW]
	(400,280) ellipse[x radius=45, y radius=45];
\draw[BLACK, solid, line join=round, line cap=round, line width=1, fill=YELLOW]
	(580,100) ellipse[x radius=55, y radius=55];
\draw[BLACK, solid, line join=round, line cap=round, line width=1, fill=YELLOW]
	(760,280) ellipse[x radius=45, y radius=45];
\definecolor{CYAN}{RGB}{0,255,255}
\draw[BLACK, solid, line join=round, line cap=round, line width=1, fill=CYAN]
	(520,240) rectangle +(120,80);
\draw[BLACK, solid, line join=round, line cap=round, line width=1, fill=CYAN]
	(540,370) rectangle +(80,60);
\draw[BLACK, solid, line join=round, line cap=round, line width=1]
	(445,280) -- (520,280);
\draw[BLACK, solid, line join=round, line cap=round, line width=1, fill=BLACK]
	(520,280) -- (510,283) -- (510,277) -- (520,280) -- cycle;
\draw[BLACK, solid, line join=round, line cap=round, line width=1]
	(437,305) -- (540,373);
\draw[BLACK, solid, line join=round, line cap=round, line width=1, fill=BLACK]
	(540,373) -- (530,371) -- (534,365) -- (540,373) -- cycle;
\draw[BLACK, solid, line join=round, line cap=round, line width=1]
	(640,280) -- (715,280);
\draw[BLACK, solid, line join=round, line cap=round, line width=1, fill=BLACK]
	(715,280) -- (705,283) -- (705,277) -- (715,280) -- cycle;
\draw[BLACK, solid, line join=round, line cap=round, line width=1]
	(580,155) -- (580,240);
\draw[BLACK, solid, line join=round, line cap=round, line width=1, fill=BLACK]
	(580,240) -- (577,230) -- (583,230) -- (580,240) -- cycle;

  \draw (400,280) node {$I_M$};
  \draw (580,283) node {$\beta_{HM}$};

\draw (580,400) node {$\mu_M$};

   \draw (760,280) node {$I_H$};
   \draw (580,103) node {$S_H$};

\end{tikzpicture}
 \end{center}
\end{minipage}

\end{example}

\begin{example}
We have already verified in Example \ref{ejVac} the next-generation matrix for the first model of vaccination is given by the following matrix: 
\begin{equation}
\begin{blockarray}{cc}
 & I \\
\begin{block}{c(c)}
 I & \frac{\beta S}{\alpha+\mu}  \\
\end{block}
\end{blockarray}\end{equation}
The unique term $I$-$I$ corresponds to the following path: 
\begin{center}
        \begin{tikzpicture}[scale=0.2,x=1pt,y=-1pt]

\definecolor{BLACK}{RGB}{0,0,0}
\definecolor{YELLOW}{RGB}{255,255,0}
\draw[BLACK, solid, line join=round, line cap=round, line width=1, fill=YELLOW]
	(400,280) ellipse[x radius=45, y radius=45];
\draw[BLACK, solid, line join=round, line cap=round, line width=1, fill=YELLOW]
	(580,100) ellipse[x radius=55, y radius=55];
\definecolor{CYAN}{RGB}{0,255,255}
\draw[BLACK, solid, line join=round, line cap=round, line width=1, fill=CYAN]
	(550,250) rectangle +(60,60);
\draw[BLACK, solid, line join=round, line cap=round, line width=1, fill=CYAN]
	(550,370) rectangle +(60,60);
\draw[BLACK, solid, line join=round, line cap=round, line width=1, fill=CYAN]
	(550,490) rectangle +(60,60);
\draw[BLACK, solid, line join=round, line cap=round, line width=1, fill=YELLOW]
	(760,280) ellipse[x radius=45, y radius=45];
\draw[BLACK, solid, line join=round, line cap=round, line width=1]
	(445,280) -- (550,280);
\draw[BLACK, solid, line join=round, line cap=round, line width=1, fill=BLACK]
	(550,280) -- (540,283) -- (540,277) -- (550,280) -- cycle;
\draw[BLACK, solid, line join=round, line cap=round, line width=1]
	(437,305) -- (550,380);
\draw[BLACK, solid, line join=round, line cap=round, line width=1, fill=BLACK]
	(550,380) -- (540,377) -- (544,372) -- (550,380) -- cycle;
\draw[BLACK, solid, line join=round, line cap=round, line width=1]
	(427,316) -- (558,490);
\draw[BLACK, solid, line join=round, line cap=round, line width=1, fill=BLACK]
	(558,490) -- (549,484) -- (554,480) -- (558,490) -- cycle;
\draw[BLACK, solid, line join=round, line cap=round, line width=1]
	(610,280) -- (715,280);
\draw[BLACK, solid, line join=round, line cap=round, line width=1, fill=BLACK]
	(715,280) -- (705,283) -- (705,277) -- (715,280) -- cycle;
\draw[BLACK, solid, line join=round, line cap=round, line width=1]
	(580,155) -- (580,250);
\draw[BLACK, solid, line join=round, line cap=round, line width=1, fill=BLACK]
	(580,250) -- (577,240) -- (583,240) -- (580,250) -- cycle;

  \draw (400,280) node {$I$};
  \draw (580,280) node {$\beta$};

\draw (580,400) node {$\alpha$};

\draw (580,520) node {$\mu$};
  
   \draw (760,280) node {$I$};
   \draw (580,100) node {$S$};
\end{tikzpicture}   
\end{center}

    Similarly, in Example \ref{ejVac}, we found the next generation matrix for the second model of vaccination 
by the matrix 
\begin{equation}
\begin{blockarray}{cc}
 & I \\
\begin{block}{c(c)}
 I & \frac{\beta S/N+\beta\delta V/N}{\gamma+\mu}  \\
\end{block}
\end{blockarray}\end{equation}
The unique term $I$-$I$ is composed of two summands corresponding to the following two paths:

\begin{minipage}[t]{0.5\textwidth} 
    \begin{center}
        \begin{tikzpicture}[scale=0.2,x=1pt,y=-1pt]

\definecolor{BLACK}{RGB}{0,0,0}
\definecolor{YELLOW}{RGB}{255,255,0}
\draw[BLACK, solid, line join=round, line cap=round, line width=1, fill=YELLOW]
	(400,280) ellipse[x radius=45, y radius=45];
\draw[BLACK, solid, line join=round, line cap=round, line width=1, fill=YELLOW]
	(580,100) ellipse[x radius=55, y radius=55];
\definecolor{CYAN}{RGB}{0,255,255}
\draw[BLACK, solid, line join=round, line cap=round, line width=1, fill=CYAN]
	(550,250) rectangle +(60,60);
\draw[BLACK, solid, line join=round, line cap=round, line width=1, fill=CYAN]
	(550,370) rectangle +(60,60);
\draw[BLACK, solid, line join=round, line cap=round, line width=1, fill=CYAN]
	(550,490) rectangle +(60,60);
\draw[BLACK, solid, line join=round, line cap=round, line width=1, fill=YELLOW]
	(760,280) ellipse[x radius=45, y radius=45];
\draw[BLACK, solid, line join=round, line cap=round, line width=1]
	(445,280) -- (550,280);
\draw[BLACK, solid, line join=round, line cap=round, line width=1, fill=BLACK]
	(550,280) -- (540,283) -- (540,277) -- (550,280) -- cycle;
\draw[BLACK, solid, line join=round, line cap=round, line width=1]
	(437,305) -- (550,380);
\draw[BLACK, solid, line join=round, line cap=round, line width=1, fill=BLACK]
	(550,380) -- (540,377) -- (544,372) -- (550,380) -- cycle;
\draw[BLACK, solid, line join=round, line cap=round, line width=1]
	(427,316) -- (558,490);
\draw[BLACK, solid, line join=round, line cap=round, line width=1, fill=BLACK]
	(558,490) -- (549,484) -- (554,480) -- (558,490) -- cycle;
\draw[BLACK, solid, line join=round, line cap=round, line width=1]
	(610,280) -- (715,280);
\draw[BLACK, solid, line join=round, line cap=round, line width=1, fill=BLACK]
	(715,280) -- (705,283) -- (705,277) -- (715,280) -- cycle;
\draw[BLACK, solid, line join=round, line cap=round, line width=1]
	(580,155) -- (580,250);
\draw[BLACK, solid, line join=round, line cap=round, line width=1, fill=BLACK]
	(580,250) -- (577,240) -- (583,240) -- (580,250) -- cycle;

  \draw (400,280) node {$I$};

    \draw (640,200) node {$1/N$};
    
  \draw (580,280) node {$\beta$};

\draw (580,400) node {$\gamma$};

\draw (580,520) node {$\mu$};
  
   \draw (760,280) node {$I$};
   \draw (580,100) node {$S$};

\end{tikzpicture}
    \end{center}
\end{minipage}
\begin{minipage}[t]{0.5\textwidth} 
\begin{center}
 \begin{tikzpicture}[scale=0.2,x=1pt,y=-1pt]

\definecolor{BLACK}{RGB}{0,0,0}
\definecolor{YELLOW}{RGB}{255,255,0}
\draw[BLACK, solid, line join=round, line cap=round, line width=1, fill=YELLOW]
	(400,280) ellipse[x radius=45, y radius=45];
\draw[BLACK, solid, line join=round, line cap=round, line width=1, fill=YELLOW]
	(580,100) ellipse[x radius=55, y radius=55];
\definecolor{CYAN}{RGB}{0,255,255}
\draw[BLACK, solid, line join=round, line cap=round, line width=1, fill=CYAN]
	(550,250) rectangle +(60,60);
\draw[BLACK, solid, line join=round, line cap=round, line width=1, fill=CYAN]
	(550,370) rectangle +(60,60);
\draw[BLACK, solid, line join=round, line cap=round, line width=1, fill=CYAN]
	(550,490) rectangle +(60,60);
\draw[BLACK, solid, line join=round, line cap=round, line width=1, fill=YELLOW]
	(760,280) ellipse[x radius=45, y radius=45];
\draw[BLACK, solid, line join=round, line cap=round, line width=1]
	(445,280) -- (550,280);
\draw[BLACK, solid, line join=round, line cap=round, line width=1, fill=BLACK]
	(550,280) -- (540,283) -- (540,277) -- (550,280) -- cycle;
\draw[BLACK, solid, line join=round, line cap=round, line width=1]
	(437,305) -- (550,380);
\draw[BLACK, solid, line join=round, line cap=round, line width=1, fill=BLACK]
	(550,380) -- (540,377) -- (544,372) -- (550,380) -- cycle;
\draw[BLACK, solid, line join=round, line cap=round, line width=1]
	(427,316) -- (558,490);
\draw[BLACK, solid, line join=round, line cap=round, line width=1, fill=BLACK]
	(558,490) -- (549,484) -- (554,480) -- (558,490) -- cycle;
\draw[BLACK, solid, line join=round, line cap=round, line width=1]
	(610,280) -- (715,280);
\draw[BLACK, solid, line join=round, line cap=round, line width=1, fill=BLACK]
	(715,280) -- (705,283) -- (705,277) -- (715,280) -- cycle;
\draw[BLACK, solid, line join=round, line cap=round, line width=1]
	(580,155) -- (580,250);
\draw[BLACK, solid, line join=round, line cap=round, line width=1, fill=BLACK]
	(580,250) -- (577,240) -- (583,240) -- (580,250) -- cycle;
  \draw (400,280) node {$I$};

    \draw (640,200) node {$1/N$};
    
  \draw (580,280) node {$\delta\beta$};

\draw (580,400) node {$\gamma$};

\draw (580,520) node {$\mu$};
  
   \draw (760,280) node {$I$};
   \draw (580,100) node {$S$};
  
\end{tikzpicture}   
\end{center}
\end{minipage}
Notice that vaccination is implemented in these paths of the Petri net as valves that adjust the flow of susceptible individuals in the infection transitions associated with $\beta$ and $\delta\beta$.
    
\end{example}

\section{Graph-theoretic method for the basic reproduction number}
\label{Graph}
Our approach with Petri nets provides an alternative to work in an epidemiological model using Petri nets instead of the system of ODEs. We developed in Section \ref{GeoNGM} the analog of the assumptions (A1)-(A5) of the next-generation matrix method for Petri nets. A Petri net defines the functions $\mathcal{F}$ and $\mathcal{V}$ in Equations \ref{equacion1} and \ref{equacion2}, giving a precise form of the matrices $F$ and $V$ in Equation \ref{eqFV1} and Equation \ref{eqFV2}. Therefore, the assumptions (G1)-(G5) assure the next-generation matrix $FV^{-1}$ have a dominant eigenvalue defining the basic reproduction number $R_0$. The calculation of this dominant eigenvalue can be realized as usual by the analytic process. However, there is a geometric procedure developed by de Camino-Beck-Lewis in \cite{Camino-Lewis}, where a digraph is associated with the matrix $F\lambda^{-1}-V$, and a geometric Gaussian elimination is performed to obtain the solution of $\det(F\lambda^{-1}-V)$ by the basic reproduction number $R_0$; see \cite{Camin-Lewis-Driessche}. 
This process comes from matrix models of structured populations \cite{Caswell} and is useful to find $R_0$ geometrically.

The matrices $F$ and $V$ conform the projection matrix $A=F+(I-V)$. The life cycle graph $G_A$ associated with $A=(a_{ij})$, is a weighted digraph with vertices $\{1,\cdots,s\}$ and for $a_{ij}\neq 0$, we have an edge from $j$ to $i$ with weight $a_{ij}$.
To calculate $R_0$, consider the life cycle graph of $F\lambda^{-1}-V$ and perform certain geometric Gaussian elimination, described in \cite{Camin-Lewis-Driessche}, to arrive at a single loop. The solution of the equation $\det(F\lambda^{-1}-V)=0$ provides the value for $R_0$. There is an algebraic counterpart by Rueffler-Metz \cite{Rueffler-Metz} where for the matrices $F=(f_{lk})$ and $I-V=(s_{lk})$, we consider loops 
as sequences $s_{lk}$, $f_{lk}$ from a state in the cycle graph to itself without passing through any stage more than once. A fertility loop is a loop that contains at least one fertility parameter $f_{lk}$. In the case of the next-generation matrix has only a single non-zero eigenvalue, there is a formula in \cite{Rueffler-Metz} for $R_0$ as follows:
\begin{equation}\label{RM}
    R_0=\sum_{\mathcal{L}_{f}}L\frac{\det (V_{\setminus \breve{L}})}{\det(V)}\,,
\end{equation}
where $\mathcal{L}_{f}$ denotes the collection of all fertility loops, $\breve{L}$ denotes all the states that are traversed by loop $L$, and $V_{\setminus \breve{L}}$ denotes the matrix $V$ dropping all the rows and columns corresponding to the states in $\breve{L}$.

If the next-generation matrix has rank one, the formula \eqref{RM} formula corresponds to Petri nets' setting to all the loops that start in the infection module traversing through the infection-process module once. The explanation comes from the characteristic polynomial where the solution is the trace. Hence, we are summing all the loops in the diagonal. We do not know a model in epidemiology with the next-generation matrix of rank $>1$ with only a single non-zero eigenvalue. In what follows, we present an example comparing the procedure with the life cycle graph and the approach with Petri nets.

\begin{example} In \cite{SB}, a SEIR model is used for the COVID-19 outbreak in India with the following system of ODEs:
$$\begin{array}{l}
S'(t)=\Lambda -(\beta+\beta')SI-\mu S,\\
E'(t)=\beta SI-(\eta +\mu )E,\\
I'(t)=\eta E-(\alpha+\mu) I+\beta'SI,\\ 
R'(t)=\alpha I-\mu R\,.
\end{array}$$
The associated Petri net is depicted in Figure \ref{SEIR-BS}.
\begin{figure}
    \centering
    \begin{tikzpicture}[scale=0.3,x=1pt,y=-1pt]

\definecolor{BLACK}{RGB}{0,0,0}
\definecolor{r255g255b10}{RGB}{255,255,10}
\draw[BLACK, solid, line join=round, line cap=round, line width=1, fill=r255g255b10]
	(200,200) ellipse[x radius=30, y radius=30];
\draw[BLACK, solid, line join=round, line cap=round, line width=1, fill=r255g255b10]
	(440,200) ellipse[x radius=30, y radius=30];
\draw[BLACK, solid, line join=round, line cap=round, line width=1, fill=r255g255b10]
	(700,200) ellipse[x radius=30, y radius=30];
\draw[BLACK, solid, line join=round, line cap=round, line width=1, fill=r255g255b10]
	(940,200) ellipse[x radius=30, y radius=30];
\definecolor{r33g255b255}{RGB}{33,255,255}
\draw[BLACK, solid, line join=round, line cap=round, line width=1, fill=r33g255b255]
	(75,55) rectangle +(50,50);
\draw[BLACK, solid, line join=round, line cap=round, line width=1, fill=r33g255b255]
	(175,295) rectangle +(50,50);
\draw[BLACK, solid, line join=round, line cap=round, line width=1, fill=r33g255b255]
	(415,295) rectangle +(50,50);
\draw[BLACK, solid, line join=round, line cap=round, line width=1, fill=r33g255b255]
	(675,295) rectangle +(50,50);
\draw[BLACK, solid, line join=round, line cap=round, line width=1, fill=r33g255b255]
	(915,295) rectangle +(50,50);
\draw[BLACK, solid, line join=round, line cap=round, line width=1, fill=r33g255b255]
	(295,75) rectangle +(50,50);
\draw[BLACK, solid, line join=round, line cap=round, line width=1, fill=r33g255b255]
	(555,175) rectangle +(50,50);
\draw[BLACK, solid, line join=round, line cap=round, line width=1, fill=r33g255b255]
	(795,175) rectangle +(50,50);
\draw[BLACK, solid, line join=round, line cap=round, line width=1]
	(121,105) -- (175,170);
\draw[BLACK, solid, line join=round, line cap=round, line width=1, fill=BLACK]
	(175,170) -- (166,164) -- (171,160) -- (175,170) -- cycle;
\draw[BLACK, solid, line join=round, line cap=round, line width=1]
	(200,230) -- (200,295);
\draw[BLACK, solid, line join=round, line cap=round, line width=1, fill=BLACK]
	(200,295) -- (197,285) -- (203,285) -- (200,295) -- cycle;
\draw[BLACK, solid, line join=round, line cap=round, line width=1]
	(440,230) -- (440,295);
\draw[BLACK, solid, line join=round, line cap=round, line width=1, fill=BLACK]
	(440,295) -- (437,285) -- (443,285) -- (440,295) -- cycle;
\draw[BLACK, solid, line join=round, line cap=round, line width=1]
	(700,230) -- (700,295);
\draw[BLACK, solid, line join=round, line cap=round, line width=1, fill=BLACK]
	(700,295) -- (697,285) -- (703,285) -- (700,295) -- cycle;
\draw[BLACK, solid, line join=round, line cap=round, line width=1]
	(940,230) -- (940,295);
\draw[BLACK, solid, line join=round, line cap=round, line width=1, fill=BLACK]
	(940,295) -- (937,285) -- (943,285) -- (940,295) -- cycle;
\draw[BLACK, solid, line join=round, line cap=round, line width=1]
	(230,175) -- (295,121);
\draw[BLACK, solid, line join=round, line cap=round, line width=1, fill=BLACK]
	(295,121) -- (289,130) -- (285,125) -- (295,121) -- cycle;
\draw[BLACK, solid, line join=round, line cap=round, line width=1]
	(345,121) -- (410,175);
\draw[BLACK, solid, line join=round, line cap=round, line width=1, fill=BLACK]
	(410,175) -- (400,171) -- (404,166) -- (410,175) -- cycle;
\draw[BLACK, solid, line join=round, line cap=round, line width=1]
	(470,200) -- (555,200);
\draw[BLACK, solid, line join=round, line cap=round, line width=1, fill=BLACK]
	(555,200) -- (545,203) -- (545,197) -- (555,200) -- cycle;
\draw[BLACK, solid, line join=round, line cap=round, line width=1]
	(605,200) -- (670,200);
\draw[BLACK, solid, line join=round, line cap=round, line width=1, fill=BLACK]
	(670,200) -- (660,203) -- (660,197) -- (670,200) -- cycle;
\draw[BLACK, solid, line join=round, line cap=round, line width=1]
	(730,200) -- (795,200);
\draw[BLACK, solid, line join=round, line cap=round, line width=1, fill=BLACK]
	(795,200) -- (785,203) -- (785,197) -- (795,200) -- cycle;
\draw[BLACK, solid, line join=round, line cap=round, line width=1]
	(845,200) -- (910,200);
\draw[BLACK, solid, line join=round, line cap=round, line width=1, fill=BLACK]
	(910,200) -- (900,203) -- (900,197) -- (910,200) -- cycle;
\draw[BLACK, solid, line join=round, line cap=round, line width=1]
	(308,75) -- (280,20) -- (280,20) -- (740,20) -- (707,170);
\draw[BLACK, solid, line join=round, line cap=round, line width=1, fill=BLACK]
	(707,170) -- (706,160) -- (712,161) -- (707,170) -- cycle;
\draw[BLACK, solid, line join=round, line cap=round, line width=1]
	(700,170) -- (700,100) -- (700,100) -- (345,100);
\draw[BLACK, solid, line join=round, line cap=round, line width=1, fill=BLACK]
	(345,100) -- (355,97) -- (355,103) -- (345,100) -- cycle;
\draw[BLACK, solid, line join=round, line cap=round, line width=1, fill=r33g255b255]
	(295,295) rectangle +(50,50);
\draw[BLACK, solid, line join=round, line cap=round, line width=1]
	(230,230) -- (295,295);
\draw[BLACK, solid, line join=round, line cap=round, line width=1, fill=BLACK]
	(295,295) -- (286,290) -- (290,286) -- (295,295) -- cycle;
\draw[BLACK, solid, line join=round, line cap=round, line width=1]
	(670,230) -- (520,380) -- (520,380) -- (320,380) -- (320,345);
\draw[BLACK, solid, line join=round, line cap=round, line width=1, fill=BLACK]
	(320,345) -- (323,355) -- (317,355) -- (320,345) -- cycle;
\draw[BLACK, solid, line join=round, line cap=round, line width=1]
	(305,345) -- (260,420) -- (260,420) -- (620,420) -- (689,230);
\draw[BLACK, solid, line join=round, line cap=round, line width=1, fill=BLACK]
	(689,230) -- (689,241) -- (683,238) -- (689,230) -- cycle;

    \draw (200,200) node {$S$};
      \draw (440,200) node {$E$};
         \draw (700,200) node {$I$};
            \draw (940,200) node {$R$};
             \draw (100,80) node {$\Lambda$};
             \draw (200,320) node {$\mu$};
             \draw (440,320) node {$\mu$};
             \draw (450,450) node {$2$};
             \draw (700,320) node {$\mu$};
             \draw (940,320) node {$\mu$};
             \draw (320,100) node {$\beta$};
             \draw (320,320) node {$\beta'$};
             \draw (580,200) node {$\eta$};
             \draw (820,200) node {$\alpha$};

\end{tikzpicture}
    \caption{The SEIR model of Saika-Bora.}
    \label{SEIR-BS}
\end{figure}
We apply the formula \eqref{RM} for the matrices $F$ and $V$:
    $$F=\begin{pmatrix}
0 & \beta S \\
0 & \beta' S
\end{pmatrix}\,\textrm{ and }
V=\begin{pmatrix}
\eta+\mu & 0\\
-\eta & \alpha+\mu
\end{pmatrix}\,.$$ 
The life cycle graph has two vertices and five edges as follows: 
\begin{equation}
    \xymatrix{*+[Fo]{1}\ar@/_{5mm}/[rr]_\eta\ar@(ul,dl)[]_{1-(\eta+\mu)}\ && *+[Fo]{2}\ar@/_{5mm}/[ll]_{\beta S}
  \ar@`{[]+/ul+5pc/,[]+/ur+5pc/}[]^{\beta' S}\ar@(ur,dr)[]^{1-(\alpha+\mu)}}
\end{equation}
Therefore, we have two fertility loops: $L_1=\beta S\eta$ and $L_2=\beta'S$. The basic reproduction number is calculated 
with $\det (V_{\setminus\{1,2\}})=1$, $\det (V_{\setminus\{2\}})=\eta+\mu$, and $\det (V)=(\eta+\mu)(\alpha+\mu)$; with the formula $$R_0=\frac{\beta S\eta}{(\eta+\mu)(\alpha+\mu)}+\frac{\beta' S}{\alpha+\mu}\,.$$

Instead, we consider the next-generation matrix 
$$\begin{blockarray}{ccc}
      & E & I \\
\begin{block}{c(cc)}
 E & \frac{\beta\eta S}{(\alpha+\mu)(\eta+\mu)}  & \frac{\beta S}{\alpha+\mu} \\
 I & \frac{\beta'\eta S}{(\alpha+\mu)(\eta+\mu)} & \frac{\beta' S}{\alpha+\mu} \\
\end{block}
\end{blockarray}$$
and the fertility loops correspond to the following loops in the Petri net: 

\begin{minipage}[t]{0.5\textwidth} 
        \begin{center}
        \begin{tikzpicture}[scale=0.2,x=1pt,y=-1pt]

\definecolor{BLACK}{RGB}{0,0,0}
\definecolor{YELLOW}{RGB}{255,255,0}
\draw[BLACK, solid, line join=round, line cap=round, line width=1, fill=YELLOW]
	(100,280) ellipse[x radius=45, y radius=45];
\draw[BLACK, solid, line join=round, line cap=round, line width=1, fill=YELLOW]
	(400,280) ellipse[x radius=45, y radius=45];
\draw[BLACK, solid, line join=round, line cap=round, line width=1, fill=YELLOW]
	(580,100) ellipse[x radius=55, y radius=55];
\draw[BLACK, solid, line join=round, line cap=round, line width=1, fill=YELLOW]
	(760,280) ellipse[x radius=45, y radius=45];
\definecolor{CYAN}{RGB}{0,255,255}
\draw[BLACK, solid, line join=round, line cap=round, line width=1, fill=CYAN]
	(210,250) rectangle +(60,60);
\draw[BLACK, solid, line join=round, line cap=round, line width=1, fill=CYAN]
	(70,390) rectangle +(60,60);
\draw[BLACK, solid, line join=round, line cap=round, line width=1, fill=CYAN]
	(550,250) rectangle +(60,60);
\draw[BLACK, solid, line join=round, line cap=round, line width=1, fill=CYAN]
	(370,390) rectangle +(60,60);
\draw[BLACK, solid, line join=round, line cap=round, line width=1]
	(145,280) -- (210,280);
\draw[BLACK, solid, line join=round, line cap=round, line width=1, fill=BLACK]
	(210,280) -- (200,283) -- (200,277) -- (210,280) -- cycle;
\draw[BLACK, solid, line join=round, line cap=round, line width=1]
	(100,325) -- (100,390);
\draw[BLACK, solid, line join=round, line cap=round, line width=1, fill=BLACK]
	(100,390) -- (97,380) -- (103,380) -- (100,390) -- cycle;
\draw[BLACK, solid, line join=round, line cap=round, line width=1]
	(270,280) -- (355,280);
\draw[BLACK, solid, line join=round, line cap=round, line width=1, fill=BLACK]
	(355,280) -- (345,283) -- (345,277) -- (355,280) -- cycle;
\draw[BLACK, solid, line join=round, line cap=round, line width=1]
	(445,280) -- (550,280);
\draw[BLACK, solid, line join=round, line cap=round, line width=1, fill=BLACK]
	(550,280) -- (540,283) -- (540,277) -- (550,280) -- cycle;
\draw[BLACK, solid, line join=round, line cap=round, line width=1]
	(400,325) -- (400,390);
\draw[BLACK, solid, line join=round, line cap=round, line width=1, fill=BLACK]
	(400,390) -- (397,380) -- (403,380) -- (400,390) -- cycle;
\draw[BLACK, solid, line join=round, line cap=round, line width=1]
	(610,280) -- (715,280);
\draw[BLACK, solid, line join=round, line cap=round, line width=1, fill=BLACK]
	(715,280) -- (705,283) -- (705,277) -- (715,280) -- cycle;
\draw[BLACK, solid, line join=round, line cap=round, line width=1]
	(580,155) -- (580,250);
\draw[BLACK, solid, line join=round, line cap=round, line width=1, fill=BLACK]
	(580,250) -- (577,240) -- (583,240) -- (580,250) -- cycle;

 \draw (100,280) node {$E$};

\draw (245,280) node {$\eta$};

\draw (100,420) node {$\mu$};

  \draw (400,280) node {$I$};
  \draw (580,280) node {$\beta$};

\draw (400,420) node {$\mu$};
  
   \draw (760,280) node {$E$};
   \draw (580,100) node {$S$};
 
\end{tikzpicture}
 \end{center}
\end{minipage}
\begin{minipage}[t]{0.5\textwidth} 
\begin{center}
\begin{tikzpicture}[scale=0.2,x=1pt,y=-1pt]

\definecolor{BLACK}{RGB}{0,0,0}
\definecolor{YELLOW}{RGB}{255,255,0}
\draw[BLACK, solid, line join=round, line cap=round, line width=1, fill=YELLOW]
	(400,280) ellipse[x radius=45, y radius=45];
\draw[BLACK, solid, line join=round, line cap=round, line width=1, fill=YELLOW]
	(580,100) ellipse[x radius=55, y radius=55];
\draw[BLACK, solid, line join=round, line cap=round, line width=1, fill=YELLOW]
	(760,280) ellipse[x radius=45, y radius=45];
\definecolor{CYAN}{RGB}{0,255,255}
\draw[BLACK, solid, line join=round, line cap=round, line width=1, fill=CYAN]
	(550,250) rectangle +(60,60);
\draw[BLACK, solid, line join=round, line cap=round, line width=1, fill=CYAN]
	(450,390) rectangle +(60,60);
\draw[BLACK, solid, line join=round, line cap=round, line width=1]
	(445,280) -- (550,280);
\draw[BLACK, solid, line join=round, line cap=round, line width=1, fill=BLACK]
	(550,280) -- (540,283) -- (540,277) -- (550,280) -- cycle;
\draw[BLACK, solid, line join=round, line cap=round, line width=1]
	(422,319) -- (463,390);
\draw[BLACK, solid, line join=round, line cap=round, line width=1, fill=BLACK]
	(463,390) -- (455,383) -- (461,380) -- (463,390) -- cycle;
\draw[BLACK, solid, line join=round, line cap=round, line width=1]
	(610,280) -- (715,280);
\draw[BLACK, solid, line join=round, line cap=round, line width=1, fill=BLACK]
	(715,280) -- (705,283) -- (705,277) -- (715,280) -- cycle;
\draw[BLACK, solid, line join=round, line cap=round, line width=1]
	(580,155) -- (580,250);
\draw[BLACK, solid, line join=round, line cap=round, line width=1, fill=BLACK]
	(580,250) -- (577,240) -- (583,240) -- (580,250) -- cycle;
\draw[BLACK, solid, line join=round, line cap=round, line width=1, fill=CYAN]
	(290,390) rectangle +(60,60);
\draw[BLACK, solid, line join=round, line cap=round, line width=1]
	(378,319) -- (337,390);
\draw[BLACK, solid, line join=round, line cap=round, line width=1, fill=BLACK]
	(337,390) -- (339,380) -- (345,383) -- (337,390) -- cycle;

\draw (320,420) node {$\mu$};
\draw (480,420) node {$\alpha$};

  \draw (400,280) node {$I$};
  \draw (580,280) node {$\beta'$};

   \draw (760,280) node {$I$};
   \draw (580,100) node {$S$};
 
\end{tikzpicture}
\end{center}
\end{minipage}
The basic reproduction number is given by the trace, and by Proposition \ref{exo}, we obtain 
$$R_0=\frac{\beta\eta S}{(\alpha+\mu)(\eta+\mu)}+\frac{\beta' S}{\alpha+\mu}\,.$$
\end{example}

\section{Conclusions}
The analysis of various disease models in epidemiology revealed that even though the model structure can be highly sophisticated, the structural background follows the basic SIR model. This was the objective of introducing the Kermack-McKendrick modules (susceptible/infection-process/infection). 
In addition, we provide the (G1)-(G5) assumptions for Petri nets that assure the basic reproduction number can be defined by the matrix of flows. 
In conclusion, we ended with a geometric version of the next-generation matrix method. 
This method provides an accessible manner of understanding the basic reproduction number by summing up all the flows of secondary infections.

Petri nets are efficient structures that can be used in epidemiological problems. The reconstruction of the associated Petri net requires describing a set of proposed disease dynamics that enable us to model our epidemic process.
By building geometric diagrams composed of substructures Petri nets, such as Malthusian models, SIR type models, mitigation strategies, and vaccination adjustments, among others, they provide a useful form of interaction in the design of models in epidemiology in a geometrical manner that recovers the system of ordinary differential equations using the rate equation. 
Reciprocally, we can start with a system of ordinary differential equations associated with an epidemiological model. 
The correspondence between ODEs and Petri nets developed in this work makes it possible to associate a Petri net in which rate equations recover the original ODEs. We found these two interactions between the diagrammatic model of a Petri net and its associated ODE beneficial.
We expect this to motivate developments in the area, like geometric interpretations of other important concepts, such as herd immunity.

Finally, the geometric next-generation matrix method produces precise forms of the matrices $F$ and $V$. Subsequently, we can find the basic reproduction number $R_0$ analytically or use the geometric method by de Camino-Beck--Lewis with the life cycle graph. If the next-generation matrix has rank one, the trace provides the sum of all the loops inside the infection modules that traverse the infection-process module once.

\noindent

{\bf Conflict of interest} The author declares no conflicts of interest.

\bibliographystyle{amsalpha}
\bibliography{biblio}

\begin{sidewaystable}\label{Tabla1}
\scriptsize
\centering
\begin{tabular}{ C{2.1cm} C{6cm} C{3.1cm} C{3.1cm} C{5.1cm} C{3.1cm} }    
\toprule
\tiny{MODEL}  &   \tiny{ODE}    &   \tiny{F-Matrix}    &   \tiny{V-Matrix}    &   \tiny{Next Generation Matrix}   &     \tiny{$R_0$}  \\   \midrule
SIR & 
$$\begin{array}{l}
S'(t)=-\beta SI,\\
I'(t)=\beta SI-\alpha I,\\
R'(t)=\alpha I\,.
\end{array}$$
& $(\beta S)$ & $(\alpha)$ & $\left(\frac{\beta S}{\alpha}\right)$ & $\frac{\beta S}{\alpha}$ \\\hline
SIS & 
$$\begin{array}{l}
S'(t)=-\beta SI+\alpha I\,,\\
I'(t)=\beta SI-\alpha I\,.
\end{array}$$
& $(\beta S)$ & $(\alpha)$ & $\left(\frac{\beta S}{\alpha}\right)$ &$\frac{\beta S}{\alpha}$ \\\hline
SIRS & 
$$\begin{array}{l}
S'(t)=-\beta SI+\gamma R,\\
I'(t)=\beta SI-\alpha I,\\
R'(t)=\alpha I-\gamma R\,.
\end{array}$$
& $(\beta S)$ & $(\alpha)$ & $\left(\frac{\beta S}{\alpha}\right)$ & $\frac{\beta S}{\alpha}$ \\\hline
 SEIR\cite{Ma} &  
 $$\begin{array}{l}
S'(t)=\Lambda -\beta SI-\mu S,\\
E'(t)=\beta SI-(\eta +\mu )E,\\
I'(t)=\eta E-(\alpha+\mu) I,\\ 
R'(t)=\alpha I-\mu R\,.
\end{array}$$
 & 
 $$\begin{pmatrix}
0 & \beta S\\
0 & 0
\end{pmatrix}$$ 
 & 
 $$\begin{pmatrix}
\eta+\mu & 0\\
-\eta & \alpha+\mu
\end{pmatrix}$$
 & 
 $$\begin{pmatrix}
\frac{\beta\eta S}{(\eta+\mu)(\alpha+\mu)} & \frac{S\beta}{\alpha+\mu}\\
0 & 0
\end{pmatrix}$$  
 & $\frac{\beta\eta S}{(\eta+\mu)(\alpha+\mu)}$ \\\hline
SEIR\cite{SB} &  
$$\begin{array}{l}
S'(t)=\Lambda -(\beta+\beta')SI-\mu S,\\
E'(t)=\beta SI-(\eta +\mu )E,\\
I'(t)=\eta E-(\alpha+\mu) I+\beta'SI,\\ 
R'(t)=\alpha I-\mu R\,.
\end{array}$$  
& 
$$\begin{pmatrix}
0 & \beta S \\
0 & \beta' S
\end{pmatrix}$$ 
& 
$$\begin{pmatrix}
\eta+\mu & 0\\
-\eta & \alpha+\mu
\end{pmatrix}$$ 
& 
$$\begin{pmatrix}
\frac{\beta\eta S}{(\alpha+\mu)(\eta+\mu)} & \frac{\beta S}{\alpha+\mu}\\
\frac{\beta'\eta S}{(\alpha+\mu)(\eta+\mu)} & \frac{\beta' S}{\alpha+\mu}
\end{pmatrix}$$
&
  $\frac{\beta\eta S}{(\alpha+\mu)(\eta+\mu)}+\frac{\beta' S}{\alpha+\mu}$\\\hline
   SEAIR\cite{Ma} &  
   $$\begin{array}{l}
S'(t)=\Lambda -\beta S(I+qA)-\mu S,\\
E'(t)=\beta S(I+qA)-(\eta +\mu )E,\\
A'(t)=(1-p)\eta E-(\gamma+\mu)A,\\
I'(t)=p\eta E-(\alpha+\mu) I,\\
R'(t)=\alpha I+\gamma A-\mu R\,.
\end{array}$$
   & $$\begin{pmatrix}
0 & \beta q S & \beta S\\
0 & 0 & 0 \\
0 & 0 & 0
\end{pmatrix}$$
   & $$\begin{pmatrix}
\eta+\mu & 0 & 0\\
\eta (p-1) & \gamma+\mu & 0 \\
-\eta p & 0 & \alpha+\mu
\end{pmatrix}$$
   & $$\begin{pmatrix}
\frac{\beta \eta (1-p)qS}{(\eta+\mu)(\gamma+\mu)}+\frac{\beta\eta p S}{(\eta+\mu)(\alpha+\mu)} & \frac{\beta q S}{\gamma+\mu} & \frac{\beta S}{\alpha+\mu}\\
0 & 0 & 0 \\
0 & 0 & 0
\end{pmatrix}$$
   & $\frac{\beta \eta (1-p)qS}{(\eta+\mu)(\gamma+\mu)}+\frac{\beta\eta p S}{(\alpha+\mu)(\eta+\mu)}$ \\\hline
  SEAIR\cite{Bar} &  
  $$\begin{array}{l}
S'(t)=-\beta_1 SA-\beta_2 SI,\\
E'(t)=\beta_1 SA + \beta_2 SI-\gamma E,\\
A'(t)=\gamma  E-(\sigma +\mu)A,\\
I'(t)=\sigma A-\mu I,\\
R'(t)=\mu A + \mu I\,.
\end{array}$$
  &  
  $$\begin{pmatrix}
0 & \beta_1 S & \beta_2 S\\
0 & 0 & 0 \\
0 & 0 & 0
\end{pmatrix}$$
  & 
  $$\begin{pmatrix}
\gamma & 0 & 0\\
-\gamma & \mu+\sigma & 0 \\
0 & -\sigma & \mu
\end{pmatrix}$$
  & 
  $$\begin{pmatrix}
\frac{\beta_1S}{\mu+\sigma}+\frac{\beta_2\sigma S}{(\mu+\sigma)\mu} & \frac{\beta_1S}{\mu+\sigma}+\frac{\beta_2\sigma S}{(\mu+\sigma)\mu} & \frac{\beta_2 S}{\mu}\\
0 & 0 & 0 \\
0 & 0 & 0
\end{pmatrix}$$
  & $\frac{\beta_1S}{\mu+\sigma}+\frac{\beta_2\sigma S}{(\mu+\sigma)\mu}$ \\\hline
\bottomrule 
\end{tabular}
\end{sidewaystable}

\begin{sidewaystable}
\scriptsize
\centering
\begin{tabular}{ C{1.9cm} C{6.5cm} C{3.1cm} C{3.1cm} C{5.1cm} C{3.1cm} }    
\toprule
\tiny{MODEL}  &   \tiny{ODE}    &   \tiny{F-Matrix}    &   \tiny{V-Matrix}    &   \tiny{Next Generation Matrix}   &     \tiny{$R_0$}  \\   \midrule
  SCIRS\cite{Ma} &  
  $$\begin{array}{l}
S'(t)=\Lambda -\beta S(I+qC)-\mu S+\rho R,\\
C'(t)=\beta S(I+qC)-(\eta +\gamma+\mu )C,\\
I'(t)=\eta C-(\alpha+\mu) I,\\
R'(t)=\alpha I+\gamma C-(\mu+\rho) R\,.
\end{array}$$
  & 
  $$\begin{pmatrix}
\beta S q & \beta S\\
0 & 0
\end{pmatrix}$$ 
  & 
  $$\begin{pmatrix}
\eta+\gamma+\mu & 0\\
-\eta & \alpha+\mu
\end{pmatrix}$$
  & 
  $$\begin{pmatrix}
\frac{\beta q S }{\eta+\gamma+\mu}+\frac{\beta\eta S}{(\alpha+\mu)(\eta+\gamma+\mu)} & \frac{\beta S}{\alpha+\mu}\\
0 & 0
\end{pmatrix}$$
  & $\frac{\beta q S }{\eta+\gamma+\mu}+\frac{\beta\eta S}{(\alpha+\mu)(\eta+\gamma+\mu)}$ \\\hline
 SIWR\cite{WBK} &  
 $$\begin{array}{l}
S'(t)=\mu N-\beta_W SW -\beta SI -\mu S\,,\\
I'(t)=\beta_W SW+\beta SI -(\gamma+\mu)I\,,\\
W'(t)=\alpha I-\xi W\,,\\
R'(t)=\gamma I-\mu R\,.
\end{array}$$
 & 
 $$\begin{pmatrix}
\beta S & \beta_W S\\
0 & 0
\end{pmatrix}$$
 & 
 $$\begin{pmatrix}
\gamma+\mu & 0\\
-\alpha & \xi
\end{pmatrix}$$
 &
 $$\begin{pmatrix}
\frac{\beta S}{\gamma+\mu} +\frac{\alpha\beta_WS}{(\gamma+\mu)\xi}& \frac{\beta_W S}{\xi}\\
0 & 0
\end{pmatrix}$$
 & $\frac{\beta S}{\gamma+\mu} +\frac{\alpha\beta_WS}{(\gamma+\mu)\xi}$\\\hline
  SIQR\cite{Ma} &  
  $$\begin{array}{l}
S'(t)=\Lambda -\beta SI/A-\mu S,\\
I'(t)=\beta SI/A-(\alpha+\gamma+\mu)I\\
Q'(t)=\gamma I-(\eta+\mu) Q,\\
R'(t)=\alpha I+\eta Q-\mu R\,.
\end{array}$$
  &
  $$\begin{pmatrix}
\beta S  & 0\\
0 & 0
\end{pmatrix}$$  
  & 
  $$\begin{pmatrix}
\alpha+\gamma+\mu & 0\\
-\gamma & \eta+\mu
\end{pmatrix}$$ 
  & 
  $$\begin{pmatrix}
\frac{\beta S}{\alpha+\gamma+\mu}  & 0\\
0 & 0
\end{pmatrix}$$
  & $\frac{\beta S}{\alpha+\gamma+\mu}$ \\\hline
  Malaria\cite{WBK}&  
  $$\begin{array}{l}
S'_H(t)=\Pi-\beta_{HM}S_HI_M-\mu_HS_H\,,\\
I'_H(t)=\beta_{HM}S_HI_M-(\mu_H+\alpha+\sigma)I_H+ \delta I_H\,,\\
R'_H(t)=\sigma I_H-\mu_HR_H\,,\\
S'_M(t)=\Lambda-\beta_{MH}S_MI_H-\mu_MS_M\,,\\
I'_M(t)=\beta_{MH}S_MI_H-\mu_MI_M\,.
\end{array}$$
  & 
  $$\begin{pmatrix}
0 & \beta_{HM} S_H\\
\beta_{MH} S_M & 0
\end{pmatrix}$$
  & 
  $$\begin{pmatrix}
\alpha-\delta+\mu_H+\sigma&0\\
0 & \mu_M
\end{pmatrix}$$
  & 
  $$\begin{pmatrix}
0 & \frac{\beta_{HM} S_H}{\mu_M}\\
\frac{\beta_{MH} S_M}{\alpha-\delta+\mu_H+\sigma} & 0
\end{pmatrix}$$
  & $\sqrt{\frac{\beta_{HM}\beta_{MH}S_HS_M}{(\alpha-\delta+\mu_H+\sigma)\mu_M}}$ \\\hline
\bottomrule 
\end{tabular}
\end{sidewaystable}

\end{document}